%% file: tools.tex
\documentclass{article}

\usepackage[T1]{fontenc}
\usepackage{textgreek}
\usepackage{a4wide}
\usepackage{amsmath,amssymb}
\usepackage{slashed}
\usepackage{xcolor}
\usepackage{graphicx}
\usepackage{xspace}
\usepackage{url}
\usepackage{cite}
\usepackage{listings}
\usepackage{underscore}
\usepackage[colorlinks=true,allcolors=darkred,pdfborder={0 0 0}]{hyperref}
\usepackage{geometry}
\def\hc{\text{h.c.}}

\newcommand\SARAH{{\tt SARAH}\xspace}

\newcommand\MG{{\tt MadGraph}\xspace}

\newcommand\FeynArts{{\tt FeynArts}\xspace}

\newcommand\CalcHep{{\tt CalcHep}\xspace}
\newcommand\CompHep{{\tt CompHep}\xspace}
\newcommand\WHIZARD{{\tt WHIZARD}\xspace}
\newcommand\OMEGA{{\tt O'Mega}\xspace}
\newcommand\SPheno{{\tt SPheno}\xspace}
\newcommand\Vevacious{{\tt Vevacious}\xspace}
\newcommand\MO{{\tt MicrOmegas}\xspace}

\newcommand\UFO{{\tt UFO}\xspace}
\newcommand\FlavorKit{{\tt FlavorKit}\xspace}

\newcommand\Susyno{{\tt Susyno}\xspace}
\newcommand\Mathematica{{\tt Mathematica}\xspace}
\newcommand\Python{{\tt Python}\xspace}
\newcommand{\Fortran}{\texttt{Fortran}\xspace}
\newcommand\FeynRules{{\tt FeynRules}\xspace}

\newcommand{\vsp}{\vspace*{5mm}}

\newcommand{\tip}[1]{\begin{center} \noindent \fbox{\begin{minipage}{0.7\textwidth} {\bf TIP:} #1 \end{minipage}} \end{center}}

\definecolor{darkred}{rgb}{0.6,0,0}

\lstdefinestyle{mathematica}{
        basicstyle=\ttfamily\mdseries,
	language=bash,
	frame=false,
	xleftmargin=.25in}   

\lstdefinestyle{terminal}{
	language=bash,
	frame=lines,
	xleftmargin=.5in,
        numbers=none}
        
\lstdefinestyle{file}{
        basicstyle=\ttfamily\mdseries,
	language=bash,
	frame=shadowbox,
        numbers=left,   
        numberstyle=\tiny} 

\lstdefinestyle{scotogenic}{
        basicstyle=\ttfamily\mdseries,
	language=bash,
	frame=shadowbox,
        title=\hspace{14cm}{\tt Scotogenic.m},	
        numbers=left,   
        numberstyle=\tiny}

\lstdefinestyle{darkbs}{
        basicstyle=\ttfamily\mdseries,
	language=bash,
	frame=shadowbox,
        title=\hspace{15cm}{\tt DarkBS.m},	
        numbers=left,   
        numberstyle=\tiny}

\lstdefinestyle{parameters}{
        basicstyle=\ttfamily\mdseries,
	language=bash,
	frame=shadowbox,
        title=\hspace{14.2cm}{\tt parameters.m},	
        numbers=left,   
        numberstyle=\tiny}

\lstdefinestyle{particles}{
        basicstyle=\ttfamily\mdseries,
	language=bash,
	frame=shadowbox,
        title=\hspace{14.4cm}{\tt particles.m},	
        numbers=left,   
        numberstyle=\tiny}

\lstdefinestyle{SPheno}{
        basicstyle=\ttfamily\mdseries,
	language=bash,
	frame=shadowbox,
        title=\hspace{14.8cm}{\tt SPheno.m},	
        numbers=left,   
        numberstyle=\tiny}

\lstdefinestyle{LesHouches}{
        basicstyle=\ttfamily\mdseries,
	language=bash,
	frame=shadowbox,
        title=\hspace{12cm}{\tt LesHouches.in.Scotogenic},	
        numbers=left,   
        numberstyle=\tiny}

\lstdefinestyle{LesHouchesDarkBS}{
        basicstyle=\ttfamily\mdseries,
	language=bash,
	frame=shadowbox,
        title=\hspace{12.8cm}{\tt LesHouches.in.DarkBS},	
        numbers=left,   
        numberstyle=\tiny}

\lstdefinestyle{LesHouchesOut}{
        basicstyle=\ttfamily\mdseries,
	language=bash,
	frame=shadowbox,
        title=\hspace{12.5cm}{\tt SPheno.spc.Scotogenic},	
        numbers=left,   
        numberstyle=\tiny}

\lstdefinestyle{LesHouchesOutDarkBS}{
        basicstyle=\ttfamily\mdseries,
	language=bash,
	frame=shadowbox,
        title=\hspace{13.3cm}{\tt SPheno.spc.DarkBS},	
        numbers=left,   
        numberstyle=\tiny}

\lstdefinestyle{mg5conf}{
        basicstyle=\ttfamily\mdseries,
	language=bash,
	frame=shadowbox,
        title=\hspace{12.8cm}{\tt mg5\underline{\space}configuration.txt},	
        numbers=left,   
        numberstyle=\tiny}

\lstdefinestyle{bashrc}{
        basicstyle=\ttfamily\mdseries,
	language=bash,
	frame=shadowbox,
        title=\hspace{14.2cm}{\tt bashrc},	
        numbers=none,   
        numberstyle=\tiny}

\lstdefinestyle{maFile}{
        basicstyle=\ttfamily\mdseries,
	language=bash,
	frame=shadowbox,
        title=\hspace{13.8cm}{\tt plotDarkBS.txt},	
        numbers=left,   
        numberstyle=\tiny}

\lstdefinestyle{runcard}{
        basicstyle=\ttfamily\mdseries,
	language=bash,
	frame=shadowbox,
        title=\hspace{14.2cm}{\tt run\underline{\space}card.dat},	
        numbers=left,   
        numberstyle=\tiny}

\begin{document}


\begin{center}
\vspace*{15mm}

\vspace{1cm}
{\Large \bf 
Computer tools in particle physics
} \\
\vspace{1cm}

{\bf A. Vicente$^{a,b}$}

 \vspace*{.5cm} 
$^{a}$ Instituto de F\'{\i}sica Corpuscular (CSIC-Universitat de Val\`{e}ncia), \\
Apdo. 22085, E-46071 Valencia, Spain
\vspace*{.2cm} 

$^{b}$ IFPA, Dep. AGO, Universit\'e de Li\`ege, \\
Bat B5, Sart-Tilman B-4000 Li\`ege 1, Belgium

\end{center}

\vspace*{10mm}

\begin{abstract}\noindent\normalsize
The field of particle physics is living very exciting times with a
plethora of experiments looking for new physics in complementary
ways. This has made increasingly necessary to obtain precise
predictions in new physics models in order to be ready for a discovery
that might be just around the corner. However, analyzing new models
and studying their phenomenology can be really challenging. Computing
mass matrices, interaction vertices and decay rates is already a
tremendous task. In addition, obtaining predictions for the dark
matter relic density and its detection prospects, computing flavor
observables or estimating the LHC reach in certain collider signals
constitutes quite a technical work due to the precision level that is
currently required. For this reason, computer tools such as \SARAH,
\MO, \MG, \SPheno or \FlavorKit have become quite popular,
and many physicists use them on a daily basis. In this course we will
learn how to use these computer tools to explore new physics models
and get robust numerical predictions to probe them in current and
future experiments.
\end{abstract}

\vspace*{2cm}

\begin{center}
\begin{minipage}{0.7\textwidth}
\centering
\it
Notes of the mini-course `Computer tools in particle physics', \\
first given at CINVESTAV, Mexico \\
C\'atedra Augusto Garc\'ia Gonz\'alez, June 22nd-26th, 2015.
\end{minipage}
\end{center}

\newpage

\tableofcontents

\newpage

\input{intro}

\input{lecture1}

\input{lecture2}

\input{lecture3}

\input{lecture4}

\input{summary}

\section*{Acknowledgements}

I thank Cinvestav for giving me the opportunity to give these lectures
and all the students and attendants to the course for their presence
and fruitful questions. I am very grateful to Omar Miranda, for his
kind invitation, and to Diego Aristiz\'abal Sierra, Florian Staub and
Nicol\'as Rojas for their collaboration on the implementation of the
models discussed in this course. I also acknowledge fruitful
discussions with Diego Restrepo, Alexander Merle and Takashi Toma on
some aspects of the scotogenic model relevant for this
course. Finally, I feel indebted to Florian Staub for his continuous
technical assistance with \SARAH and related issues.

\newpage

\input{appendix}

\bibliographystyle{utphys}

\providecommand{\href}[2]{#2}\begingroup\raggedright

\endgroup

\end{document}

%% file: intro.tex
\section{Introduction}
\label{sec:intro}

When Wolfgang Pauli addressed his famous letter to the
\emph{Radioactive Ladies and Gentlemen} who met in T\"ubingen to
discuss beta decay and other hot topics of the moment, inventing new
particles was not a popular solution for a fundamental problem. In
fact, a feeling of unease was clearly present in the text. In 1930,
only three types of particles were already discovered: electrons,
protons and photons. Adding a new particle, especially one with so
weak interactions, was seen as a \emph{desperate remedy}. However,
these particles, later known as \emph{neutrinos}, were indeed the
solution for the problem of the continuous beta decay spectrum.

The situation is completely different nowadays. New particles are
proposed every day in order to solve current problems in particle
physics. Bosons or fermions, light or heavy, weakly or strongly
interacting, these new hypothetical particles appear on arXiv on a
daily basis. They are part of new models, some of them quite involved,
proposed to address some of the issues that our current theoretical
ideas, materialized in the Standard Model (SM), cannot explain.

With such an explosion of new models and particles, a systematic study
of new physics scenarios has become necessary. In order to achieve
this goal, many computer tools have been developed. They replace the
\emph{dirty work} traditionally given to PhD students, who spent long
and tedious periods performing lengthy calculations or typing
numerical codes. Instead, many of these tasks can nowadays be
automatized, allowing physicists to concentrate on new ideas, rather
than on technical details of complex computations. Furthermore, this
also makes possible to run precise calculations, as currently required
in order to test our theoretical expections at the accuracy level
delivered by the experiments.

\subsubsection*{What is this course about?}

In this course we will learn how to use several computer tools widely
employed nowadays for phenomenological studies: \SARAH, \SPheno, \MO
and \MG. We do not aim at a complete review of these tools, but just
intend to provide a basic introduction. Nevertheless, after the course
has been finished, we will be able to explore new physics models,
study their properties analytically, obtain numerical results for many
physical observables and run dark matter and collider studies.

\subsubsection*{Versions used for this course}

These notes were elaborated using the following versions of the
computer codes presented in the course:

\begin{itemize}

\item {\tt SARAH-4.11.0}
\item {\tt SPheno-4.0.2}
\item {\tt micromegas_4.2.5}
\item {\tt MG5_aMC_v2_5_4}
\item {\tt MadAnalysis5_v1.2 (patch 4)}

\end{itemize}

\subsubsection*{Two messages}

Before we get started, let me tell you two important messages:

\begin{itemize}
\item {\bf It is not so hard!}
\end{itemize}

One of the main goals of this course is to convince you that most of
the current computer tools in particle physics are easy to use. Of
course, it takes a lot of training to become a master, but getting
started with them is actually quite easy. We will see that in some
cases it suffices with a basic introduction to be able to produce
valuable results.
 
\begin{itemize}
\item {\bf Do not trust (too much) in codes!}
\end{itemize}

We must be careful when working with complicated computer codes. There
is no such thing as a bug free software, and this also applies to
particle physics tools. Therefore, we must analyze the results we
obtain with a critical eye, trying to make sense of them using our
physical intuition. Otherwise, we might make terrible mistakes simply
by relying on buggy codes.\\

And now let us get started!


\newpage

%% file: lecture1.tex
\section{Lecture 1: Exploring new models with \SARAH}
\label{sec:lecture1}

In the first lecture we will learn how to use \SARAH to explore a new
model, study its properties and obtain input files for other computer
tools.

\subsection{What is \SARAH?}
\label{subsec:SARAH}

\SARAH is a \Mathematica package for building and analyzing particle
physics models. Although originally \SARAH was designed to work only
with supersymmetric models, after version 3 non-supersymmetric ones
can be implemented as well. Once a model is defined in \SARAH, the
user can get all sorts of details about it: all vertices, mass
matrices, tadpoles equations, 1-loop corrections for tadpoles and
self-energies, and 2-loop renormalization group equations (RGEs). All
this information about the model is derived by \SARAH analytically and
the user can simply handle it in \Mathematica and use it for his own
purposes. Furthermore, \SARAH can export these analytical expressions
into \LaTeX\ files which, after the usual \LaTeX\ compilation, result
in pdf files containing all the details of the model.

\tip{The \LaTeX\ output given by \SARAH is quite handy when preparing
  a paper. One can simply copy the relevant information, thus avoiding
  the tedious task of typing analytical formulas into \LaTeX.}

\SARAH would already be a very useful tool just with the features
described so far. However, there is more. \SARAH writes input model
files for \FeynArts \cite{Hahn:2000kx}, \CalcHep/\CompHep
\cite{Pukhov:2004ca,Boos:1994xb} (which can also be used as input for
\MO), the \UFO format which is supported by \MG, as well as for
\WHIZARD \cite{Kilian:2007gr} and \OMEGA \cite{Moretti:2001zz}. As we
will see in Sec. \ref{subsec:SARAH-output}, this will save us a lot of
time when we want to implement our favourite model in \MO or \MG,
since \SARAH can produce the required input files without us having to
write them by hand.

Finally, let us briefly comment on some other interesting
possibilities \SARAH offers. It is well-known that \Mathematica is not
very efficient when dealing with heavy numerical calculations. For
this reason, \SARAH creates source code for \SPheno, a code written in
\Fortran that allows for an efficient numerical evaluation of all the
analytical expressions derived with \SARAH. Other interesting features
include the \FlavorKit functionality for the calculation of flavor
observables, the output for \Vevacious, which can be used to check for
the global minimum of the scalar potential of a given model, and the
link to \Susyno for the handling of group theoretical functions. For a
complete and detailed descrition of \SARAH and its many
functionalities we refer to the manual and the recent pedagogical
review \cite{Staub:2015kfa}.

\subsection{\SARAH: Technical details, installation and load}
\label{subsec:SARAH-details}

\begin{itemize}

\item {\bf Name of the tool:} \SARAH

\item {\bf Author:} Florian Staub (florian.staub@cern.ch)

\item {\bf Type of code:} \Mathematica package

\item {\bf Website:} \url{http://sarah.hepforge.org/}

\item {\bf Manual:}
  \cite{Staub:2008uz,Staub:2009bi,Staub:2010jh,Staub:2012pb,Staub:2013tta}. For
  related functionalities see \cite{Staub:2011dp,Porod:2014xia},
  whereas for a pedagogical overview we recommend
  \cite{Staub:2015kfa}.

\end{itemize}

\SARAH does not require any compilation. After downloading the
package, one can simply copy the {\tt tar.gz} file to the directory
{\tt \$PATH}, where it can be extracted:

\vsp
\begin{lstlisting}[style=terminal]
$ cp Download-Directory/SARAH_X.Y.Z.tar.gz $PATH/
$ cd $PATH
$ tar -xf SARAH_X.Y.Z.tar.gz
\end{lstlisting}
\vsp

Here {\tt X.Y.Z} must be replaced by the \SARAH version which has been
downloaded. \SARAH can be used with any \Mathematica version between 7
and 10.

In order to load \SARAH one has to run in \Mathematica the following
command:

\vsp
\begin{lstlisting}[style=mathematica]
<<$PATH/SARAH-X.Y.Z/SARAH.m;
\end{lstlisting}
\vsp

And now we are ready to use \SARAH. For example, in order to
initialize a model we just have to use the command

\vsp
\begin{lstlisting}[style=mathematica]
Start[model];
\end{lstlisting}
\vsp

Here {\tt model} is the name of the specific particle physics model we
want to explore. However, before we do that, let us see how to define
a new model in \SARAH.

\subsection{Defining a model in \SARAH}
\label{subsec:SARAH-model}

There are already many models fully implemented in \SARAH. A complete
list is provided in Appendix \ref{sec:appendix1}. If you want to study
one of those, you can skip this section and go directly to
Sec. \ref{subsec:SARAH-exploring}.

We will now learn how to define a new model in \SARAH. For this
purpose, we have to visit the directory {\tt
  \$PATH/SARAH-X.Y.Z/Models}, where each folder contains the model
files for a specific model. For example, the folder {\tt SM} contains
the model files for the Standard Model, the folder {\tt THDM-II} those
for the Two Higgs Doublet Model of type-II and the folder {\tt MSSM}
those for the Minimal Supersymmetric Standard Model. In each of these
directories one finds four files:

\begin{itemize}

\item {\tt model.m} (where {\tt model} must be replaced by the name of
  the specific model): This file contains the basic definitions of the
  model.

\item {\tt parameters.m}: In this file we provide additional
  information about the parameters of the model.

\item {\tt particles.m}: This one is devoted to the particles in the
  model, with some details not present in {\tt model.m}.

\item {\tt SPheno.m}: This file is only required if we want to create
  a \SPheno module for our model, as shown in
  Sec. \ref{subsec:SARAH-output}.

\end{itemize}

We will now show how to prepare these files for an example model: the
\emph{scotogenic model}~\cite{Ma:2006km}. This popular particle
physics model is described in detail in Appendix
\ref{sec:appendix3}. For other physics conventions and basic
definitions we also refer to the SM description in Appendix
\ref{sec:appendix2}.

Before we move on with the scotogenic model let us make a comment on
Supersymmetry. In this course we will concentrate on
non-supersymmetric models. The implementation of supersymmetric models
is not very different, but it has a few details which would make the
explanation a little more involved. In case you are interested in
supersymmetric models, we recommend the guide \cite{Staub:2015kfa}.

Let us then implement the scotogenic model in \SARAH. This model does
not have too many ingredients beyond the SM, and thus this will be an
easy task. First of all, we must create a new folder in {\tt
  \$PATH/SARAH-X.Y.Z/Models} called {\tt Scotogenic}. Then we can copy
the files from the SM implementation (located in the directory {\tt
  \$PATH/SARAH-X.Y.Z/Models/SM}), rename {\tt SM.m} to {\tt
  Scotogenic.m}, and then add the new elements (particles and
interactions) to the four model files. We will now show the result of
doing this for the scotogenic model.

\tip{It is convenient not to create a new model from scratch. Instead,
  it is highly recommended to use a model that is already implemented
  in \SARAH as basis for the implementation and simply add the new
  ingredients (particles and parameters). This way we avoid making
  unnecessary typos. Moreover, most of the new fields and interaction
  terms that we may consider for our own models are already introduced
  in the models distributed with \SARAH. In most cases we can simply
  copy these elements to our model.}

\tip{All models files are of the type *.m. We can edit them using
  \Mathematica, but I personally prefer to use a conventional text
  editor (like {\tt emacs} or {\tt gedit}).}

All \SARAH model files for the scotogenic model can be found in
Appendix \ref{sec:SARAH-scotogenic}.

\subsubsection*{Scotogenic.m}

This is the central model file. Here we define the particles of the
model, the Lagrangian, the gauge and global symmetries and the way
they get broken. If this model file is properly written, \SARAH can
already make lots of useful computations for us, making the other
files optional.

The first lines of the file contain some general \Mathematica commands
that might be useful. In our case we have

\begin{lstlisting}[style=scotogenic,firstnumber=1]
Off[General::spell]
\end{lstlisting}

which simply switches off \Mathematica warnings when the name of a new
symbol is similar to the name of existing internal symbols. This is of
course optional, but it is usually useful to get rid of these unwanted
messages. It follows some general information about the model: its
name, the authors of the implementation and the date of the
implementation:

\begin{lstlisting}[style=scotogenic,firstnumber=3]
Model`Name      = "Scotogenic";
Model`NameLaTeX = "Scotogenic Model";
Model`Authors   = "N. Rojas, A. Vicente";
Model`Date      = "2015-04-28";

(* "28-04-2015 (first implementation)" *)
(* "25-05-2015 (removed mixings in scalar sector)" *)
(* "10-06-2015 (fixed conventions)" *)
\end{lstlisting}

The first name ({\tt Model`Name}) is the internal name that will be
used in \SARAH and should not contain any special character. The
second name ({\tt Model`NameLaTeX}) is the complete name of the model
in \LaTeX\ syntax. Notice that we have added three comments (not
relevant for \SARAH) just to keep track of the last modification in
the model files. Comments are of course accepted in the model files
and they can be introduced as usual in \Mathematica by using {\tt (*
  comment *)}. As for any code, they are welcome, since they clarify
the different parts of the model files and help us when we try to
understand the code. After these basic details of the model we must
define the symmetries of the model: global and gauge. \SARAH supports
$\mathbb{Z}_N$ as well as $U(1)$ global symmetries. These are defined
by means of the array {\tt Global}. The first element of the array is
the type of symmetry, whereas the second element is the name. In the
case of the scotogenic model we have a $\mathbb{Z}_2$
parity. Therefore,

\begin{lstlisting}[style=scotogenic,firstnumber=16]
Global[[1]] = {Z[2], Z2};
\end{lstlisting}

It is the turn for the gauge symmetry of the model. In the scotogenic
model this is just the SM gauge symmetry, defined in \SARAH as

\begin{lstlisting}[style=scotogenic,firstnumber=19]
Gauge[[1]]={B,   U[1], hypercharge, g1, False, 1};
Gauge[[2]]={WB, SU[2], left,        g2, True , 1};
Gauge[[3]]={G,  SU[3], color,       g3, False, 1};
\end{lstlisting}

Each gauge group is given with an array called {\tt Gauge}. The first
element of the array is the name of the gauge boson, the second
element is the gauge group, the third element is the name of the group
and the fourth one is the name of the gauge coupling. The fifth entry
in the array can be either {\tt True} or {\tt False}, and sets whether
the gauge indices for this group must be expanded in the analytical
expressions. For $U(1)_Y$ this is not relevant, and we just set it to
{\tt False}. For $SU(2)_L$ it is convenient to expand the analytical
expressions in terms of the elements of the $SU(2)_L$ multiplets, and
thus we use {\tt True}. Since $SU(3)_c$ will not get broken, the
elements in the multiplets will always appear together, and thus it is
preferable to use {\tt False}. Finally, the last entry of the array
sets the global charge of the gauge bosons. In this case all gauge
bosons are positively charged under $\mathbb{Z}_2$.

The next step is the definition of the particle content of the
model. First, the fermion fields:

\begin{lstlisting}[style=scotogenic,firstnumber=24]
FermionFields[[1]] = {q , 3, {uL, dL},     1/6, 2,  3, 1};
FermionFields[[2]] = {l , 3, {vL, eL},    -1/2, 2,  1, 1};
FermionFields[[3]] = {d , 3, conj[dR],     1/3, 1, -3, 1};
FermionFields[[4]] = {u , 3, conj[uR],    -2/3, 1, -3, 1};
FermionFields[[5]] = {e , 3, conj[eR],       1, 1,  1, 1};
FermionFields[[6]] = {n , 3, conj[nR],       0, 1,  1,-1};
\end{lstlisting}

Each {\tt FermionFields} array corresponds to a fermionic gauge
multiplet. The first entry is the name of the fermion, the second the
number of generations and the third the name of the $SU(2)_L$
components. The rest of entries are the charges under the gauge and
global symmetries. For example, the first fermion multiplet, {\tt
  FermionFields[[1]]} is the SM quark doublet $q$, with three
generations and decomposed in $SU(2)_L$ components as
\begin{equation}
q = \left( \begin{array}{c}
u_L \\
d_L \end{array} \right) \, .
\end{equation}
The charges under $U(1)_Y \times\, SU(2)_L \times\, SU(3)_c$ are
$(\frac{1}{6},{\bf 2},{\bf 3})$, and the charge under the global
$\mathbb{Z}_2$ is $+1$. Note that the only fermion negatively charged
under $\mathbb{Z}_2$ is the right-handed neutrino, $n$. It is also
important to notice that all fermions have to be defined as
left-handed. For example, the $SU(2)_L$ singlets are identified as $d
\equiv d_R^\ast$, $u \equiv u_R^\ast$, $e \equiv e_R^\ast$ and $n
\equiv \nu_R^\ast$.

We now introduce the scalar fields of the model,

\begin{lstlisting}[style=scotogenic,firstnumber=31]
ScalarFields[[1]] =  {H,  1, {Hp, H0},     1/2, 2,  1,  1};
ScalarFields[[2]] =  {Et, 1, {etp,et0},    1/2, 2,  1, -1};
\end{lstlisting}

\noindent which follow exactly the same conventions as for the
fermions. With these two lines we have defined the SM Higgs doublet
$H$ and the inert doublet $\eta$.

After the matter content of the model is introduced, we must define
two sets of states: {\tt GaugeES} and {\tt EWSB}.

\begin{lstlisting}[style=scotogenic,firstnumber=36]
NameOfStates={GaugeES, EWSB};
\end{lstlisting}

The first set is composed by the gauge eigenstates, whereas the second
is composed by the mass eigenstates after electroweak symmetry
breaking (EWSB). For the scotogenic model these two sets are
sufficient, but in some models we may consider some intermediate
basis. Therefore, the array {\tt NameOfStates} can be longer if
necessary.

The time has come to define the Lagrangian. In \SARAH, all kinetic
terms are supposed to be canonical and thus there is no need to define
them. For the rest, the mass and interaction terms, there is. In the
scotogenic model this can be done as follows:

\begin{lstlisting}[style=scotogenic,firstnumber=40]
DEFINITION[GaugeES][LagrangianInput]= 
{
  {LagFer   ,      {AddHC->True}},
  {LagNV    ,      {AddHC->True}},
  {LagH     ,      {AddHC->False}},
  {LagEt    ,      {AddHC->False}},
  {LagHEt   ,      {AddHC->False}},
  {LagHEtHC ,      {AddHC->True}}
};

LagFer   = Yd conj[H].d.q + Ye conj[H].e.l + Yu H.u.q + Yn Et.n.l;
LagNV    = Mn/2 n.n;
LagH     = -(- mH2 conj[H].H     + 1/2 lambda1 conj[H].H.conj[H].H );
LagEt    = -(+ mEt2 conj[Et].Et  + 1/2 lambda2 conj[Et].Et.conj[Et].Et );
LagHEt   = -(+ lambda3 conj[H].H.conj[Et].Et + lambda4 conj[H].Et.conj[Et].H );
LagHEtHC = -(+ 1/2 lambda5 conj[H].Et.conj[H].Et );
\end{lstlisting}

First, we have split the Lagrangian in different pieces: {\tt LagFer},
{\tt LagNV}, {\tt LagH}, {\tt LagEt}, {\tt LagHEt} and {\tt
  LagHEtHC}. This is done for convenience. For each piece, we must set
the option {\tt AddHC} to either {\tt True} or {\tt False}. This
option is use to decide whether the Hermitian conjugate of the
Lagrangian piece must be added as well or not. In our case, we have
used {\tt False} for the self-conjugated terms and {\tt True} for the
rest. Then the different terms are defined as well. For this purpose
we can use {\tt conj[X]} in order to denote the Hermitian conjugate of
the {\tt X} multiplet. For fermions, \emph{barred spinors} are
automatically introduced. For example, {\tt LagFer} is the Yukawa
Lagrangian for the fermions,
\begin{equation}
{\tt LagFer} \equiv \mathcal{L}_Y =
Y_d H^\dagger \, \bar d \, q + Y_e H^\dagger \, \bar e \, \ell 
+ Y_u H \, \bar u \, q + Y_N \, \eta \, \overline{N} \, \ell \, ,
\end{equation}
\noindent which requires the addition of the Hermitian
conjugate. Notice that all scalar terms are defined with a global
sign. This is simply convenient to better identify the scalar
potential of the model ($\mathcal{L} \supset - \mathcal{V}$).

Now we find several definitions related to the breaking of the gauge
symmetry and the resulting mass eigenstates. First, for the gauge
sector, we have

\begin{lstlisting}[style=scotogenic,firstnumber=59]
DEFINITION[EWSB][GaugeSector] =
{ 
  {{VB,VWB[3]},{VP,VZ},ZZ},
  {{VWB[1],VWB[2]},{VWp,conj[VWp]},ZW}
};
\end{lstlisting}

\noindent In these lines we are defining the mixing of the gauge
bosons. In line $60$ we do it for the neutral gauge bosons, $\left\{ B
, W_3 \right\} \, \to \, \left\{ \gamma , Z \right\}$, using as name
for the mixing matrix {\tt ZZ} (the unitary matrix $Z^Z$ in Appendix
\ref{sec:appendix2}), whereas in line $61$ we do it for the charged
ones, with $\left\{ W_{1} , W_{2} \right\} \, \to \, \left\{ W^+ ,
\left(W^+\right)^\ast \right\}$, using as name for the mixing matrix
     {\tt ZW} (the unitary matrix $Z^W$ in Appendix
     \ref{sec:appendix2}). Next, we define the decomposition of the
     scalar fields into their CP-even and CP-odd components, including
     also the possibility of having VEVs. This is simply done with

\begin{lstlisting}[style=scotogenic,firstnumber=67]
DEFINITION[EWSB][VEVs]= 
{
  {H0,  {v, 1/Sqrt[2]}, {Ah, \[ImaginaryI]/Sqrt[2]}, {hh, 1/Sqrt[2]}},
  {et0, {0, 0}, {etI, \[ImaginaryI]/Sqrt[2]}, {etR, 1/Sqrt[2]}}
};
\end{lstlisting}

\noindent where we take the neutral components in $H$ and $\eta$,
$H^0$ and $\eta^0$, and split them into several pieces,
\begin{eqnarray}
H^0 &=& \frac{1}{\sqrt{2}} \left( v + h + i A \right) \, , \\
\eta^0 &=& \frac{1}{\sqrt{2}} \left( \eta_R + i \eta_I \right) \, .
\end{eqnarray}
Here $v/\sqrt{2}$ is the Higgs VEV, $h$ ({\tt hh} in the code)
and $\eta_R$ ({\tt etR} in the code) are the CP-even components and
$A$ ({\tt Ah} in the code) and $\eta_I$ ({\tt etI} in the
code) the CP-odd ones. It is worth noticing that we have set the
$\eta^0$ VEV to zero.

We are almost done. The next step is the definition of the mass
eigenstates in terms of the gauge eigenstates. This is completely
equivalent to the definition of the mixings in the model. In the
scotogenic model this is accomplished by means of the following lines
of code:

\begin{lstlisting}[style=scotogenic,firstnumber=73]
DEFINITION[EWSB][MatterSector]=
{
  {{conj[nR]},{X0, ZX}},
  {{vL}, {VL, Vv}},
  {{{dL}, {conj[dR]}}, {{DL,Vd}, {DR,Ud}}},
  {{{uL}, {conj[uR]}}, {{UL,Vu}, {UR,Uu}}},
  {{{eL}, {conj[eR]}}, {{EL,Ve}, {ER,Ue}}}
};
\end{lstlisting}

Here we have, on line $74$ the right-handed neutrinos (which do not
mix with any other field), on line $75$ the left-handed neutrinos
(which do not have other mixings either), on line $76$ the down-type
quarks, on line $77$ the up-type quarks and on line $78$ the charged
leptons. As can be seen from the previous lines, there are several
ways to make this definition, depending on the type of states:

\begin{itemize}

\item {\bf For scalars and Majorana fermions:}
\begin{center}
\{\{gauge eigenstate 1, gauge eigenstate 2, \dots\}, \{mass matrix, mixing matrix\}\}
\end{center}

\item {\bf For Dirac fermions:}
\begin{center}
\{\{\{gauge eigenstate left 1, gauge eigenstate left 2, \dots\}, \{mass matrix left, mixing matrix left\},
\{gauge eigenstate right 1, gauge eigenstate right 2, \dots\}, \{mass matrix right, mixing matrix right\}\}\}
\end{center}

\end{itemize}

For example, the singlet neutrinos are Majorana fermions and can mix
among themselves. We denote the mass eigenstates as {\tt X0} and the
mixing matrix as {\tt ZX}. The charged leptons, on the other hand, are
Dirac fermions: the left-handed states are transformed with a matrix
called {\tt Ve} leading to the states {\tt EL} and the right-handed
ones are transformed with the matrix {\tt Ue} leading to the states
{\tt ER}. All these transformations are unitary matrices, and are the
usual \emph{rotations} that connect the gauge and mass basis.

We have not included in this list the mass eigenstates that do not mix
(and thus are equal to the gauge eigenstates). This is the case of
$h$, $A$, $H^+$, $\eta_R$, $\eta_I$ and $\eta^+$, whose mixings are
forbidden by the $\mathbb{Z}_2$ parity of the scotogenic model. These
mass eigenstates have to be properly defined in the file {\tt
  particles.m}, but should not be included in {\tt
  DEFINITION[EWSB][MatterSector]}.

\tip{We have to be careful when defining the mixings. We may forget
  about some of them or introduce mixing among particles which do not
  really mix. One way to realize about these potential mistakes is to
  run the command {\tt CheckModel[model]} after loading the model in
  \SARAH. This \SARAH command checks the model files trying to find
  inconsistencies or missing definitions. In some cases it might be
  able to detect undefined mixings.}

Finally, the last part of the {\tt Scotogenic.m} file is used to
define Dirac spinors. This is because so far all the fermions we have
considered are 2-components Weyl spinors, since this is the way they
are internally handled by \SARAH. Therefore, we must tell \SARAH how
to combine them to form 4-component Dirac fermions, more common in
particle physics calculations. This is done for the mass eigenstates,

\begin{lstlisting}[style=scotogenic,firstnumber=86]
DEFINITION[EWSB][DiracSpinors]=
{
  Fd  -> {  DL, conj[DR]},
  Fe  -> {  EL, conj[ER]},
  Fu  -> {  UL, conj[UR]},
  Fv  -> {  VL, conj[VL]},
  Chi -> {  X0, conj[X0] }
};
\end{lstlisting}

\noindent as well as for the gauge eigenstates,

\begin{lstlisting}[style=scotogenic,firstnumber=95]
DEFINITION[EWSB][GaugeES]=
{
  Fd1 ->{  FdL, 0},
  Fd2 ->{  0, FdR},
  Fu1 ->{  Fu1, 0},
  Fu2 ->{  0, Fu2},
  Fe1 ->{  Fe1, 0},
  Fe2 ->{  0, Fe2}
};
\end{lstlisting}

For the gauge eigenstates there is no need to be exhaustive, since
these Dirac fermions are not used for practical calculations (always
performed in the mass basis), but for the mass eigenstates we must
include in the list all the possible Dirac fermions after EWSB. Notice
that in this case we have used the definitions of the mass eigenstates
previously done in {\tt DEFINITION[EWSB][MatterSector]} and that {\tt
  Fv} and {\tt Chi} are Majorana fermions, since the 2-component
spinors that form them are conjugate of each other.

This concludes our review of the file {\tt Scotogenic.m}.

\subsubsection*{parameters.m}

In this file we provide additional information about the parameters of
the model. This includes the original Lagrangian parameters as well as
mixing matrices and angles. The proper preparation of this file is
actually optional, since it will only required when some special
\SARAH outputs are obtained.

\tip{Again, it is convenient to use one of the existing models and
  adapt the corresponding {\tt parameters.m} file. Since most of the
  parameters are common to all models (gauge couplings, SM Yukawa
  matrices, \dots), this will save a lot of time and avoid typos.}

\tip{Since {\tt SARAH-4.6.0} one can also use the \Mathematica command
  {\tt WriteTemplatesParFiles;} after creating the central model
  file. This will create templates for the {\tt parameters.m} and {\tt
    particles.m} files, including all new parameters and particles,
  generating \LaTeX\ and output names and \FeynArts numbers and
  extending or generating PDG numbers. However, notice that some
  physical information, such as electric charges and the matching
  between Goldstone and gauge bosons, has to be adjusted by the user.}

Before we explain the content of this file, please note that there is
a file placed in {\tt \$PATH/SARAH-X.Y.Z/Models} also called {\tt
  parameters.m}. This is a general file with the most common
parameters already defined. For example, in this file one can find the
definition of the SM gauge couplings ($g_1$, $g_2$ and $g_3$), some of
the usual mixing matrices (like the one for the left-handed neutrinos,
called {\tt Neutrino-Mixing-Matrix} in the code) and some derived
parameters like the weak mixing angle $\theta_W$ (called {\tt ThetaW}
in the code). These definitions have been made to simplify our life
when defining a new model that shares some of them. In this case,
although they have to be defined in our new {\tt parameters.m} file,
it suffices to point to one of these definitions for \SARAH to know
all the details of the parameter. We will see how to do this when we
discuss the option {\tt Description} below.

The structure of the file (of both {\tt parameters.m} files, actually)
is as follows:

\begin{lstlisting}[style=parameters,firstnumber=3]
ParameterDefinitions = { 

{Parameter,   {Option 1 -> "value option 1",
               Option 2 -> "value option 2",
               ... }},

...

};
\end{lstlisting}

For each parameter several options can be defined. Most of them are
optional and rarely necessary. For the implementation of the
scotogenic model we will only need the following options:

\begin{itemize}

\item {\bf Description:} this is a string that identifies the
  parameter if this has been previously defined in the general file
  {\tt \$PATH/SARAH-X.Y.Z/Models/parameters.m}. As explained above, we
  do not need to redefine the usual parameters each time we implement
  a new model. We just have to write in {\tt Description} the same
  string name that is given to the parameter in the general {\tt
    \$PATH/SARAH-X.Y.Z/Models/parameters.m}. Furthermore, even if this
  parameter is not defined in the general file, this option can be
  used as a way to give a human readable name to the parameter, so
  that we can easily identify it when we open the {\tt parameters.m}
  long after the implementation.

\item {\bf Real:} this option can be either {\tt True} or {\tt
  False}. If the option is set to {\tt True}, \SARAH assumes that the
  parameter is real. By default, all parameters are considered
  complex.

\item {\bf OutputName:} this is a string to be used to identify the
  parameters in several \SARAH outputs (like \MO). In order to be
  compatible with all computer languages, this string should not
  contain any special characters.

\item {\bf LaTeX:} this option defines the way the parameter is shown
  in \LaTeX. As we will see below
  (Sec. \ref{subsec:SARAH-exploring}), \SARAH can export all the
  derived model properties in \LaTeX\ format. Properly defining this
  option for each parameter guarantees a nice and readable text. In
  doing this one should take into account that '$\backslash$' is
  interpreted by \Mathematica as escape sequence. Therefore,
  '$\backslash$' has to be replaced by '$\backslash\backslash$' in all
  \LaTeX\ commands.

\item {\bf LesHouches:} this option defines the position of the
  parameter in a LesHouches spectrum file. This will be important
  when we run numerical studies, since most input and output files
  follow this standard. If the parameter is a matrix we have to give
  the name of the block, whereas the name of the block and the entry
  have to be provided for parameters which are numbers.

\end{itemize}

Since the list of parameters is quite long, we will not review here
all the definitions for the scotogenic model. Nevertheless, a few
examples will be useful to see how it works. First, the neutrino
Yukawa couplings $Y_N$ are defined with the lines

\begin{lstlisting}[style=parameters,firstnumber=75]
{Yn,   {LaTeX -> "Y_N",
	LesHouches -> YN,
	OutputName->Yn }},
\end{lstlisting}

\indent No additional information is required. For the neutrino mixing
matrix we have something even simpler

\begin{lstlisting}[style=parameters,firstnumber=87]
{Vv, {Description ->"Neutrino-Mixing-Matrix"}}
\end{lstlisting}

\noindent Since this mixing matrix is common to other models with
non-zero neutrino mixings, it is already defined in the general file
{\tt \$PATH/SARAH-X.Y.Z/Models/parameters.m}. Therefore, including the
option {\tt Description} suffices for all the options to be properly
defined. For example, this way we set {\tt OutputName} to {\tt UV},
{\tt LaTeX} to $U^V$ and {\tt LesHouches} to {\tt UVMIX}. Finally, the
$\lambda_5$ coupling is defined by the lines

\begin{lstlisting}[style=parameters,firstnumber=68]
{lambda5,   {Real -> True,
	     LaTeX -> "\\lambda_5",
	     LesHouches -> {HDM,6},
	     OutputName-> lam5 }},
\end{lstlisting}

\noindent Note that in the {\tt LesHouches} option we have provided a
block name ({\tt HDM}) as well as an entry number. In addition, the
parameter has been defined as real. This justifies the splitting of
the scalar fields into CP-even and CP-odd states. In the presence of a
complex $\lambda_5$ parameter this would not be possible since both
states would mix.

Just in case you find some additional requirements when implementing
another model, here you have two other useful options:

\begin{itemize}

\item {\bf Dependence:} this option should be used when we want \SARAH
  to replace the parameter by a particular expression in all
  analytical calculations. For example, in the SM the neutral gauge
  boson ($\gamma, Z$) mixing matrix is parameterized in terms of one
  angle, the so-called weak or Weinberg angle $\theta_W$, see
  Eq. \eqref{eq:ZZ}. With this option we would tell \SARAH to use this
  parameterization in all analytical computations.

\item {\bf DependenceNum:} this option is similar to {\tt Dependence},
  with the only exception that the replacement is only used in
  numerical calculations. For example, we probably want to obtain all
  analytical results in terms of the SM gauge couplings $g_i$, but
  replace them in the numerical calculations by their expressions in
  terms of $\alpha_s$ (in case of the strong coupling constant),
  $\theta_W$ and $e$ (the electron charge).

\end{itemize}

Finally, it is worth clarifying what happens in case of
conflicts. Whenever an option is defined in the general file {\tt
  \$PATH/SARAH-X.Y.Z/Models/parameters.m} and later included
explicitly in the specific {\tt parameters.m} file for our model,
\SARAH will take the value of the option given in the specific file.

\subsubsection*{particles.m}

This file is devoted to the particle content of the model. Although
the basic information is already given in {\tt Scotogenic.m}, there
are some additional details that must be given in {\tt
  particles.m}. As for {\tt parameters.m}, this is an optional file
that will only be required when producing some special \SARAH outputs.

The particles (or more precisely, \emph{states}) of the model are
distributed into three categories: gauge and mass eigenstates and
intermediate states. We have already mentioned the first two
categories. The third one is composed by states which are neither
gauge eigenstates nor mass ones, but appear in intermediate
calculations. For instance, the 2-component Weyl fermions {\tt X0}
belong to this class, since the gauge eigenstates are {\tt nR} and the
mass eigenstates (used in practical calculations) are the 4-component
Dirac fermions {\tt Chi}.

As for the parameters, there is a general file where the definitions
of the most common particles are already given. This file is located
in {\tt \$PATH/SARAH-X.Y.Z/Models/particles.m} and its practical use
is again similar: we can simply point to one of the existing
definitions in case our model has a particle that is already in the
general file.

The structure of the file (again, of both {\tt particles.m} files)
is as follows:

\begin{lstlisting}[style=particles,firstnumber=3]
ParticleDefinitions[states] = {

{Particle,   {Option 1 -> "value option 1",
              Option 2 -> "value option 2",
               ... }},

...

};
\end{lstlisting}

In practice, it is not necessary to provide definitions for gauge
eigenstates and intermediate states since these do not participate in
the calculation of physical observables. The only option that should
be defined for these states is {\tt LaTeX}, which will be helpful to
get a readable \LaTeX\ output. In contrast, the properties of the mass
eigenstates are crucial, since they must be read by other tools such
as \MO or \MG.

Let us show a couple of illustrative examples in the scotogenic
models. The definitions of the $\eta$ scalars (mass eigenstates) is as
follows

\begin{lstlisting}[style=particles,firstnumber=60]
      {etR,   {  Description -> "CP-even eta scalar", 
		 PDG -> {1001},
		 Mass -> LesHouches,
		 ElectricCharge -> 0,
		 LaTeX -> "\\eta_R",
		 OutputName -> "etR" }}, 
      {etI,   {  Description -> "CP-odd eta scalar", 
		 PDG -> {1002},
		 Mass -> LesHouches,
		 ElectricCharge -> 0,
		 LaTeX -> "\\eta_I",
		 OutputName -> "etI" }}, 
      {etp,   {  Description -> "Charged eta scalar", 
		 PDG -> {1003},
		 Mass -> LesHouches,
		 ElectricCharge -> 1,
		 LaTeX -> "\\eta^+",
                 OutputName -> "etp" }}, 
\end{lstlisting}

We have used the option {\tt Description} to give simple names to the
three mass eigenstates. Although they are not used (since these
descriptions are not present in the general file {\tt
  \$PATH/SARAH-X.Y.Z/Models/particles.m}), they are helpful for future
reference. We have also given {\tt PDG} codes to the three states,
using high numbers not reserved for other particles. This is necessary
for \MG, which uses these codes to identify the particles. Moreover,
we have defined the {\tt ElectricCharge} and the way the particles
should be shown in \LaTeX\ format. The option {\tt OutputName} is
completely analogous to the same option in the case of
parameters. Finally, we have set the option {\tt Mass} to {\tt
  LesHouches}. This option defines the way in which \MG should obtain
the value of this mass. With {\tt LesHouches}, we are telling \SARAH
that we would like \MG to take this value from a LesHouches input file
(probably obtained with a spectrum generator like \SPheno, see
Sec. \ref{subsec:SARAH-output}).

When there are several generations of a given mass eigenstate, some of
the options have to be given as arrays. An example can be found in the
definition of the fermion singlets

\begin{lstlisting}[style=particles,firstnumber=94]
{Chi,  { Description -> "Singlet Fermions",
	       PDG -> {1012,1014,1016},
	       Mass -> LesHouches,
	       ElectricCharge -> 0,
	       LaTeX -> "N",
	       OutputName -> "N" }}
\end{lstlisting}

In contrast to the definitions of the $\eta$ scalars, in this case the
option {\tt PDG} must be an array of three elements, since there are
three generations of singlet fermions.

Finally, an option that is crucial for the proper implementation of
the model is {\tt Goldstone}. This option should be included in the
definition of every massive gauge boson. It tells \SARAH where to find
the corresponding Goldstone boson that becomes its longitudinal
component. For example, for the $Z$ boson one has

\begin{lstlisting}[style=particles,firstnumber=80]
{VZ,   { Description -> "Z-Boson", Goldstone -> Ah }}, 
\end{lstlisting}

Note that the only option that we must add is {\tt Goldstone}. The
rest of options for the $Z$ boson are given in the general file {\tt
  \$PATH/SARAH-X.Y.Z/Models/particles.m}.

\subsubsection*{SPheno.m}

Finally, the file {\tt SPheno.m} is only necessary if we plan to
create a \SPheno module for our model. This is explained in more
detail in Sec. \ref{subsec:SARAH-output}, and thus we postpone the
description of this file until we reach that point of the course.

\subsection{Exploring a model}
\label{subsec:SARAH-exploring}

The model is implemented and the time has come to see what \SARAH can
do for us.

First, we have to load \SARAH and the scotogenic model. As shown
already, we can do that with these \Mathematica commands:

\vsp
\begin{lstlisting}[style=mathematica]
<<$PATH/SARAH-X.Y.Z/SARAH.m;
Start["Scotogenic"];
\end{lstlisting}
\vsp

After a few seconds all the initial \SARAH computations will be
finished and we will be ready to execute all kinds of commands to get
analytical information about the model.

\subsubsection*{Tadpole equations}

The minimization of the scalar potential proceeds via the
\emph{tadpole equations}~\footnote{Note that we are assuming here CP
  conservation in the scalar sector.},
\begin{equation} \label{eq:tadpoleGeneral}
\frac{\partial \mathcal V}{\partial v_i} = 0 \, ,
\end{equation}
where $v_i$ are the VEVs of the scalar fields. One has as many
equations as VEVs. In the scotogenic model there is only one non-zero
VEV, the VEV of the SM Higgs doublet. Therefore, we just need to solve
one equation to minimize the scalar potential. This is
\begin{equation} \label{eq:tadpoleScotogenic}
\frac{\partial \mathcal V}{\partial v} = 0 \, .
\end{equation}
\SARAH can provide the analytical form of this equation. This is
obtained with the command

\vsp
\begin{lstlisting}[style=mathematica]
TadpoleEquation[v]
\end{lstlisting}
\vsp

We find the result
\begin{equation} \label{eq:tadpoleScotogenic2}
\frac{1}{2} \lambda_1 v^3 - m_H^2 v = 0 \, ,
\end{equation}
which, using the \Mathematica command

\vsp
\begin{lstlisting}[style=mathematica]
Solve[TadpoleEquation[v], mH2]
\end{lstlisting}
\vsp

\noindent gives the well-known minimization condition
\begin{equation} \label{eq:min}
m_H^2 = \frac{1}{2} \lambda_1 v^2 \, .
\end{equation}

Finally, we point out that the command

\vsp
\begin{lstlisting}[style=mathematica]
TadpoleEquations[EWSB]
\end{lstlisting}
\vsp

\noindent can be used to obtain the complete list of tadpole equations
of a model.

\subsubsection*{Masses}

Next, we can print some masses. There are two ways to do this,
depending on whether the mass eigenstate we are interested in is a
mixture of gauge eigenstates or not. When it is, we must print the
mass matrix of the complete set of mass eigenstates. This is done with

\vsp
\begin{lstlisting}[style=mathematica]
MassMatrix[state]
\end{lstlisting}
\vsp

\noindent where {\tt state} must be replaced by the name of the mass
eigenstate. For example, we can run the command

\vsp
\begin{lstlisting}[style=mathematica]
MassMatrix[Fe]
\end{lstlisting}
\vsp

\noindent and \SARAH would return the well-known form of the charged
lepton mass matrix,
\begin{equation}
\left( \begin{array}{ccc}
- \frac{v \left(Y_e\right)_{11}}{\sqrt{2}} & - \frac{v \left(Y_e\right)_{21}}{\sqrt{2}} & - \frac{v \left(Y_e\right)_{31}}{\sqrt{2}} \\
- \frac{v \left(Y_e\right)_{12}}{\sqrt{2}} & - \frac{v \left(Y_e\right)_{22}}{\sqrt{2}} & - \frac{v \left(Y_e\right)_{32}}{\sqrt{2}} \\
- \frac{v \left(Y_e\right)_{13}}{\sqrt{2}} & - \frac{v \left(Y_e\right)_{23}}{\sqrt{2}} & - \frac{v \left(Y_e\right)_{33}}{\sqrt{2}}
\end{array} \right) \, .
\end{equation}
We can also print the mass matrix for the singlet fermions. By
executing the command

\vsp
\begin{lstlisting}[style=mathematica]
MassMatrix[Chi]
\end{lstlisting}
\vsp

\noindent we obtain
\begin{equation}
\left( \begin{array}{ccc}
- \left(M_N\right)_{11} & -\frac{1}{2} \left(M_N\right)_{12}-\frac{1}{2} \left(M_N\right)_{21} & -\frac{1}{2} \left(M_N\right)_{13}-\frac{1}{2} \left(M_N\right)_{31} \\
-\frac{1}{2} \left(M_N\right)_{12}-\frac{1}{2} \left(M_N\right)_{21} & - \left(M_N\right)_{22} & -\frac{1}{2} \left(M_N\right)_{23}-\frac{1}{2} \left(M_N\right)_{32} \\
-\frac{1}{2} \left(M_N\right)_{13}-\frac{1}{2} \left(M_N\right)_{31} & -\frac{1}{2} \left(M_N\right)_{23}-\frac{1}{2} \left(M_N\right)_{32} & - \left(M_N\right)_{33}
\end{array} \right) \, .
\end{equation}
Notice that \SARAH does not know that the matrix $M_N$ is
symmetric. If required, we can simplify the expression with the
command

\vsp
\begin{lstlisting}[style=mathematica]
MassMatrix[Chi] /. Mn[i_, j_] :> If[i > j, Mn[j, i], Mn[i, j]]
\end{lstlisting}
\vsp

\noindent obtaining the more standard form
\begin{equation}
\left( \begin{array}{ccc}
- \left(M_N\right)_{11} & - \left(M_N\right)_{12} & - \left(M_N\right)_{13}\\
- \left(M_N\right)_{12} & - \left(M_N\right)_{22} & - \left(M_N\right)_{23} \\
- \left(M_N\right)_{13} & - \left(M_N\right)_{23} & - \left(M_N\right)_{33}
\end{array} \right) \, .
\end{equation}

In case we want to print the mass of a state that does not mix with
other fields we must use the command

\vsp
\begin{lstlisting}[style=mathematica]
Mass[state] /. Masses[EWSB]
\end{lstlisting}
\vsp

\noindent where, again, {\tt state} must be replaced by the name of
the specific mass eigenstate. As a prime example, let us consider the
Higgs boson. Its mass can be printed with the command

\vsp
\begin{lstlisting}[style=mathematica]
Mass[hh] /. Masses[EWSB]
\end{lstlisting}
\vsp

\noindent leading to
\begin{equation} \label{eq:mh}
\frac{3}{2} \lambda_1 v^2 - m_H^2 \, .
\end{equation}
Two things should be noted: (1) we actually got the Higgs boson
squared mass, and (2) the minimization condition in Eq. \eqref{eq:min}
has not been applied. To obtain the resulting Higgs boson mass after
applying the tadpole equations, we can simply run

\vsp
\begin{lstlisting}[style=mathematica]
solTadpole = Solve[TadpoleEquation[v], mH2];
Mass[hh] /. Masses[EWSB] /. solTadpole
\end{lstlisting}
\vsp

\noindent obtaining the final expression for the Higgs boson mass
\begin{equation} \label{eq:mh2}
\lambda_1 v^2 \, .
\end{equation}

\subsubsection*{Vertices}

One of the most powerful features of \SARAH is the calculation of
interaction vertices. In order to obtain the a vertex one must execute
the command

\vsp
\begin{lstlisting}[style=mathematica]
Vertex[{state 1, state 2, state 3}]
\end{lstlisting}
\vsp

\noindent or

\vsp
\begin{lstlisting}[style=mathematica]
Vertex[{state 1, state 2, state 3, state 4}]
\end{lstlisting}
\vsp

\noindent depending on the number of particles involved in the
vertex. Here {\tt state 1}, {\tt state 2}, {\tt state 3} and {\tt
  state 4} are mass eigenstates. The result of this command is an
array that includes all possible Lorentz structures appearing in the
interaction vertex and the corresponding coefficients. For example,
the $\ell_i^+ - \ell_j^- - h$ vertex, where $i,j = 1,2,3$ are flavor
indices, is obtained with

\vsp
\begin{lstlisting}[style=mathematica]
Vertex[{bar[Fe], Fe, hh}]
\end{lstlisting}
\vsp

\noindent and gives, 
\begin{equation}
\frac{i}{\sqrt{2}} \sum_{m,n=1}^3 \left(V_e\right)^\ast_{jn} \left(Y_e\right)_{mn} \left(U_e\right)^\ast_{im} \, P_L + \frac{i}{\sqrt{2}} \sum_{m,n=1}^3 \left(V_e\right)_{in} \left(Y_e\right)^\ast_{mn} \left(U_e\right)_{jm} \, P_R \, ,
\end{equation}
where $P_{L,R} = \frac{1}{2} \left( 1 \mp \gamma_5 \right)$ are the
usual chirality projectors. The unitary matrices $V_e$ and $U_e$ are
defined in the model file ({\tt Scotogenic.m}) as the matrices that
transform between the gauge and mass basis for the left- and
right-handed charged leptons, respectively.

We can now consider the $\ell_i^+ - \nu_j - W_\mu^-$ vertex, where again
$i,j = 1,2,3$ are flavor indices. This is obtained with

\vsp
\begin{lstlisting}[style=mathematica]
Vertex[{bar[Fe], Fv, conj[VWp]}]
\end{lstlisting}
\vsp

\noindent and we find, 
\begin{equation} \label{eq:charged-current}
- i \frac{g_2}{\sqrt{2}} \sum_{m=1}^3 \left(V_e\right)_{im} \left(V_\nu\right)^\ast_{jm} \, \gamma_\mu P_L \, ,
\end{equation}
where $V_\nu$ is the unitary matrix that transforms the left-handed
neutrinos from the gauge to the mass
basis. Eq. \eqref{eq:charged-current} is nothing but the standard
charged current interaction in the lepton sector. It is commonly
written in terms of the so-called Pontecorvo-Maki-Nakagawa-Sakata
(PMNS) matrix, $K$, defined as~\footnote{A technical note for the
  experts that might be surprised by Eq. \eqref{eq:PMNS}. The PMNS
  matrix, also known as the leptonic mixing matrix, is usually defined
  as $K = U_\ell^\dagger U_\nu$, where the matrices $U_\ell$ and
  $U_\nu$ connect gauge ($e_L$,$\nu_L$) and mass eigenstates
  ($E_L$,$\nu$) as $e_L = U_\ell E_L$ and $\nu_L = U_\nu
  \nu$. Therefore, according to \SARAH's conventions, $U_\ell =
  V_e^\dagger$ and $U_\nu = V_\nu^\dagger$ (see Appendix
  \ref{sec:appendix2}), and this is how we get Eq. \eqref{eq:PMNS}.}
\begin{equation} \label{eq:PMNS}
K = V_e V_\nu^\dagger \, ,
\end{equation}
thus leading to the vertex
\begin{equation} \label{eq:charged-current2}
i \frac{g_2}{\sqrt{2}} \sum_{m=1}^3 K_{ij} \, \gamma_\mu P_L \, .
\end{equation}
Notice that \SARAH also identifies when a vertex does not exist. One
can see this by computing, for example, the $\nu_i - \chi_j - h$
vertex, with $i,j = 1,2,3$ are flavor indices, with the command

\vsp
\begin{lstlisting}[style=mathematica]
Vertex[{Fv, Chi, hh}]
\end{lstlisting}
\vsp

\noindent which simply returns zero due to $\mathbb{Z}_2$
conservation. Instead, if we compute the $\nu_i - \chi_j - \eta_R$
vertex with

\vsp
\begin{lstlisting}[style=mathematica]
Vertex[{Fv, Chi, etR}]
\end{lstlisting}
\vsp

\noindent we find
\begin{equation}
- \frac{i}{\sqrt{2}} \sum_{m,n=1}^3 \left(V_\nu\right)^\ast_{in} \left(Y_N\right)_{mn} \left(Z_X\right)^\ast_{jm} \, P_L - \frac{i}{\sqrt{2}} \sum_{m,n=1}^3 \left(V_\nu\right)_{in} \left(Y_N\right)^\ast_{mn} \left(Z_X\right)_{jm} \, P_R \, .
\end{equation}

\subsubsection*{Renormalization group equations}

\SARAH also obtains the renormalization group equations (RGEs) for all
the parameters of the model. More precisely, the $\beta$ functions of
all parameters are computed in $R_\xi$ gauge at the 1- and 2-loop
level. The 1- and 2-loop $\beta$ functions of the parameter $c$ are
defined as
\begin{equation}
\frac{dc}{dt} = \beta_c = \frac{1}{16 \pi^2} \beta_c^{(1)} + \frac{1}{(16 \pi^2)^2} \beta_c^{(2)} \, ,
\end{equation}
where $t = \log \mu$, $\mu$ being the energy scale, and
$\beta_c^{(1)}$ and $\beta_c^{(2)}$ are the 1- and 2-loop $\beta$
functions, respectively.

\tip{All calculations in \SARAH are performed in $R_\xi$ gauge. This
  is useful to check that all physical observables are gauge
  independent.}

The full 2-loop RGEs are computed with the command

\vsp
\begin{lstlisting}[style=mathematica]
CalcRGEs[]
\end{lstlisting}
\vsp

For non-supersymmetric models this command might take quite a long
time to finish. For this reason, and in case one is interested only in
the 1-loop $\beta$ functions, the option {\tt TwoLoop} turns out to be
useful. By setting the option to the value {\tt False}

\vsp
\begin{lstlisting}[style=mathematica]
CalcRGEs[TwoLoop -> False]
\end{lstlisting}
\vsp

\noindent the calculation becomes much faster. The analytical results
for the RGEs are saved in several arrays. In case of
non-supersymmetric models (like the one we are studying), these are

\begin{itemize}

\item {\tt Gij}: Anomalous dimensions for all fermions and scalars
\item {\tt BetaGauge}: Gauge couplings
\item {\tt BetaMuij}: Bilinear fermion terms
\item {\tt BetaBij}: Bilinear scalar terms
\item {\tt BetaTijk}: Cubic scalar couplings
\item {\tt BetaLijkl}: Quartic scalar couplings
\item {\tt BetaYijk}: Yukawa couplings
\item {\tt BetaVEVs}: VEVs

\end{itemize}

Each entry in these arrays contain three elements: the name of the
parameter and the 1- and 2-loops $\beta$ functions. For example, in
the array {\tt BetaGauge} the RGEs for the gauge couplings are
saved. In the scotogenic model these are the same as in the SM. Simply
by running

\vsp
\begin{lstlisting}[style=mathematica]
BetaGauge
\end{lstlisting}
\vsp

\noindent we find
\begin{equation} \label{eq:RGEsgi}
\beta_{g_i}^{(1)} = \left( \frac{21}{5} g_1^3 , -3 g_2^3 , -7 g_3^3 \right) \, ,
\end{equation}
with $i=1,2,3$. Two comments are in order: (1) the running $g_1$
coupling already includes the usual \emph{GUT normalization factor}
$\sqrt{5/3}$, and (2) the 2-loop RGEs are zero simply because we
decided not to compute them. We note that the GUT normalization factor
is not hardcoded, but can be changed by the user when $g_1$ is defined
in the {\tt parameters.m} file. The RGEs for the scalar squared masses
are saved in the array {\tt BetaBij}. Therefore, we can execute

\vsp
\begin{lstlisting}[style=mathematica]
BetaBij
\end{lstlisting}
\vsp

\noindent to find, for example, that the 1-loop running of $m_\eta^2$
is given by the $\beta$ function
\begin{equation}
\beta_{m_\eta^2}^{(1)} = - \frac{9}{2} \left( \frac{1}{5} g_1^2 + g_2^2 \right) m_\eta^2 + 6 \lambda_2 m_\eta^2 - 2 \left( 2 \lambda_3 + \lambda_4 \right) m_H^2 + 2 \, m_\eta^2 \text{Tr}\left( Y_N Y_N^\dagger \right) - 4 \text{Tr}\left( M_N M_N^\ast Y_N Y_N^\dagger \right) \, .
\end{equation}
Here $\text{Tr}$ denotes the conventional matrix trace.

These arrays are also saved in external files, so that they can be
loaded in other \Mathematica sessions without the need to compute them
again. These files are placed in the directory {\tt
  \$PATH/SARAH-X.Y.Z/Output/Scotogenic/RGEs} and one can easily load
them with commands such as

\vsp
\begin{lstlisting}[style=mathematica]
BetaGauge = <<$PATH/SARAH-X.Y.Z/Output/Scotogenic/RGEs/BetaGauge.m;
\end{lstlisting}
\vsp

\subsubsection*{Writing all information in \LaTeX\ format}

Finally, we can export all this information to \LaTeX\ format so that,
after the usual compilation, we obtain a pdf file with all the
analytical results derived by \SARAH. The \LaTeX\ output of \SARAH is
very useful. In addition to being visual and easy to read, we can copy
the \LaTeX\ code to our own publications, saving time and getting rid
of the tedious task of typing analytical formulas.

We can generate the \LaTeX\ output for our model with the commands

\vsp
\begin{lstlisting}[style=mathematica]
ModelOutput[EWSB]
CalcRGEs[TwoLoop -> False]
MakeTeX[]
\end{lstlisting}
\vsp

The first line tells \SARAH to run a long list of computations, saving
the results in several directories in {\tt
  \$PATH/SARAH-X.Y.Z/Output/Scotogenic/EWSB}. The second line ({\tt
  CalcRGEs[TwoLoop -> False]}) is optional. If we do not include it,
the resulting \LaTeX\ files will not contain the RGEs. Moreover, if
this line has been executed already in the \Mathematica session we are
in, there is no need to execute it again since the RGEs are already
computed.

The results of these commands are put in the directory {\tt
  \$PATH/SARAH-X.Y.Z/Output/Scotogenic/EWSB/TeX}. In order to generate
the pdf file with all the information, we just have to go to that
directory and execute a shell script that is already provided by
\SARAH. This is done with

\vsp
\begin{lstlisting}[style=terminal]
$ cd $PATH/SARAH-X.Y.Z/Output/Scotogenic/EWSB/TeX
$ sh MakePDF.sh
\end{lstlisting}
\vsp

Two problems might be encountered when running these commands:

\begin{enumerate}

\item In some cases (this is computer-dependent) it might be necessary
  to make the script executable by assigning the required permissions
  before we can run it. This is done with the terminal command

\vsp
\begin{lstlisting}[style=terminal]
$ chmod 755 MakePDF.sh
\end{lstlisting}
\vsp

\item By default, the pdf file will contain all vertices in the
  model. These are shown graphically with a Feynman diagram for each
  vertex. For this purpose, \SARAH makes use of the \LaTeX\ package
  {\tt feynmf}. This package must be installed in case it is not. For
  instance, in Debian based systems, this is done with

\vsp
\begin{lstlisting}[style=terminal]
$ sudo apt-get install feynmf
\end{lstlisting}
\vsp

\end{enumerate}

As a result of these commands, a pdf file with the name {\tt
  Scotogenic-EWSB.pdf} is created in the folder {\tt
  \$PATH/SARAH-X.Y.Z/Output/Scotogenic/EWSB/TeX}. Now you can simply
open it with your favourite pdf viewer.

\subsection{Creating input for other computer tools}
\label{subsec:SARAH-output}

We have finished our overview of the possibilities offered by \SARAH
concerning analytical calculations and thus we will start discussing
how to obtain reliable numerical results for observables of interest
in our model.

Being a \Mathematica package, \SARAH is not suited for heavy numerical
studies. However, it can be used to generate the required input files
for other popular computer tools, which can then be used for that
purpose. In this course we will focus on three tools:

\begin{itemize}

\item \SPheno: To compute the mass spectrum, decay rates and flavor observables 
\item \MO: To compute the dark matter relic density and other related observables
\item \MG: To run Monte Carlo simulations for collider studies

\end{itemize}

\tip{Of course, the preparation of the input files for \MO and \MG
  does not require \SARAH. There are other ways, including the direct
  writing \emph{by hand}, to create these files. However, I recommend
  \SARAH for three reasons: (1) it saves a lot of time, and (2) it is
  reliable (since it is automatized), and (3) it is a good idea to
  have a \emph{central} code to generate all the input files, since
  this guarantees that all definitions and conventions will be
  consistent.}

\subsubsection*{\SPheno}

\SPheno is a spectrum calculator: it takes the values of the input
parameters and computes, numerically, all masses and mixing matrices
in the model. Besides, with this information it also computes the
vertices, decay rates and many flavor observables. Although it was
originally designed to cover just a few specific supersymmetric
models, now it has become available for many other models (including
non-supersymmetric ones) thanks to \SARAH. The code is written in
\Fortran. See Sec. \ref{subsec:SPheno} for details.

The strategy is simple. \SPheno has many model-independent numerical
routines to perform standard numerical operations such as matrix
diagonalization or resolution of differential equations. In order to
study our own model, we just have to provide some additional routines
with the details of the model. And this is what \SARAH can do for
us. It creates a group of \Fortran files (what we call a \emph{\SPheno
  module}) with all the information of our model. We just have to add
these files to the ones already in \SPheno, and the resulting code
will become a numerical spectrum calculator for our model.

In order to create a \SPheno module with \SARAH we have to do two
things: (1) create a {\tt SPheno.m} file (already mentioned in
Sec. \ref{subsec:SARAH-model}), and (2) run a command. Let us first
show how to prepare the {\tt SPheno.m} file.

The user can create two \SPheno versions:

\begin{itemize}

\item {\bf `GUT' version:} the user defines input values for the
  parameters of the model at some high-energy scale and then they are
  evolved using the RGEs down to the electroweak scale where the
  spectrum is computed. This is common practice in supersymmetric
  models, with boundary conditions at the grand unification (GUT)
  scale, $m_{\text{GUT}} \simeq 2 \cdot 10^{16}$ GeV.

\item {\bf Low scale version:} the user defines all input values at
  the electroweak scale (or SUSY scale in case of supersymmetric
  models). There is no need for RGE running.

\end{itemize}

In this course we will show how to create a low scale version of
\SPheno, since this is the most common practice with
non-supersymmetric models. As you will see, this is actually
simpler. For instructions about how to create a GUT version we refer
to the \SARAH manual or to \cite{Staub:2015kfa}.

Since we are interested in a low scale \SPheno, the first line of the
{\tt SPheno.m} file must be

\begin{lstlisting}[style=SPheno,firstnumber=1]
OnlyLowEnergySPheno = True;
\end{lstlisting}

When the \SPheno code is ready, it will require an input file in
LesHouches format to run. The LesHouches format
\cite{Skands:2003cj,Allanach:2008qq} distributes the parameters in
\emph{blocks}. Some blocks are devoted to lists of parameters, all of
them numbers, whereas some blocks are devoted to arrays. In
particular, there must be a block called {\tt MINPAR} where the
\emph{minimal parameters} are listed. This may include parameters of
the scalar potential, some specific terms or even derived parameters
not present in the Lagrangian. Usually, the parameters in the scalar
potential are listed here.

When preparing the {\tt SPheno.m} file we must tell \SARAH the list of
parameters in the {\tt MINPAR} block. In the case of the scotogenic
model this is done with the following lines

\begin{lstlisting}[style=SPheno,firstnumber=3]
MINPAR={
  {1,lambda1Input},
  {2,lambda2Input},
  {3,lambda3Input},
  {4,lambda4Input},
  {5,lambda5Input},
  {6,mEt2Input}
};
\end{lstlisting}

Note that the only parameter of the scalar potential that we have not
included in this list is $m_H^2$. We will see the reason in the next
step. We also point out that matrices (like the Yukawa coupling $Y_N$
or the right-handed neutrino Majorana mass $M_N$) are not listed in
the {\tt MINPAR} block.

Now we must tell \SARAH what parameters should be used to solve the
tadpole equations. These parameters will not have input values, but
instead they will be initialized to the value resulting from the
solution of the minimum conditions. In principle, several parameters
can be used for this purpose, but it is important to select one that
will lead to a simple solution of the tadpole equations. In the
scotogenic model the best choice is $m_H^2$, and this is indicated in
the {\tt SPheno.m} file as

\begin{lstlisting}[style=SPheno,firstnumber=12]
ParametersToSolveTadpoles = {mH2};
\end{lstlisting}

\tip{\SARAH makes use of the \Mathematica command {\tt Solve} to solve
  analytically the tadpole equations. If no solution is found, all the
  subsequent steps in \SPheno will contain errors. For this reason, it
  is crucial to choose the parameters that will be used to solve the
  tadpole equations properly. Usually, the best choice is to select
  bilinear scalar terms, like $m_H^2$ in the scotogenic model. The
  reason is that the tadpole equations are linear in them.}

In the next step we link the model parameters to the input parameters
introduced in the first lines of the file. This is done with

\begin{lstlisting}[style=SPheno,firstnumber=14]
BoundaryLowScaleInput={
  {lambda1,lambda1Input},
  {lambda2,lambda2Input},
  {lambda3,lambda3Input},
  {lambda4,lambda4Input},
  {lambda5,lambda5Input},
  {mEt2,mEt2Input},
  {Yn, LHInput[Yn]},
  {Mn, LHInput[Mn]}
};
\end{lstlisting}

We notice that here we also indicated that the matrices $Y_N$ and
$M_N$ should have input values. However, the syntaxis is different
with respect to the rest of inputs since they are matrices.

Next, we match the SM parameters to the parameters in the scotogenic
model. For this trivial extension of the SM this is simply given by

\begin{lstlisting}[style=SPheno,firstnumber=25]
DEFINITION[MatchingConditions]= 
{{v, vSM}, 
 {Ye, YeSM},
 {Yd, YdSM},
 {Yu, YuSM},
 {g1, g1SM},
 {g2, g2SM},
 {g3, g3SM}};
\end{lstlisting}

Finally, we have to define two lists. These contain the mass
eigenstates for which \SPheno will compute decay widths and branching
ratios. The first list ({\tt ListDecayParticles}) is for 2-body
decays, whereas the second list ({\tt ListDecayParticles3B}) is for
3-body decays. This is indicated in the {\tt SPheno.m} file as

\begin{lstlisting}[style=SPheno,firstnumber=34]
ListDecayParticles = {Fu,Fe,Fd,Fv,VZ,VWp,hh,etR,etI,etp,Chi};
ListDecayParticles3B = {{Fu,"Fu.f90"},{Fe,"Fe.f90"},{Fd,"Fd.f90"}};
\end{lstlisting}

And with this we conclude the {\tt SPheno.m} file. Now we just have to
run, after loading \SARAH and initializing our model, the following
command in \Mathematica~\footnote{After {\tt SARAH-4.11.0}, released
  in March 2017, all 2-body decay widths can be computed by \SPheno at
  full 1-loop level following \cite{Goodsell:2017pdq}. In case the
  user wants to disable this feature, the {\tt MakeSPheno} option {\tt
    IncludeLoopDecays} must be set to {\tt False}, resulting in the
  \Mathematica command {\tt MakeSPheno[IncludeLoopDecays -> False]}.}

\vsp
\begin{lstlisting}[style=mathematica]
MakeSPheno[]
\end{lstlisting}
\vsp

After some minutes (depending on the model this can be quite a lengthy
process) the \SPheno module will be created. This will be located in
the directory {\tt \$PATH/SARAH-X.Y.Z/Output/Scotogenic/EWSB/SPheno}.

The installation and usage of \SPheno, as well as what to do with the
generated \SPheno module, will be explained in
Sec. \ref{subsec:SPheno}.\\

\subsubsection*{\MO}

The most popular public code for dark matter studies is \MO. With \MO
one can compute several dark matter properties, including the relic
density, as well as direct and indirect detection rates. This can be
done in many particle physics models. For this purpose, the user just
needs to provide model files in \CalcHep format
\cite{Belyaev:2012qa}. This can be obtained with \SARAH via the
command

\vsp
\begin{lstlisting}[style=mathematica]
MakeCHep[]
\end{lstlisting}
\vsp

The folder {\tt \$PATH/SARAH-X.Y.Z/Output/Scotogenic/EWSB/CHep}
contains all the files generated with this command. We will have an
introduction to \MO in the third lecture, see Sec. \ref{sec:lecture2},
where we will learn how to use them.\\

\subsubsection*{\MG}

\MG is a well-known computer tool to run Monte Carlo simulations for
collider studies. This code can handle the computation of tree-level
and next-to-leading order cross-sections and their matching to parton
shower simulations.

The input for \MG must use the Universal \FeynRules Output (UFO) format
\cite{Degrande:2011ua}. This format can be generated with \FeynRules
\cite{Alloul:2013bka}, a \Mathematica package for the calculation of
Feynman rules in models defined by the user. However, we will again
use \SARAH, which can also export the model information into the
required UFO format via the command

\vsp
\begin{lstlisting}[style=mathematica]
MakeUFO[]
\end{lstlisting}
\vsp

The resulting UFO files are located in {\tt
  \$PATH/SARAH-X.Y.Z/Output/Scotogenic/EWSB/UFO}. We will learn how to
use them with \MG in the third lecture, see
Sec. \ref{sec:lecture3}.\\

\subsection{Brief introduction to \SPheno}
\label{subsec:SPheno}

In this Section we will have a brief introduction to \SPheno. A
complete overview of all possibilities with this code is beyond the
scope of this course, where we will only learn how to use it to obtain
the spectrum and compute some flavor observables. For a complete
review we refer to the manual \cite{Porod:2003um,Porod:2011nf}. For
the calculation of flavor observables we recommend having a look at
the \FlavorKit manual \cite{Porod:2014xia}.

\subsubsection*{\SPheno: Technical details, installation and load}

\begin{itemize}

\item {\bf Name of the tool:} \SPheno

\item {\bf Author:} Werner Porod (porod@physik.uni-wuerzburg.de) and Florian Staub (florian.staub@cern.ch)

\item {\bf Type of code:} \Fortran

\item {\bf Website:} \url{http://spheno.hepforge.org/}

\item {\bf Manual:} The original manual can be found in
  \cite{Porod:2003um}. For a newer version see \cite{Porod:2011nf}.

\end{itemize}

Installing \SPheno is easy. First, we download the code from the indicated url. Then we copy the {\tt tar.gz} file to the directory {\tt
  \$PATH}, where it can be extracted:

\vsp
\begin{lstlisting}[style=terminal]
$ cp Download-Directory/SPheno-X.Y.Z.tar.gz $PATH/
$ cd $PATH
$ tar -xf SPheno-X.Y.Z.tar.gz
\end{lstlisting}
\vsp

Here {\tt X.Y.Z} must be replaced by the \SPheno version which has
been downloaded. Once this is done, we must copy the \SPheno module
created with \SARAH (see Sec. \ref{subsec:SARAH-output}). First, we
create a directory in {\tt \$PATH/SPheno-X.Y.Z} with the name of the
model, {\tt Scotogenic} in this case,

\vsp
\begin{lstlisting}[style=terminal]
$ cd $PATH/SPheno-X.Y.Z
$ mkdir Scotogenic
\end{lstlisting}
\vsp

Then, we copy the \SPheno module to this directory

\vsp
\begin{lstlisting}[style=terminal]
$ cp -r $PATH/SARAH-X.Y.Z/Output/Scotogenic/EWSB/SPheno/* $PATH/SPheno-X.Y.Z/Scotogenic
\end{lstlisting}
\vsp

Notice that we had to add {\tt -r} because the \SPheno module was
created by \SARAH using sub-directories. Finally, in order to compile
the \Fortran code we run

\vsp
\begin{lstlisting}[style=terminal]
$ make Model=Scotogenic
\end{lstlisting}
\vsp

\noindent and wait until the compilation has finished. Once this
happens, we are ready to use our \SPheno code for the scotogenic
model.

\subsubsection*{Using \SPheno}

As explained above, \SPheno reads an input file in LesHouches
format. When the \SPheno module is created, \SARAH produces two
templates in {\tt
  \$PATH/SARAH-X.Y.Z/Output/Scotogenic/EWSB/SPheno/Input_Files}. Since we have
copied them to the \SPheno directory, they will also be located in
{\tt \$PATH/SPheno-X.Y.Z/Scotogenic/Input_Files}. The files are named {\tt
  LesHouches.in.Scotogenic} and {\tt
  LesHouches.in.Scotogenic_low}. For non-supersymmetric models there
is no difference between them. Therefore, we will just take the first
one for our numerical studies in the scotogenic model.

It is convenient to place ourselves in the root directory of \SPheno
and copy the selected input file to this directory. This is done with
the terminal commands

\vsp
\begin{lstlisting}[style=terminal]
$ cd $PATH/SPheno-X.Y.Z
$ cp Scotogenic/Input_Files/LesHouches.in.Scotogenic .
\end{lstlisting}
\vsp

If we open the input file we will see that the information (model
parameters and \SPheno flags) is distributed in blocks. The first
block that is relevant to us is {\tt MINPAR}. We defined this block in
the {\tt SPheno.m} file we created for \SARAH, and now we can find
here the input values for the parameters $\lambda_i$, with $i=1, \dots
, 5$, and $m_\eta^2$. All dimensionful parameters (like $m_\eta^2$)
are assumed to be given in GeV (or powers of GeV).



The block called {\tt SPhenoInput} includes many \SPheno options. Each
option is given in terms of a flag and value. Comments are also
included to help the user identify the meaning of each option. For
example, flag number {\tt 11} determines whether \SPheno should
calculate decay rates (if we choose the value {\tt 1}) or not (if we
choose the value {\tt 0}). For reasons that will become clear in
lecture 2 (devoted to \MO), we will modify the value of flag number
{\tt 50} to {\tt 0}:

\vsp
\begin{lstlisting}[style=LesHouches,firstnumber=33]
50 0               # Majorana phases: use only positive masses
\end{lstlisting}
\vsp

The rest of the options will be left as given by default for this
course, but we encourage to take a look at the \SPheno manual to learn
about their meanings. For simplicity, we will not discuss the rest of
blocks. The input values for all the parameters of the model are
introduced in the three blocks we have mentioned, {\tt MINPAR}, {\tt
  MNIN} and {\tt YNIN}, whereas the \SPheno options are given in the
block {\tt SPhenoInput}.

We are ready to introduce input values and run the code. Let us
consider the benchmark point {\bf BS1} (Benchmark Scotogenic 1)
defined by the input parameters

\begin{center}
\fbox{
\begin{minipage}[c]{0.8\textwidth}
\vspace*{0.2cm}
\centering
\begin{tabular}{lll}
$\lambda_1 = 0.26 \qquad $ & $\lambda_2 = 0.5 \qquad $ & $\lambda_3 = 0.5 \qquad $ \\
$\lambda_4 = -0.5 \qquad $ & $\lambda_5 = 8 \cdot 10^{-11} \qquad $ & $m_\eta^2 = 1.85 \cdot 10^5 \, \text{GeV}^2 \qquad $
\end{tabular}
\begin{equation*}
M_N = \left( \begin{array}{ccc} 
345 \, \text{GeV} & 0 & 0 \\
0 & 4800 \, \text{GeV} & 0 \\
0 & 0 & 6800 \, \text{GeV}
\end{array} \right)
\end{equation*}
\begin{equation*}
Y_N = \left( \begin{array}{ccc} 
0.0172495 & 0.300325 & 0.558132 \\
-0.891595 & 1.00089 & 0.744033 \\
-1.39359 & 0.207173 & 0.253824
\end{array} \right)
\end{equation*}
\vspace*{0.1cm}
\end{minipage}
}
\end{center}

As we will see later, this benchmark point is experimentally excluded,
but it will serve to illustrate how to use the computer tools we are
interested in. We must introduce these input values in the
corresponding entries in the {\tt LesHouches.in.Scotogenic} file. This
results in

\vsp
\begin{lstlisting}[style=LesHouches,firstnumber=12]
Block MINPAR      # Input parameters 
1   0.260000E+00    # lambda1Input
2   0.500000E+00    # lambda2Input
3   0.500000E+00    # lambda3Input
4  -0.500000E+00    # lambda4Input
5   8.000000E-11    # lambda5Input
6   1.850000E+05    # mEt2Input
\end{lstlisting}
\vsp

\noindent and

\vsp
\begin{lstlisting}[style=LesHouches,firstnumber=47]
Block MNIN    #  
1 1   3.450000E+02         # Mn(1,1)
1 2   0.000000E+00         # Mn(1,2)
1 3   0.000000E+00         # Mn(1,3)
2 1   0.000000E+00         # Mn(2,1)
2 2   4.800000E+03         # Mn(2,2)
2 3   0.000000E+00         # Mn(2,3)
3 1   0.000000E+00         # Mn(3,1)
3 2   0.000000E+00         # Mn(3,2)
3 3   6.800000E+03         # Mn(3,3)
Block YNIN    #  
1 1   1.724950E-02         # Yn(1,1)
1 2   3.003250E-01         # Yn(1,2)
1 3   5.581320E-01         # Yn(1,3)
2 1  -8.915950E-01         # Yn(2,1)
2 2   1.000890E-00         # Yn(2,2)
2 3   7.440330E-01         # Yn(2,3)
3 1  -1.393590E-00         # Yn(3,1)
3 2   2.071730E-01         # Yn(3,2)
3 3   2.538240E-01         # Yn(3,3)
\end{lstlisting}
\vsp

The syntaxis is clear. For example, the first line in the block {\tt
  YNIN} is the value of the real part of $\left(Y_N\right)_{11}$. If
only these two blocks are present, \SPheno will assume that $M_N$ and
$Y_N$ are real. In principle one could introduce additional blocks for
the imaginary parts, but we will not do so in this course. Notice also
that all dimensionful parameters in the LesHouches input file are
given in GeV or GeV$^2$.

And now we can execute the code. This is done with

\vsp
\begin{lstlisting}[style=terminal]
$ bin/SPhenoScotogenic
\end{lstlisting}
\vsp

And \SPheno is running! After a few seconds it will be finished and a
few output files will be generated in the \SPheno root directory. We
are only interested in the file called {\tt SPheno.spc.Scotogenic},
where all the output information is saved. This type of file is
usually called \emph{spectrum file}, since it contains all the details
of the mass spectrum of the model.

Again, we can simply open the output file with a text editor and read
it. The output values are distributed in blocks, following the
LesHouches format. The name of the blocks and the comments added in
the output are quite intuitive, making unnecessary a long explanation
about what we have in this file. However, let us comment on some
particularly important blocks.

After some blocks with the values of all parameters in the model
(scalar couplings, Yukawa matrices, \dots), we find the block {\tt
  MASS}.

\vsp
\begin{lstlisting}[style=LesHouchesOut,firstnumber=117]
Block MASS  # Mass spectrum
#   PDG code      mass          particle
        25     1.25548349E+02  # hh
      1001     4.30116263E+02  # etR
      1002     4.30116263E+02  # etI
      1003     4.47388134E+02  # etp
        23     9.11887000E+01  # VZ
        24     8.03497269E+01  # VWp
         1     5.00000000E-03  # Fd_1
         3     9.50000000E-02  # Fd_2
         5     4.18000000E+00  # Fd_3
         2     2.50000000E-03  # Fu_1
         4     1.27000000E+00  # Fu_2
         6     1.73500000E+02  # Fu_3
        11     5.10998930E-04  # Fe_1
        13     1.05658372E-01  # Fe_2
        15     1.77669000E+00  # Fe_3
        12    -9.04451554E-13  # Fv_1
        14    -1.00553015E-11  # Fv_2
        16    -4.81551646E-11  # Fv_3
      1012    -3.45000000E+02  # Chi_1
      1014    -4.80000000E+03  # Chi_2
      1016    -6.80000000E+03  # Chi_3
\end{lstlisting}
\vsp

We can read in this block the masses for all mass eigenstates in the
model. For example, we can see that in the benchmark point {\bf BS1}
the Higgs boson mass is found to be $m_h = 125.5$ GeV. Another detail
that is remarkable about these results is the presence of non-zero
masses for the neutrinos. Even though we have computed the rest of
masses at tree-level, \SPheno has also included 1-loop corrections to
neutrino masses. We see that the lightest neutrino is actually
massless (note the vanishing row in $Y_N$ for {\bf BS1}), although
\SPheno returns a tiny (non-zero) numerical value. Finally, we find
that some of the Majorana fermion masses are negative. This is just a
convention that will be explained in Sec. \ref{subsec:MO-running}.

Next, we find several blocks with mixing matrices. For example:

\vsp
\begin{lstlisting}[style=LesHouchesOut,firstnumber=150]
Block UVMIX Q=  1.60000000E+02  # ()
  1  1     6.24218732E-02   # Real(UV(1,1),dp)
  1  2     7.13483380E-01   # Real(UV(1,2),dp)
  1  3    -6.97886077E-01   # Real(UV(1,3),dp)
  2  1     6.67236147E-01   # Real(UV(2,1),dp)
  2  2     4.90182507E-01   # Real(UV(2,2),dp)
  2  3     5.60818183E-01   # Real(UV(2,3),dp)
  3  1     7.42225999E-01   # Real(UV(3,1),dp)
  3  2    -5.00662138E-01   # Real(UV(3,2),dp)
  3  3    -4.45463791E-01   # Real(UV(3,3),dp)
\end{lstlisting}
\vsp

This is the neutrino mixing matrix. Note that the resulting values for
the mixing angles are in good agreement with current oscillation
data. \SPheno can also compute many quark and lepton flavor
observables thanks to the \FlavorKit extension. The numerical results
can be found in the blocks {\tt FlavorKitQFV} and {\tt
  FlavorKitLFV}. For example, the branching ratios for the radiative
lepton decays $\ell_i \to \ell_j \gamma$ are

\vsp
\begin{lstlisting}[style=LesHouchesOut,firstnumber=320]
Block FlavorKitLFV # lepton flavor violating observables 
     701    1.02041308E-10  # BR(mu->e gamma)
     702    5.89179996E-12  # BR(tau->e gamma)
     703    1.27947536E-09  # BR(tau->mu gamma)
\end{lstlisting}
\vsp

Therefore, this point is actually excluded, since it predicts a
branching ratio for the radiative decay $\mu \to e \gamma$ above the
current experimental limit ($4.2 \cdot 10^{-13}$) established by the
MEG experiment. Finally, \SPheno also computes decay rates. The
results are located by the end of the output file. For instance, the
Higgs boson branching ratios are found to be

\vsp
\begin{lstlisting}[style=LesHouchesOut,firstnumber=692]
DECAY        25     3.94503471E-03   # hh
#    BR                NDA      ID1      ID2
     2.66302509E-03    2           22         22   # BR(hh -> VP VP )
     8.72461559E-02    2           21         21   # BR(hh -> VG VG )
     2.48408517E-02    2           23         23   # BR(hh -> VZ VZ )
     2.26599428E-01    2          -24         24   # BR(hh -> VWp^* VWp_virt )
     2.11242982E-04    2           -3          3   # BR(hh -> Fd_2^* Fd_2 )
     5.65822420E-01    2           -5          5   # BR(hh -> Fd_3^* Fd_3 )
     2.27603074E-04    2          -13         13   # BR(hh -> Fe_2^* Fe_2 )
     6.56985485E-02    2          -15         15   # BR(hh -> Fe_3^* Fe_3 )
     2.66900197E-02    2           -4          4   # BR(hh -> Fu_2^* Fu_2 )
\end{lstlisting}
\vsp

Therefore, we find that the dominant Higgs boson decay in the {\bf
  BS1} benchmark point is to $b \bar b$, as expected for a Higgs mass
in the $125$ GeV ballpark.

We conclude our brief review of \SPheno emphasizing that many other
things can be done with this code. In combination with \SARAH, one can
create \SPheno modules for many models and use them to run all kinds
of numerical computations.

\subsection{Summary of the lecture}
\label{subsec:SARAH-summary}

In this lecture we learned how to use \SARAH to study the properties
of our favourite model and produce input files for other computer
tools. We also had a brief overview of \SPheno, the numerical tool
that complements \SARAH perfectly. These two codes will be central in
the rest of the course and we will make use of the input files
produced in this lecture as basis to run \MO and \MG. Furthermore, we
used the scotogenic model as a simple example that allows for an easy
introduction to the basics. In lectures 2 and 3 we will continue
applying computer tools to this model.


\newpage

%% file: lecture2.tex
\section{Lecture 2: Computing dark matter properties with \MO}
\label{sec:lecture2}

\subsection{What is \MO?}
\label{subsec:MO}

\MO is probably the most popular computer tool for the study of dark
matter. First developed to compute the relic density of a stable
massive particle, the code also computes the rates for direct and
indirect detection rates of dark matter. Nowadays, many
phenomenologists and dark matter model builders use it on a daily
basis.

\subsection{\MO: Technical details, installation and load}
\label{subsec:MO-details}

\begin{itemize}

\item {\bf Name of the tool:} \MO

\item {\bf Author:} Genevi\`eve B\'elanger, Fawzi Boudjema, Alexander
  Pukhov and Andrei Semenov. They can be contacted via
  micromegas@lapth.cnrs.fr.

\item {\bf Type of code:} {\tt C} and \Fortran

\item {\bf Website:} \url{https://lapth.cnrs.fr/micromegas/}

\item {\bf Manual:} The manual for the latest version of \MO can be
  found in \cite{Belanger:2014vza}. For previous versions see
  \cite{Belanger:2006is,Belanger:2008sj,Belanger:2010gh,Belanger:2013oya}.

\end{itemize}

After downloading the package, one should copy the {\tt tar.gz} file
into the {\tt \$PATH} folder and extract its contents,

\vsp
\begin{lstlisting}[style=terminal]
$ cp Download-Directory/micromegas_X.Y.Z.tar.gz $PATH/
$ cd $PATH
$ tar -xf micromegas_X.Y.Z.tar.gz
\end{lstlisting}
\vsp

Here {\tt X.Y.Z} must be replaced by the \MO version which has been
downloaded. The next step is the compilation of the code, performed
with the commands

\vsp
\begin{lstlisting}[style=terminal]
$ cd micromegas_X.Y.Z
$ make
\end{lstlisting}
\vsp

And \MO will be ready to run our own dark matter studies.

\subsection{General usage and description of the input files}
\label{subsec:MO-input}

Before we describe the input files generated by \SARAH, it is
convenient to explain the general usage of \MO. This will clarify the
role of each file.

Strictly speaking, \MO is not a code, but a collection of routines for
the evaluation of dark matter properties. It contains many
model-independent functions and routines, which can be used for the
specific models we are studying. The way this is done is quite
simple. The user must write a short \emph{steering file}, or main
program, that (i) defines options for the DM calculations and output,
and (ii) calls the built-in routines in \MO that run the desired DM
calculations. Hence, the user does not need to enter into the details
of the \MO routines, but just call them with the proper options. \MO
will then read the details of the input model (contained in external
{\tt mdl} files, to be provided for each model) and execute the
routines, returning the DM properties (such as relic density and
detection rates) required by the user in the steering file.

Let us now describe the input files. As already explained in
Sec. \ref{subsec:SARAH-output}, \SARAH can produce model files for
\MO. Thanks to this feature, the user gets rid of the most tedious
task when working with \MO. Once generated, these files are located in
the directory {\tt
  \$PATH/SARAH-X.Y.Z/Output/Scotogenic/EWSB/CHep}. One finds the
following files:

\begin{itemize}

\item {\tt CalcOmega.cpp}
\item {\tt CalcOmega_with_DDetection_MOv3.cpp}
\item {\tt CalcOmega_with_DDetection_MOv4.cpp}
\item {\tt CalcOmega_with_DDetection_MOv4.2.cpp}
\item {\tt func1.mdl}
\item {\tt lgrng1.mdl}
\item {\tt prtcls1.mdl}
\item {\tt vars1.mdl}

\end{itemize}

As explained above, the {\tt mdl} files define the input model, with
details such as particle content, interactions and parameters. On the
other hand, the {\tt cpp} files are steering files that tell \MO what
dark matter properties we are interested in. Although we will not have
to get into the details of these files (since \SARAH did it for us),
let us briefly review their content.

The {\tt mdl} files contain information about the model. In the file
{\tt vars1.mdl} one finds the definition of several decay widths of
the particles in the model (including the SM ones) and standard
parameters like the Fermi constant $G_F$. The file {\tt func1.mdl} is
devoted to the \emph{constrained variables} of the model, this is, to
all masses and vertices. The Feynman rules are given in the file {\tt
  lgrng1.mdl}. Notice that each interaction vertex is given in terms
of the interacting states, the Lorentz structure and the value of the
vertex itself, using for the former the list of vertices defined in
{\tt func1.mdl}. For example, we see that the $\eta_I-\eta_I-h$
Feynman rule is given by the {\tt v0002} vertex, which is defined to
be equal to $-(\lambda_3+\lambda_4-\lambda_5) \, v$ in {\tt
  func1.mdl}. Also note that the parameters of the model use the names
we introduced in the {\tt parameters.m} file using the {\tt
  OutputName} option. Finally, the file {\tt prtcls1.mdl} contains
information about the particles in the model.

The {\tt cpp} files are the main programs, in this case {\tt C}
programs, containing all the calculations we want \MO to perform. The
file {\tt CalcOmega.cpp} only computes the dark matter relic density,
$\Omega_{\text{DM}} h^2$, whereas the other files (written by \SARAH
to be used with different \MO versions) also include the calculation
of direct detection rates. We will talk a little about this
possibility in Sec. \ref{subsec:MO-other}. However, for our first
example we will use the file {\tt CalcOmega.cpp} to obtain
$\Omega_{\text{DM}} h^2$.

\subsection{Running \MO}
\label{subsec:MO-running}

In order to implement our model in \MO, we must create a new project
and copy the files to the corresponding folder. This is done with

\vsp
\begin{lstlisting}[style=terminal]
$ cd $PATH/micromegas_X.Y.Z
$ ./newProject Scotogenic
$ cd Scotogenic
$ cp $PATH/SARAH-X.Y.Z/Output/Scotogenic/EWSB/CHep/* work/models
\end{lstlisting}
\vsp

The next step is the compilation of the selected {\tt cpp}
file. However, before we do that, let us notice one thing. In the
directory {\tt \$PATH/micromegas_X.Y.Z/Scotogenic} one can find two
files, {\tt main.c} and {\tt main.F}, with example programs for
\MO. They are equivalent to the {\tt cpp} files generated by \SARAH
and contain examples of calculations one can perform with
\MO. Therefore, although we will not use them in this course, it might
be helpful to take a look at them in order to see the different
options and how to turn on specific calculations and outputs in \MO.

In order to compile our own \MO code for the scotogenic model we
execute

\vsp
\begin{lstlisting}[style=terminal]
$ mv work/models/CalcOmega.cpp .
$ make main=CalcOmega.cpp
\end{lstlisting}
\vsp

This will create the binary file {\tt CalcOmega} in the {\tt
  Scotogenic} folder. In order to run it and get our results there is
only one thing missing: input parameters. For this purpose we will
make use of a very convenient feature of \MO: it can read a spectrum
file in LesHouches format. Therefore, we can use \SPheno, run with the
input values of our choice, and pass the resulting output file to \MO,
which can then read it and compute the DM observables.

\tip{It has become common nowadays to combine different codes. This
  makes the study of a model a more efficient task, since using the
  output of a code as input for another code solves many conversion
  and formatting issues. For this reason, it is convenient to choose
  computer tools which can be easily combined.}

We already learned how to run \SPheno in
Sec. \ref{subsec:SPheno}. There is only one detail that we must take
into account in case we want to use the \SPheno output file as input
for \MO. \MO cannot handle rotation matrices with complex
entries. Since these may appear in some calculations with Majorana
fermions (like the neutrinos in the scotogenic model), we must tell
\SPheno that we want numerical results without them. Indeed, it is
common to find purely imaginary rotation matrices, or rows of them, in
models with Majorana fermions even in the absence of CP violating
phases. This type of complex phases can be absorbed by adding a
negative sign to the mass of the Majorana fermion. This can be easily
understood by looking at the transformation between the mass matrix in
the gauge basis ($M$) and the diagonal mass matrix in the mass basis
($\hat M$). For a Majorana fermion, this transformation is of the form
\begin{equation} \label{eq:nurot}
V \, M \, V^T = \hat M \, ,
\end{equation}
where $V$ is a unitary matrix. It is clear that multiplying a row of
the $V$ matrix by the imaginary unit $i$ is equivalent to a change of
sign in one eigenvalue of $\hat M$. Therefore, it is just a matter of
convention whether we present the results with complex rotation
matrices and positive masses or with real rotation matrices and
negative masses. \MO can only understand the input if we take the
second option, and thus we must tell \SPheno to produce an output file
with this choice. This is done by setting the flag {\tt 50} in the
LesHouches input file of \SPheno to the value {\tt 0}, as we already
did in Sec. \ref{subsec:SPheno}.

After this comment, we can proceed to run \MO in the benchmark point
{\bf BS1} of the scotogenic model. In order to do this, we must copy
the \SPheno spectrum file to the {\tt Scotogenic} folder in \MO and
execute the binary file we just created

\vsp
\begin{lstlisting}[style=terminal]
$ cp $PATH/SPheno-X.Y.Z/SPheno.spc.Scotogenic .
$ ./CalcOmega
\end{lstlisting}
\vsp

The first time we run the binary it can take some time, even up to
several hours depending on the computer power, since \MO has to
compile all necessary annihilation channels of the DM candidate for
that particular parameter point. All further evaluations of similar
points are done in a second or less.

When the run is finished, we get the results on the screen:

\vsp
\begin{lstlisting}[style=terminal]
Masses of odd sector Particles:
~N1  : MN1   =   345.0 || ~etI : MetI  =   430.1 || ~etR : MetR  =   430.1 
~etp : Metp  =   447.4 || ~N2  : MN2   =  4800.0 || ~N3  : MN3   =  6800.0 

Xf=2.41e+01 Omega h^2=1.08e+00

# Channels which contribute to 1/(omega) more than 1%.
# Relative contributions in % are displayed
   28% ~N1 ~N1 ->e3 E3 
   21% ~N1 ~N1 ->nu2 nu3 
   15% ~N1 ~N1 ->nu2 nu2 
    8% ~N1 ~N1 ->e2 E3 
    8% ~N1 ~N1 ->E2 e3 
    7% ~N1 ~N1 ->nu3 nu3 
    4% ~N1 ~N1 ->nu1 nu2 
    3% ~N1 ~N1 ->nu1 nu3 
    2% ~N1 ~N1 ->e2 E2
\end{lstlisting}
\vsp

First, \MO writes the masses (in GeV) of all particles charged under
the $\mathbb{Z}_2$ parity. Note that their names are written including
a tilde (\textasciitilde), in contrast to the names of the
$\mathbb{Z}_2$-even particles, which do not have it. Since the
lightest $\mathbb{Z}_2$-odd particle in the {\bf BS1} point is the
lightest right-handed neutrino, $N_1$, it is stable and constitutes
the dark matter of the universe.

Next, \MO gives us two quantities. $x_f = m_{N_1}/T_f$ characterizes
the freeze-out temperature, $T_f$, and $\Omega_{\text{DM}} h^2$ is the
dark matter relic density. We see that in the benchmark point {\bf
  BS1} we obtain $\Omega_{\text{DM}} h^2 = 1.08$. This relic density
is too high, since the Planck observations prefer a value in the
$\Omega_{\text{DM}} h^2 \sim 0.11$ ballpark. Therefore, this parameter
point is also excluded due to DM constraints.

\MO also gives a list with the annihilation channels that give the
most relevant contributions to the DM relic density. In the {\bf BS1}
point, the most important one is
\begin{equation}
N_1 N_1 \to \tau^+ \tau^-
\end{equation}
which constitutes $28\%$ of the total annihilation cross-section. Note
also that flavor violating channels are present as well in the
list. For example, we find that the channels
\begin{equation}
N_1 N_1 \to \mu^\pm \tau^\mp
\end{equation}
contribute with $16\%$ of the annihilation cross-section.

Finally, note that this information is exported to the external files
{\tt omg.out} and {\tt channels.out}. These two files are written
following the LesHouches format: each entry is defined by a flag (or a
few of them) and a numerical value.

\subsection{Other computations in \MO}
\label{subsec:MO-other}

In the previous Section we learned how to use the {\tt CalcOmega.cpp}
file which is automatically provided by \SARAH. With the aid of this
file we can easily compute the DM relic density. However, it is easy
to modify these files to (i) change some details of these
calculations, (ii) change the amount of information that is shown as
output, and (iii) compute additional observables.

We already saw that we can compute direct detection rates with the
files {\tt CalcOmega_with_DDetection_MOvX.cpp}, where {\tt X} refers
to the \MO version. These files~\footnote{In this course we are using
  {\tt micromegas_4.2.5} and then the chosen file should be {\tt
    CalcOmega_with_DDetection_MOv4.2.cpp}.} are the main files for a
{\tt C} program that uses \MO to calculate the DM relic density
$\Omega_{\text{DM}} h^2$ as well as some direct detection rates: (i)
spin independent cross-section with proton and neutron in pb, (ii)
spin dependent cross-section with proton and neutron in pb, (iii)
recoil events in the 10 - 50 keV region at $^{73}Ge$, $^{131}Xe$,
$^{23}Na$ and $^{127}I$ nuclei. We decided not to use this file in
benchmark point {\bf BS1} because, for this parameter point, the DM
scattering cross-sections with nucleons is zero at
tree-level. Therefore, we would have obtained vanishing direct
detection rates.

Just to see how direct detection rates are obtained, let us consider a
slight modification of the {\bf BS1} benchmark point. The reason why
the benchmark point {\bf BS1} leads to vanishing direct detection
rates at tree-level is because the DM particle in this point is the
lightest right-handed neutrino and this state does not couple directly
to the nucleons. Instead, in a parameter point with scalar DM
($\eta_I$), the tree-level scattering cross-section with the nucleons
does not vanish. Therefore, let us define a new benchmark point, {\bf
  BS2} (Benchmark Scotogenic 2), with a lighter $\eta_I$ state. The
only change with respect to the {\bf BS1} point is:

\begin{center}
\fbox{
\begin{minipage}[c]{0.8\textwidth}
\vspace*{0.2cm}
\begin{equation*}
m_{\eta}^2 = 5 \cdot 10^4 \, \text{GeV}^2
\end{equation*}
\vspace*{0.1cm}
\end{minipage}
}
\end{center}

Using this parameter point is straightforward. We just have to modify
a single line in the {\tt MINPAR} block of the {\tt
  LesHouches.in.Scotogenic} input file:

\vsp
\begin{lstlisting}[style=LesHouches,firstnumber=18]
6   5.000000E+04    # mEt2Input
\end{lstlisting}
\vsp

After running \SPheno with this modification in the input file, we
generate a new {\tt SPheno.spc.Scotogenic} output file that we can use
with \MO. We can easily check that this parameter indeed leads to a
much lighter $\eta_I$ state:

\vsp
\begin{lstlisting}[style=LesHouchesOut,firstnumber=121]
1002     2.23606798E+02  # etI
\end{lstlisting}
\vsp

Now we can create the {\tt CalcOmega_with_DDetection_MOvX} binary. This is
completely analogous to what we did for the {\tt CalcOmega} binary:

\vsp
\begin{lstlisting}[style=terminal]
$ mv work/models/CalcOmega_with_DDetection_MOvX.cpp .
$ make main=CalcOmega_with_DDetection_MOvX.cpp
\end{lstlisting}
\vsp

Once compiled, we copy our new {\tt SPheno.spc.Scotogenic} file and
run the {\tt CalcOmega_with_DDetection_MOvX} binary file,

\vsp
\begin{lstlisting}[style=terminal]
$ cp $PATH/SPheno-X.Y.Z/SPheno.spc.Scotogenic .
$ ./CalcOmega_with_DDetection_MOvX
\end{lstlisting}
\vsp

This is the result that is printed on the screen:

\vsp
\begin{lstlisting}[style=terminal]
Masses of odd sector Particles:
~etI : MetI  =   223.6 || ~etR : MetR  =   223.6 || ~etp : Metp  =   255.3 
~N1  : MN1   =   345.0 || ~N2  : MN2   =  4800.0 || ~N3  : MN3   =  6800.0 

Xf=2.93e+01 Omega h^2=2.69e-03

# Channels which contribute to 1/(omega) more than 1%.
# Relative contributions in % are displayed
   23% ~etI ~etR ->nu3 nu3 
   13% ~etR ~etR ->Wp Wm 
   13% ~etI ~etI ->Wp Wm 
   11% ~etR ~etR ->nu3 nu3 
   11% ~etI ~etI ->nu3 nu3 
    5% ~etI ~etR ->nu2 nu2 
    4% ~etI ~etR ->nu2 nu3 
    3% ~etR ~etR ->Z Z 
    3% ~etI ~etI ->Z Z 
    3% ~etR ~etR ->nu2 nu2 
    3% ~etI ~etI ->nu2 nu2 
    2% ~etR ~etR ->nu2 nu3 
    2% ~etI ~etI ->nu2 nu3 

==== Calculation of CDM-nucleons amplitudes  =====
         TREE LEVEL
PROCESS: QUARKS,~etI->QUARKS,~etI{d1,D1,d2,D2,d3,D3,u1,U1,u2,U2,u3,U3
Delete diagrams with _S0_!=1,_V5_,A
CDM-nucleon micrOMEGAs amplitudes:
proton:  SI  -1.235E-18  SD  0.000E+00
neutron: SI  -1.253E-18  SD  0.000E+00
         BOX DIAGRAMS
CDM-nucleon micrOMEGAs amplitudes:
proton:  SI  -1.235E-18  SD  0.000E+00
neutron: SI  -1.253E-18  SD  0.000E+00
CDM-nucleon cross sections[pb]:
 proton  SI 6.616E-28  SD 0.000E+00
 neutron SI 6.802E-28  SD 0.000E+00

======== Direct Detection ========
73Ge: Total number of events=1.41E-22 /day/kg
Number of events in 10 - 50 KeV region=7.61E-23 /day/kg
131Xe: Total number of events=2.35E-22 /day/kg
Number of events in 10 - 50 KeV region=1.18E-22 /day/kg
23Na: Total number of events=1.40E-23 /day/kg
Number of events in 10 - 50 KeV region=7.66E-24 /day/kg
I127: Total number of events=2.30E-22 /day/kg
Number of events in 10 - 50 KeV region=1.18E-22 /day/kg
\end{lstlisting}
\vsp

We note that this scenario leads to a tiny dark matter relic density,
of the order of $2.69 \cdot 10^{-3}$, due to the large annihilation
cross-sections into pairs of neutrinos and gauge bosons ($W^+ W^-$ and
$Z Z$). In fact, these are not only annihilations, but also
co-annihilations with the $\eta_R$ state, which is almost degenerate
in mass. Regarding the calculation of the direct detection
cross-sections, the most relevant information is given after {\tt
  CDM-nucleon cross sections[pb]}. These are the (spin independent and
dependent) cross-sections with proton and neutron in pb. It is usually
convenient to multiply these values by a factor $10^{-36}$ to get the
cross-sections in cm$^2$, the units commonly employed by the
experimental collaborations. We find that in the {\bf BS2} point these
cross-sections are tiny.

Finally, \MO also computes the number of recoil events per day in the
10 - 50 keV region for a kg of $^{73}Ge$, $^{131}Xe$, $^{23}Na$ and
$^{127}I$. Again, and due to the small direct detection
cross-sections, these numbers are tiny in the {\bf BS2} point. The
largest number of events would be obtained in $^{131}Xe$, but even in
this case we would expect only $\sim 10^{-22} \, \text{events} \,
\text{kg}^{-1} \, \text{day}^{-1}$.

Before concluding the lecture, we emphasize that many other dark
matter related observables can be computed using \MO. For a detailed
list see the \MO manual \cite{Belanger:2014vza}.

\subsection{Summary of the lecture}
\label{subsec:MO-summary}

In this lecture we learned how to use \MO to compute observables
related to dark matter physics. Since we had produced the input files
with \SARAH, we did not have to worry about how to write
them. Instead, we focused on their practical use to obtain reliable
predictions for the DM relic density and direct and indirect detection
rates.


\newpage

%% file: lecture3.tex
\section{Lecture 3: LHC physics with \MG}
\label{sec:lecture3}

\subsection{What is \MG?}
\label{subsec:MG}

\MG is a Monte Carlo event generator for collider studies, nowadays
widely used to simulate events at the LHC. Before the LHC era, this
tool was used to obtain future predictions in new physics
models. Currently, one can also recast the results of the searches
published by the LHC collaborations and interpret these analysis in
specific models.

The \MG software can be extended to incorporate several programs: a
random event generator, the code {\tt Pythia}, used for parton
showering and hadronization, and two detector simulators ({\tt PGS}
and {\tt Delphes}). This suit allows for a complete simulation at the
LHC, from events at the parton level to detector response.

Depending on the type of simulation we are interested in, some of
these additional pieces might be unnecessary. For example, in case we
just want to compute a cross-section at the parton level, it suffices
to use the basic \MG software. However, if we want to go beyond and
include hadronization or detector simulation we will have to use {\tt
  Pythia} and {\tt PGS} or {\tt Delphes} as well. This may sound a
little bit complicated, but in practice the combination of these tools
is straightforward, and in fact the \MG suit is prepared to do it
automatically.

\subsection{\MG: Technical details, installation and load}
\label{subsec:MG-details}

\begin{itemize}

\item {\bf Name of the tool:} \MG (more precisely, in this course we
  will use {\tt MadGraph5_aMC@NLO})

\item {\bf Author:} \emph{The MadTeam}, composed by Johan Alwall,
  Rikkert Frederix, Stefano Frixione, Michel Herquet, Valentin
  Hirschi, Fabio Maltoni, Olivier Mattelaer, Hua-Sheng Shao, Timothy
  J. Stelzer, Paolo Torrielli and Marco Zaro. They can be contacted
  through the \MG website.

\item {\bf Type of code:} \Python

\item {\bf Website:} \url{http://madgraph.hep.uiuc.edu/}

\item {\bf Manual:} See Refs. \cite{Alwall:2011uj,Alwall:2014hca}.

\end{itemize}

The version of \MG ({\tt MadGraph5_aMC@NLO}) we are going to use needs
\Python version 2.6 or 2.7 and {\tt gfortran/gcc} 4.6 or higher (in
case of NLO calculations). Besides those two requirements, which will
be fulfilled in most computers, there is no need for an installation
of \MG. The only \emph{special} requirement is to register in the \MG
website before being able to download the latest version of the
tool. Once this registration is done and the file is downloaded, we
just untar it as usual

\vsp
\begin{lstlisting}[style=terminal]
$ cp Download-Directory/MG5_aMC_vX_Y_Z.tar.gz $PATH/
$ cd $PATH
$ tar -xf MG5_aMC_vX_Y_Z.tar.gz
\end{lstlisting}
\vsp

Here {\tt X_Y_Z} is the version that has been downloaded. And then, in
order to load \MG we just get into the untarred folder and run the
binary file

\vsp
\begin{lstlisting}[style=terminal]
$ cd $PATH/MG5_aMC_vX_Y_Z
$ bin/mg5_aMC
\end{lstlisting}
\vsp

This opens \MG. In principle, we would be ready to use it. However,
before we do so let us configure some details and install additional
tools that can be added to the \MG suit.

\subsubsection*{Configuration and installation of additional tools}

Let us first comment on how to configure some options in \MG. We can
see that in the folder {\tt input} there is a file called {\tt
  mg5_configuration.txt}. This file contains the configuration details
of \MG, including options such as the prefered text editor (which can
be user to edit some input files, the so-called \emph{cards}, before
starting the simulation), the prefered web browser (some of the
results obtained in our \MG runs are shown in a user-friendly way with
a web browser) or the time given to the user to answer questions by
the code (some optoinal calculations can be switched on or off at some
intermediate steps in the runs). By default, these are {\tt vi}, {\tt
  Firefox} and 60 seconds. In case you do not like these choices, you
can simply modify them by removing the {\tt \#} symbol in front of
{\tt text_editor}, {\tt web_browser} and {\tt timeout}. For example,
many people will prefer {\tt emacs} instead of {\tt vi}.

\vsp
\begin{lstlisting}[style=mg5conf,firstnumber=42]
text_editor = emacs
\end{lstlisting}
\vsp

Once configured, we can consider adding further pieces to the \MG
puzzle. With the \MG code that we just downloaded, untarred and
configured we can run simulations at the parton level. This means that
we are interested in \emph{core} processes, such as $e^+ e^- \to q
\bar q$. This is already sufficient for many collider studies, for
example those aiming at the determination of an approximate number of
expected events for a given process in a given new physics
model. However, reality is way more complicated. The products of the
process will suffer many complex processes after the collision, such
as hadronization and showering, and they must be taken into account in
realistic simulations. Furthermore, detectors are not perfect, and
their simulation must also take into account many factors, such as
inefficiencies.

\tip{Although we will not use some of these additional codes for the
  moment, it is convenient to install them as soon as possible. This
  way we will be able to detect incompatibility issues which later,
  when we have already configured and run \MG many times, might be
  problematic.}

For this reason, we may want to install {\tt Pythia} and {\tt
  PGS}. These two additional codes are necessary if you are interested
in hadronization and detector response. otherwise, you do not need to
install them. However, before we install these two codes we must
install a prerequisite tool: {\tt ROOT}. This popular code is an
object oriented framework for large scale data analysis, and has been
developed by CERN. In order to take advantage of the rest of the
tools, we should install {\tt ROOT} before. The way this is done is
explained in Appendix \ref{sec:root}.

Once {\tt ROOT} is already installed, the installation of {\tt Pythia}
and {\tt PGS} is trivial. We just have to open \MG and execute a
command:

\vsp
\begin{lstlisting}[style=terminal]
$ bin/mg5_aMC
$ install pythia-pgs
\end{lstlisting}
\vsp

This way, \MG makes sure that these two codes are installed in the
correct locations and makes the required links to combine them in
future simulations. However, note that you must be connected to the
internet for this command to work.

Last but not least, there is another useful code: {\tt
  MadAnalysis}~\cite{Conte:2012fm}. This independent code, which
combines nicely with the \MG suit, is devoted to the analysis of our
simulations and is usually employed for plotting purposes. The output
of our simulations is given in several formats and we must use a tool
to \emph{extract} the relevant information (this can be done, for
example, with {\tt ROOT}). {\tt MadAnalysis} simplifies our life. One
can use it to read the \MG output and present the results in a
graphical way. Since we will interested in such graphical
representations, we are going to incorporate it to our \MG suit.

Like \MG, {\tt MadAnalysis} does not require any special
installation. It only requires to have {\tt ROOT} properly installed.

\vsp
\begin{lstlisting}[style=terminal]
$ cp Download-Directory/MadAnalysis5_vX.tar.gz $PATH/
$ cd $PATH
$ tar -xf MadAnalysis5_vX.tar.gz
\end{lstlisting}
\vsp

Here {\tt X} is the {\tt MadAnalysis} version that we have
downloaded. Once this is done, \MG will be absolutely ready to run our
own collider studies.

\subsection{General usage and description of the input files}
\label{subsec:MG-input}

As already explained, the \MG suit allows for different levels of
sofistication in the simulation:

\begin{equation*}
\begin{array}{c}
\text{Events at the parton level} \\
\text{\small \tt MadEvent}
\end{array} \quad \Rightarrow \quad \begin{array}{c}
\text{Showering and hadronization} \\
\text{\small \tt{Pythia}}
\end{array} \quad \Rightarrow \quad \begin{array}{c}
\text{Detector response} \\
\text{\small \tt{PGS} or \tt{Delphes}}
\end{array}
\end{equation*}

A complete review of all the possibilities is clearly beyond this
course. Fortunately, there are many resources to learn about \MG in
the internet: courses, guides and presentations. The first thing we
can do is to run the tutorial provided with \MG:

\vsp
\begin{lstlisting}[style=terminal]
$ cd $PATH/MG5_aMC_vX_Y_Z
$ bin/mg5_aMC
MG5_aMC > tutorial
\end{lstlisting}
\vsp

This tutorial shows the basic steps to simulate $t \bar t$ production
in the SM. In fact, we should note that when \MG loads the model that
is loaded by default is the SM: we can read on the screen {\tt Loading
  default model: sm}. The first command of the tutorial is the
generation of a new process

\vsp
\begin{lstlisting}[style=terminal]
MG5_aMC > generate p p > t t~
\end{lstlisting}
\vsp

This tells \MG that we want to simulate
\begin{equation*}
p p \to t \bar t
\end{equation*}
Next, we must export the process to several formats. This is obtained
with the \MG command

\vsp
\begin{lstlisting}[style=terminal]
MG5_aMC > output MY_FIRST_MG5_RUN
\end{lstlisting}
\vsp

This command will do all sorts of things. Among them, it will create a
new folder called {\tt MY_FIRST_MG5_RUN}. This is where the
information about this process the output of all future runs are
saved. There is a sub-folder we should visit. It is the one named {\tt
  Cards}. It is full of {\tt dat} files with information about the
process and about the simulation we are about to begin. There are two
files of special relevance to us: {\tt param_card.dat} and {\tt
  run_card.dat}. The first one contains the input values for all
parameters of the model, whereas the second one sets the parameters of
the run itself. For example, the user can determine in the {\tt
  run_card.dat} file the beam type, the energy, the renormalization
scale or the number of random events to launch in the Monte Carlo.

The next step in the tutorial is to launch the event generator
itself. For this we must execute the command

\vsp
\begin{lstlisting}[style=terminal]
MG5_aMC > launch MY_FIRST_MG5_RUN
\end{lstlisting}
\vsp

After we push enter, a question will appear on the screen. \MG needs
to know the type of run. We will have 60 seconds to answer (unless we
changed this option in {\tt mg5_configuration.txt}. In order to run
a complete simulation including hadronization and showering (with {\tt
  Pythia}) and detector response (with {\tt PGS}) we must press {\tt
  2}. Since {\tt PGS} requires {\tt Pythia} to run before, \MG will
automatically add it. Then we can simply push enter and a new question
will appear. \MG wants to know if we want to modify any of the cards
or just prefer to use the ones by default. For the moment we will
simply stick to the ones by default in the {\tt Cards}
sub-folder. Therefore, we just go ahead by pushing enter. And then the
web browser will open. This is a very user-friendly feature of \MG. As
soon as the simulation begins, all information is nicely displayed
using a web browser. After some time (not too long) \MG will be
done. We will be able to read the results on the terminal or on the
web page shown by the web browser. Either way, we find that the
cross-section for $t \bar t$ production is about $504.9 \pm 0.8$
pb. Of course, the exact number might be different (it has been
obtained by Monte Carlo methods), but it should agree within the
error.

We can also explore these results using the information on the
generated web page. The first thing we can see is that \MG shows the
Feynman diagrams used for the computation of the events. These can be
found in {\tt Process information}. For $t \bar t$ production in the
SM these can be classified into two types: gluon-induced and
quark-induced diagrams. Moreover, in the first case there are three
sub-diagrams. If we click on {\tt Results and event database} we can
see the numerical results obtained with the simulation. By clicking on
the cross-section result, we can even read the individual
contributions given by the different Feynman diagrams. In our case,
the dominant production channel is the gluon-induced one.

This concludes the tutorial, where we already run an interesting
calculation.

\subsection{Computing a cross-section}
\label{subsec:MG-xsection}

As an example of what we can do with \MG, we are going to compute a
production cross-section in the scotogenic model. For more fancy
simulations we refer to the next lecture, where we will make use of
all the tools installed in Sec. \ref{subsec:MG-details}.

Since the scotogenic model is not among the models provided with \MG,
we must add it. In the \MG folder there is a sub-folder named {\tt
  models}, where all the model definitions are saved in UFO format. In
order to implement the scotogenic model, we just have to copy the
files that we produced with \SARAH in the first lecture. These are
located in {\tt
  \$PATH/SARAH-X.Y.Z/Output/Scotogenic/EWSB/UFO}. Therefore, we must
execute the following terminal commands:

\vsp
\begin{lstlisting}[style=terminal]
$ cd $PATH/MG5_aMC_vX_Y_Z/models
$ mkdir Scotogenic
$ cp $PATH/SARAH-X.Y.Z/Output/Scotogenic/EWSB/UFO/* Scotogenic
\end{lstlisting}
\vsp

Next, we go back to the main \MG folder, open \MG and load the model:

\vsp
\begin{lstlisting}[style=terminal]
$ cd ..
$ bin/mg5_aMC
MG5_aMC > import model Scotogenic -modelname
MG5_aMC > define p d1 d1bar d2 d2bar u1 u1bar u2 u2bar g
\end{lstlisting}
\vsp

Note that we added the option {\tt -modelname}. This is used to keep
the names of the particles as given in the \SARAH model
files. Although sometimes this will not be necessary, it is convenient
to use this option when loading models created with \SARAH. Next, we
defined the multiparticle {\tt p}, including the gluon ({\tt g}) and
all the SM quarks. This is required because we are not using \MG's
naming conventions, and thus the multiparticle states must be
redefined.

Now we can run a simple simultation where we compute the cross-section
for the production of a pair of $\eta$ scalars, a CP-even neutral one
and a charged one,
\begin{equation*}
p \, p \to \eta_R \, \eta^+
\end{equation*}
We do this with the command

\vsp
\begin{lstlisting}[style=terminal]
MG5_aMC > generate p p > etr etp
\end{lstlisting}
\vsp

Next, we can create the output folder, which we will call {\tt
  SimScotogenic}, and launch the simulation

\vsp
\begin{lstlisting}[style=terminal]
MG5_aMC > output SimScotogenic
MG5_aMC > launch SimScotogenic
\end{lstlisting}
\vsp

To the first question we will simply press enter (equivalent to option
{\tt 0}), since we just want to compute the cross-section at partonic
level. Then we will get the second question, asking about whether we
want to modify any cards. And in this case we cannot simply press
enter. The {\tt param_card.dat} by default does not correspond to our
{\bf BS1} benchmark point. In fact, all the scotogenic model
parameters are zero in the file by default. Therefore, we must replace
the file by one with the correct input values for the model
parameters. Since the {\tt param_card.dat} file uses the standard
LesHouches format, we can simply use the {\tt SPheno.spc.Scotogenic}
file that we generated with \SPheno as input for \MG. This is as
simple as typing in the terminal

\vsp
\begin{lstlisting}[style=terminal]
$ cp $PATH/SPheno-X.Y.Z/SPheno.spc.Scotogenic $PATH/MG5_aMC_vX/SimScotogenic/Cards/param_card.dat
\end{lstlisting}
\vsp

Here, remember, {\tt X.Y.Z} and {\tt X} are the versions of the two
codes. This way, we have replaced the {\tt param_card.dat} file by
default by one of our own, with the correct parameter values in the
{\bf BS1} benchmark point. Once this is done, we can press enter and
continue with the simulation. \MG will print some warning
messages. These are due to some format issues in the {\tt
  SPheno.spc.Scotogenic} file. However, they are not relevant at
all. After a few minutes, we will get a result for the
cross-section. This is found to be $\sigma = 0.7943 \pm 0.0008$ fb (or
something very similar, since this is a random simulation).

We actually expected to get a small cross-section. The $\eta_R \,
\eta^+$ states are produced at the LHC by electroweak
interactions. This is shown in the Feynman diagrams produced by \MG,
see {\tt Process Information} in the web page, where we see that $p \,
p \to \eta_R \, \eta^+$ is induced by $W^+$ s-channel
exchange. Therefore, it is not surprising that the resulting
cross-section turns out to be much lower than the usual QCD induced
cross-sections at the LHC. With an integrated luminosity of $300$
fb$^{-1}$, as expected by the end of the LHC Phase I, we would obtain
about $240$ events of this type.

\MG can be also used to generate all possible chains (Feynman
diagrams) leading to a specific final state. For example, in the case
we just studied, the $\eta_R$ and $\eta^+$ scalars are not stable, but
they decay to final states including the lightest right-handed
neutrino $N_1$. For example, one can have the decays
\begin{align*}
\eta_R \to & N_1 \, \nu \\
\eta^+ \to & N_1 \, \mu^+
\end{align*}

where $\nu$ is any of the light neutrinos (undistinguishable at the
LHC). Therefore, we eventually get the process
\begin{equation}
p \, p \to N_1 \, N_1 \, \mu^+ \, \nu
\end{equation}
wich would be seen as an antimuon plus missing energy at the
LHC. However, the final state $N_1 \, N_1 \, \mu^+ \, \nu$ can be
reached by other production mechanisms (not only through intermediate
$\eta_R \, \eta^+$). How can we obtain all possible topologies in the
scotogenic model leading to $N_1 \, N_1 \, \mu^+ \, \nu$? \MG can do
it for us.

In fact, in order to see all possible diagrams leading to a specific
final state we do not even need to launch the simultation. We just
have to generate it. In this case we just need open \MG and run a few
commands

\vsp
\begin{lstlisting}[style=terminal]
$ bin/mg5_aMC
MG5_aMC > import model Scotogenic -modelname
MG5_aMC > define p d1 d1bar d2 d2bar u1 u1bar u2 u2bar g
MG5_aMC > generate p p > n1 n1 e2bar vl
MG5_aMC > output SimScotogenic2
\end{lstlisting}
\vsp

Note that in the generation of the process we have used the
multiparticle state {\tt vl}, containing all light neutrinos. This
simulation would take quite a long time before is finished. However,
we can already see the Feynman diagrams in the web page generated by
\MG. We can load it with

\vsp
\begin{lstlisting}[style=terminal]
MG5_aMC > open index.html
\end{lstlisting}
\vsp

Then, we just have to browse via {\tt Process Information} until we
find all the Feynman diagrams that participate in this process. The
Feynman diagram with intermediate $\eta_R \, \eta^+$ scalars is {\tt
  diagram 5}, and it is just one among many. We also find diagrams
where only one of them, $\eta_R$ or $\eta^+$, participates.

\tip{By default, \MG uses $10^4$ events for a simulation. This number
  will be sufficient in many cases, but not for rare LHC events. We
  must then use a larger number of events ($10^5$ or $10^6$) when we
  simulate events with small cross-sections. Otherwise, the resulting
  simulation will contain large statistical errors.}

\subsection{Summary of the lecture}
\label{subsec:MG-summary}

\MG is one of the most popular computer tools in particle physics, due
to its high level of sofistication. In this lecture we learned how to
use it to run simple LHC simulations. Since this code offers many more
possibilities, clearly beyond the scope of this introductory course,
we strongly encourage to explore its potential capabilities. The
internet is full of resources to learn how to use \MG, from slides to
tutorials, including many detailed guides.


\newpage

%% file: lecture4.tex
\section{Lecture 4: Final exercise}
\label{sec:lecture4}

\subsection{What is this lecture about?}
\label{subsec:lecture4}

In the final lecture we are going to review our previous lectures by
going through the whole process for a new model. For this purpose we
will choose the model introduced in \cite{Sierra:2015fma}, described
in detail in Sec. \ref{sec:appendix4}. As we will see, this model has
a few additional complications that will be helpful to learn a few
more features and possibilities of the computer tools presented in
this course.

\subsection{Implementing the model in \SARAH}
\label{subsec:implementing}

First of all, we must implement the model in \SARAH. We know already
that, in addition to useful analytical results, \SARAH can also
produce input files for the rest of the codes. Therefore, implementing
the model in \SARAH is always a practical approach.

The \SARAH name of the model will be {\tt DarkBS}. Since most of the
definitions are analogous to the ones in lecture 1, we will only
highlight those that require further refinements. All \SARAH model
files for the DarkBS model can be found in Appendix
\ref{sec:SARAH-darkbs}.

\subsubsection*{DarkBS.m}

The model is based on the extended gauge symmetry $U(1)_Y \times
SU(2)_L \times SU(3)_c \times U(1)_X$. Although the $\mathbb{Z}_2$
parity obtained after symmetry breaking is automatic, we must tell
\SARAH so that identifies the dark matter candidate,

\begin{lstlisting}[style=darkbs,firstnumber=14]
Global[[1]] = {Z[2], Z2};
\end{lstlisting}

The definition of
the gauge groups must contain an additional piece:

\begin{lstlisting}[style=darkbs,firstnumber=18]
Gauge[[1]]={B,   U[1], hypercharge, g1,False,1};
Gauge[[2]]={WB, SU[2], left,        g2,True,1};
Gauge[[3]]={G,  SU[3], color,       g3,False,1};
Gauge[[4]]={Bp,  U[1], Uchi,        gX,False,1};
\end{lstlisting}

The additional gauge group must also appear in the definition of the
particles in the model. For example, the vector-like fermions are
defined as

\begin{lstlisting}[style=darkbs,firstnumber=32]
FermionFields[[6]] = {lL, 1, {v4, e4},    -1/2, 2,  1,  2, 1};
FermionFields[[7]] = {lR, 1, {e5, v5},     1/2, 2,  1, -2, 1};
FermionFields[[8]] = {qL, 1, {u4, d4},     1/6, 2,  3,  2, 1};
FermionFields[[9]] = {qR, 1, {d5, u5},    -1/6, 2, -3, -2, 1};
\end{lstlisting}

Notice that all right-handed fields have opposite gauge charges to
those for the left-handed ones. This is equivalent to identifying, for
example, {\tt lR} with $\overline{L_R}$. As a consequence of this, we
must write the components of the right-handed doublets as in a {\tt
  -2} of $U(1)_X$. This practical choice simplifies the writing of the
Lagrangian, which takes a more transparent form. For instance, for the
Yukawa terms one has

\begin{lstlisting}[style=darkbs,firstnumber=58]
LagHC =  -(Yd conj[H].d.q + Ye conj[H].e.l + Yu H.u.q + mQ qL.qR + mL lL.lR + lamQ Phi.qR.q + lamL Phi.lR.l );
\end{lstlisting}

The additional $U(1)_X$ factor implies the existence of an additional
neutral gauge boson. Since this vector will get a mass after the
spontaneous breaking of the gauge symmetry, we can identify it with
the $Z^\prime$ boson. This is important when defining the gauge
sector:

\begin{lstlisting}[style=darkbs,firstnumber=63]
DEFINITION[EWSB][GaugeSector] =
{ 
  {{VB,VWB[3],VBp},{VP,VZ,VZp},ZZ},
  {{VWB[1],VWB[2]},{VWp,conj[VWp]},ZW}
};  
\end{lstlisting}

The rest of the {\tt DarkBS.m} file goes along the same lines of the
{\tt Scotogenic.m} file described in the first lecture. The only
detail that should be pointed out is the resulting fermion mixings,

\begin{lstlisting}[style=darkbs,firstnumber=80]
{{{dL,d4}, {conj[dR],d5}}, {{DL,Vd}, {DR,Ud}}},
{{{uL,u4}, {conj[uR],u5}}, {{UL,Vu}, {UR,Uu}}},
{{{eL,e4}, {conj[eR],e5}}, {{EL,Ve}, {ER,Ue}}},
\end{lstlisting}

It is important to note that the field that should be written in the
basis definition for the charged leptons is {\tt e5}, and not {\tt
  conj[e5]}. One can easily understand this fact by having a look at
the way {\tt eR} and {\tt e5} are defined in the Matter Fields
section.

\subsubsection*{parameters.m}

$U(1)$ mixing is a general feature in models with several $U(1)$
factors. \SARAH can perfectly handle this property, but we must define
the mixed gauge couplings in the {\tt parameters.m} file

\begin{lstlisting}[style=parameters,firstnumber=76]
{g1X,       {LaTeX -> "\\tilde{g}",
             LesHouches -> {GAUGE,10},
             OutputName -> g1X}},
{gX1,       {LaTeX -> "\\bar{g}",
             LesHouches -> {GAUGE,11},
             OutputName -> gX1}},
\end{lstlisting}

For the mixing matrices of the CP-even and CP-odd neutral scalars we
have to add the lines

\begin{lstlisting}[style=parameters,firstnumber=109]
{ZH, { Description->"Scalar-Mixing-Matrix", 
       LaTeX -> "Z^H",
       Real -> True, 
       DependenceOptional ->   {{-Sin[\[Alpha]],Cos[\[Alpha]]},
                                {Cos[\[Alpha]],Sin[\[Alpha]]}}, 
       Value -> None, 
       LesHouches -> SCALARMIX,
       OutputName-> ZH     }},
             
{ZA, { Description->"Pseudo-Scalar-Mixing-Matrix", 
       LaTeX -> "Z^A",
       Real -> True,
       DependenceOptional -> {{-Cos[\[Beta]],Sin[\[Beta]]},
                              {Sin[\[Beta]],Cos[\[Beta]]}}, 
       Value -> None, 
       LesHouches -> PSEUDOSCALARMIX,
       OutputName-> ZA      }},
\end{lstlisting}

The inclusion of the {\tt Description} options is crucial. This is
because it is necessary to properly identify these two matrices since
they play a role in some specific calculations (for example, the
calculation of the Higgs boson flavor violating decay rate to a pair
of leptons, $h \to \ell^+_i \ell_j^-$). Without this, \SARAH would not
know how to identify these matrices among all the mixing matrices in
the model. Notice also that these two mixing matrices have been
expressed in terms of the angles $\alpha$ and $\beta$.

Finally, the mixing matrix in the neutral gauge sector is also defined
in terms of two angles: $\theta_W$ and $\theta_W^\prime$:

\begin{lstlisting}[style=parameters,firstnumber=139]
{ZZ, { Description -> "Photon-Z Mixing Matrix",
       Dependence->  {{Cos[ThetaW],-Sin[ThetaW] Cos[ThetaWp], Sin[ThetaW] Sin[ThetaWp]},                             
                      {Sin[ThetaW],Cos[ThetaW] Cos[ThetaWp],-Cos[ThetaW] Sin[ThetaWp]},
                      {0, Sin[ThetaWp], Cos[ThetaWp]}} }},
\end{lstlisting}

\subsubsection*{particles.m}

There are just a few details worth pointing out in the {\tt
  particles.m} file. They all have to do with the same feature in this
model: many existing sets of mass eigenstates are now extended to
included additional particles. For example, the model has two CP-even
neutral scalars,

\begin{lstlisting}[style=particles,firstnumber=33]
{hh   ,  {  Description -> "Higgs",
            PDG -> {25,35},
            PDG.IX -> {101000001,101000002} }}, 
\end{lstlisting}

and, for example, four charged leptons,

\begin{lstlisting}[style=particles,firstnumber=78]
{Fe,   { Description -> "Leptons",
         PDG -> {11,13,15,17},
         PDG.IX -> {-110000601,-110000602,-110000603,-110000604} }},
\end{lstlisting}

It is also very important to define the new $Z^\prime$ boson. This is
done with the lines

\begin{lstlisting}[style=particles,firstnumber=55]
  {VZp,   { Description -> "Z'-Boson",
            Goldstone -> Ah[{2}] }},
\end{lstlisting}

Notice that the option {\tt Description} has been used to take
advantage of the general definition of $Z^\prime$ bosons in the file
{\tt \$PATH/SARAH-X.Y.Z/Models/particles.m}. Moreover, we must
indicate the Goldstone boson that constitutes the longitudinal part of
the massive $Z^\prime$. In this case this is given by the second
CP-odd neutral scalar, the first one being the Goldstone boson of the
SM $Z$ boson.

\subsubsection*{SPheno.m}

Finally, the last model file is {\tt SPheno.m}. We have decided to use
again a low scale version of \SPheno. There are only two details which
differ slightly from the {\tt SPheno.m} file we prepared for the
scotogenic model. Let us comment on them.

The first comment is about the tadpole equations. In this model there
are two scalar fields acquiring a VEV. Therefore, we must solve two
tadpole equations and hence select two parameters to solve them:

\begin{lstlisting}[style=SPheno,firstnumber=17]
  ParametersToSolveTadpoles = {mH2, mPhi2};
\end{lstlisting}

The second comment is about a feature that we can exploit to make our
life simpler when we are targeting a specific value of a derived
parameter. \SPheno must have input values for all the parameters of
the model in order to run properly. However, we can choose between
giving this input \emph{directly} or \emph{indirectly}. In the first
case, we introduce the values for the fundamental parameters in the
Lagrangian. In the second, we introduce the values for some derived
parameters which do not appear directly in the Lagrangian, like a
gauge boson mass, and tell \SPheno (and \SARAH) how to obtain the
fundamental parameters from them. This is useful when we are
interested in a parameter point with a specific value for a derived
parameter.

In the model under consideration, the $Z^\prime$ mass is an important
derived parameter, since the phenomenology strongly depends on its
precise value. It depends on two quantities, the new gauge coupling
$g_X$ and the $SU(2)_X$ breaking VEV, $v_\phi$, via
\begin{equation}
m_{Z^\prime} = 2 \, g_X \, v_\phi \, .
\end{equation}
Given the relevance of the $Z^\prime$ mass, it is useful to replace
$v_\phi$ by $m_{Z^\prime}$ as input parameter. We begin by introducing
the $Z^\prime$ mass as one of the input parameters in the {\tt MINPAR}
block,

\begin{lstlisting}[style=SPheno,firstnumber=13]
  {20, gXInput},
  {21, MZpMass}
\end{lstlisting}

And then, among the definitions in {\tt BoundaryLowScaleInput} we
establish the relation with $v_\phi$,

\begin{lstlisting}[style=SPheno,firstnumber=31]
  {gX,  gXInput},
  {g1X, 0},
  {gX1, 0},
  {vP,  MZpMass/(2*gX)}
\end{lstlisting}

This way \SPheno will have input values for all the relevant
parameters of the model and we will make sure that $m_{Z^\prime}$ has
exactly the value we are interested in. Note also that we have
considered a scenario with vanishing $U(1)$ mixing by setting the
mixed gauge couplings to zero.

\subsection{Generating input files for the other tools}
\label{subsec:generating}

Once the model is implemented in \SARAH we can generate input files
for the other tools. Instead of generating input for the different
tools one by one, we can make use of the {\tt MakeAll[]} command to
generate input files for all the tools at once. This will generate
automatically the \SPheno module as well as input files for \MO and
\MG. Therefore, we just have to execute the following three lines in
\Mathematica:

\vsp
\begin{lstlisting}[style=mathematica]
<<$PATH/SARAH-X.Y.Z/SARAH.m;
Start["DarkBS"];
MakeAll[]
\end{lstlisting}
\vsp

The results will be saved in different sub-folders of the {\tt
  \$PATH/SARAH-X.Y.Z/Output/DarkBS/EWSB} folder. Notice that {\tt
  MakeAll[]} also includes the generation of the \LaTeX\ files with
all the model details.

\subsection{Benchmark point and numerical results}
\label{subsec:benchmark}

The first thing we can do after executing {\tt MakeAll[]} is to
compile our new \SPheno code. This step was explained in
Sec. \ref{subsec:SPheno} and the process in this case is completely
analogous:

\vsp
\begin{lstlisting}[style=terminal]
$ cd $PATH/SPheno-X.Y.Z
$ mkdir DarkBS
$ cp -r $PATH/SARAH-X.Y.Z/Output/DarkBS/EWSB/SPheno/* $PATH/SPheno-X.Y.Z/DarkBS
$ make Model=DarkBS
\end{lstlisting}
\vsp

Next, let us consider a benchmark point for the DarkBS model. We will
call it benchmark point {\bf BDarkBS1} (Benchmark DarkBS 1), and it is
defined by the input parameters

\begin{center}
\fbox{
\begin{minipage}[c]{0.8\textwidth}
\vspace*{0.2cm}
\centering
\begin{tabular}{lll}
$\lambda = 0.26 \qquad $ & $\lambda_\phi = 0.1 \qquad $ & $\lambda_\chi = 10^{-5} \qquad $ \\
$\lambda_{\phi \chi} = 10^{-5} \qquad $ & $\lambda_{H \phi} = 0 \qquad $ & $\lambda_{H \chi} = 0 \qquad $ \\
$m_\chi^2 = 3 \cdot 10^6 \, \text{GeV}^2 \qquad $ & $m_Q = 1 \, \text{TeV} \qquad $ & $m_L = 1 \, \text{TeV} \qquad $ \\
$g_X = 1 \qquad $ & $m_{Z^\prime} = 4 \, \text{TeV} \qquad $ &
\end{tabular}
\begin{equation*}
\lambda_Q = \left( \begin{array}{c} 
0 \\
3 \cdot 10^{-3} \\
3 \cdot 10^{-3}
\end{array} \right) \quad  \quad
\lambda_L = \left( \begin{array}{c} 
0 \\
1 \\
0
\end{array} \right)
\end{equation*}
\vspace*{0.1cm}
\end{minipage}
}
\end{center}

In order to use this parameter point we must include the following
lines in the {\tt LesHouches.in.DarkBS} input file:

\vsp
\begin{lstlisting}[style=LesHouchesDarkBS,firstnumber=12]
Block MINPAR      # Input parameters 
1   0.260000E+00   # LambdaInput
2   0.100000E+00    # LPInput
3   0.000010E+00    # LCInput
4   0.000010E+00    # LCPInput
5   0.000000E+00    # LHPInput
6   0.000000E+00    # LHCInput
10   3.000000E+06   # mChi2Input
11   1.000000E+03   # mQInput
12   1.000000E+03   # mLInput
20   1.000000E+00   # gXInput
21   4.000000E+03   # MZpMass
Block LAMQIN #
1 0.0 #
2 3.0E-3 #
3 3.0E-3 #
Block LAMLIN #
1 0.0 #
2 1.0 #
3 0.0 #
\end{lstlisting}
\vsp

By running \SPheno we can see that the {\bf BDarkBS1} benchmark point
has a Higgs mass consistent with the observed value by ATLAS and
CMS. Moreover, the {\tt MASS} block also shows that the light
neutrinos are massless in this model, whereas the two heavy neutral
leptons form a Dirac pair,

\vsp
\begin{lstlisting}[style=LesHouchesOutDarkBS,firstnumber=146]
        12     0.00000000E+00  # Fv_1
        14     0.00000000E+00  # Fv_2
        16     0.00000000E+00  # Fv_3
        18    -1.73205081E+03  # Fv_4
        20     1.73205081E+03  # Fv_5
\end{lstlisting}
\vsp

\subsection{Calculating the dark matter relic density}
\label{subsec:dmlecture4}

As the next step in our phenomenological study, we can compute the
dark matter relic density in the {\bf BDarkBS1} benchmark point using
\MO. As explained in Appendix \ref{sec:appendix4}, the spontaneous
breaking of the continuous $U(1)_X$ gauge symmetry leaves a remnant
$\mathbb{Z}_2$ that stabilizes the $\chi$ scalar. This is therefore
the dark matter particle in this model.

The calculation of the dark matter relic density is straightforward
and follows the same procedure as for the scotogenic model. Using the
files in the {\tt \$PATH/SARAH-X.Y.Z/Output/DarkBS/EWSB/CHep} folder,
we can proceed in exactly the same way. We find $\Omega_{\text{DM}}
h^2 = 0.132$, in reasonable agreement with the observed value. The
most important annihilation channels are $\chi \, \chi \to d_4 \, \bar
d_4$ and $\chi \, \chi \to u_4 \, \bar u_4$, this is, to final states
including heavy vector-like quarks.

\subsection{Signatures at the LHC}
\label{subsec:LHClecture4}

Finally, we can use \MG for some simple (but illustrative) LHC
simulations. This model has a $Z^\prime$ boson with a relatively large
branching ratio into a pair of muons, $\mu^+ \mu^-$. Therefore, let us
consider
\begin{equation*}
p \, p \to \mu^+ \, \mu^-
\end{equation*}
at the LHC. This process will receive many different
contributions. Among them, the one induced by s-channel $Z^\prime$
exchange, $p \, p \to Z^\prime \to \mu^+ \, \mu^-$. However, note that
in the benchmark point {\bf BDarkBS1} the $Z^\prime$ production at the
LHC is strongly suppressed, since we have taken $\lambda_Q^1 =
0$. Given that protons have very little content of second and third
family quarks, this leads to tiny production cross-sections for the
$Z^\prime$. Moreover, we had a $Z^\prime$ mass of $4$ TeV, which again
suppresses its production. Therefore, let us consider a new benchmark
point, called {\bf BDarkBS2} (Benchmark DarkBS 2), where these are
changed. The only changes with respect to the {\bf BDarkBS1} point
are:

\begin{center}
\fbox{
\begin{minipage}[c]{0.8\textwidth}
\vspace*{0.2cm}
\begin{equation*}
m_{Z^\prime} = 300 \, \text{GeV} \qquad
\lambda_Q = \left( \begin{array}{c} 
1 \\
3 \cdot 10^{-3} \\
3 \cdot 10^{-3}
\end{array} \right)
\end{equation*}
\vspace*{0.1cm}
\end{minipage}
}
\end{center}

This parameter point is of course experimentally excluded, since such
a light and strongly coupled $Z^\prime$ boson would have been observed
already at the LHC. However, it serves as an academic illustration. In
order to use this parameter point we must modify some lines in the
{\tt MINPAR} and {\tt LAMQIN} blocks of the {\tt LesHouches.in.DarkBS}
input file:

\vsp
\begin{lstlisting}[style=LesHouchesDarkBS,firstnumber=12]
Block MINPAR      # Input parameters 
1   0.260000E+00   # LambdaInput
2   0.100000E+00    # LPInput
3   0.000010E+00    # LCInput
4   0.000010E+00    # LCPInput
5   0.000000E+00    # LHPInput
6   0.000000E+00    # LHCInput
10   3.000000E+06   # mChi2Input
11   1.000000E+03   # mQInput
12   1.000000E+03   # mLInput
20   1.000000E+00   # gXInput
21   3.000000E+02   # MZpMass
Block LAMQIN #
1 1.0 #
2 3.0E-3 #
3 3.0E-3 #
\end{lstlisting}
\vsp

After running \SPheno with this point we generate a new {\tt
  SPheno.spc.DarkBS} file that we can now use with \MG. We will follow
the same procedure as for the scotogenic model:

\vsp
\begin{lstlisting}[style=terminal]
$ cd ..
$ bin/mg5_aMC
MG5_aMC > import model DarkBS -modelname
MG5_aMC > define p d1 d1bar d2 d2bar u1 u1bar u2 u2bar g
MG5_aMC > generate p p > e2 e2bar
MG5_aMC > output SimDBS
MG5_aMC > launch SimDBS
\end{lstlisting}
\vsp

In this simulation we will also include hadronization, showering and
detector response effects. This requires running {\tt Pythia} and {\tt
  PGS}. Therefore, we will answer {\tt 2} to the first question. To
the second question we will also answer {\tt 2} in order to modify the {\tt
  run_card.dat} file. Instead of $10^4$ events, we want to generate
$10^5$. This will imply a more precise simulation. In order to
increase the number of events we just have to modify the option {\tt
  nevents}, which now will read

\vsp
\begin{lstlisting}[style=runcard,firstnumber=32]
  100000 = nevents ! Number of unweighted events requested 
\end{lstlisting}
\vsp

We should also modify the {\tt param_card.dat} file before we save and
close the {\tt run_card.dat} file. Again, we will simply copy the file
we generated with \SPheno for the {\bf BDarkBS2} benchmark point,

\vsp
\begin{lstlisting}[style=terminal]
$ cp $PATH/SPheno-X.Y.Z/SPheno.spc.DarkBS $PATH/MG5_aMC_vX/SimDBS/Cards/param_card.dat
\end{lstlisting}
\vsp

Then we are ready to save and close the {\tt run_card.dat} file and
press enter in \MG. The simulation starts, running the standard
simulation at parton level, hadronization and detector response. After
some minutes (remember, we are using $10^5$ random events) it will be
finished, showing a cross-section of about $831$ pb. We will now use
{\tt MadAnalysis} to plot the results. This powerful tool has many
features for the analysis of the \MG results. One can in principle use
it in a similar way as \MG, opening the code and executing some
commands one by one in the therminal. Instead, in this case we will
use an external file with the collection of commands we want {\tt
  MadAnalysis} to execute. This file can be placed in the {\tt
  MadAnalysis} folder:

\vsp
\begin{lstlisting}[style=maFile]
import $PATH/MG5_aMC_vX/SimDBS/Events/run_01/tag_1_pgs_events.lhco.gz
plot M(mu+ mu-) 100 50 500 [logX logY]
submit DarkBSPlot
\end{lstlisting}
\vsp

As usual, {\tt X} must be replaced in the first command by the
specific \MG version we are using. With these commands, we are just
telling {\tt MadAnalysis} to import the results of our simulation and
to an histogram with the number of $\mu^+ \, \mu^-$ events as a
function of the invariant mass of the muon pair, $m_{\mu \mu}$. We
have also decided to use logarithmic scales in both axes, and show the
invariant mass between $50$ GeV and $500$ GeV distributed in $100$
bins. Finally, the result of the analysis should be saved in a folder
called {\tt DarkBSPlot}.

Therefore, we just have to call {\tt MadAnalysis} with this input file

\vsp
\begin{lstlisting}[style=terminal]
$ cd $PATH/madanalysis5
$ bin/ma5 -R plotDarkBS.txt
\end{lstlisting}
\vsp

The flag {\tt -R} is used because we are going to read the dataset
produced by {\tt PGS} including detector response. If we wanted to use
the dataset without hadronization and detector response, instead of
importing the file {\tt tag_1_pgs_events.lhco.gz} we would have to
import {\tt unweighted_events.lhe.gz}, in lhe format, which would not
require the {\tt -R} flag.

{\tt MadAnalysis} will ask us the number of cores we want to use for
the analysis. After answering the question it will proceed to the
compilation of some parts of the code and the analysis of our
results. Eventually, it will be finished and the folder {\tt
  \$PATH/madanalysis5/DarkBSPlot} will be created. In this folder we
can find the results in several formats. We can open a pdf file where
these are nicely presented using the {\tt MadAnalysis} command

\begin{figure}[t]
\centering
\includegraphics[scale=0.6]{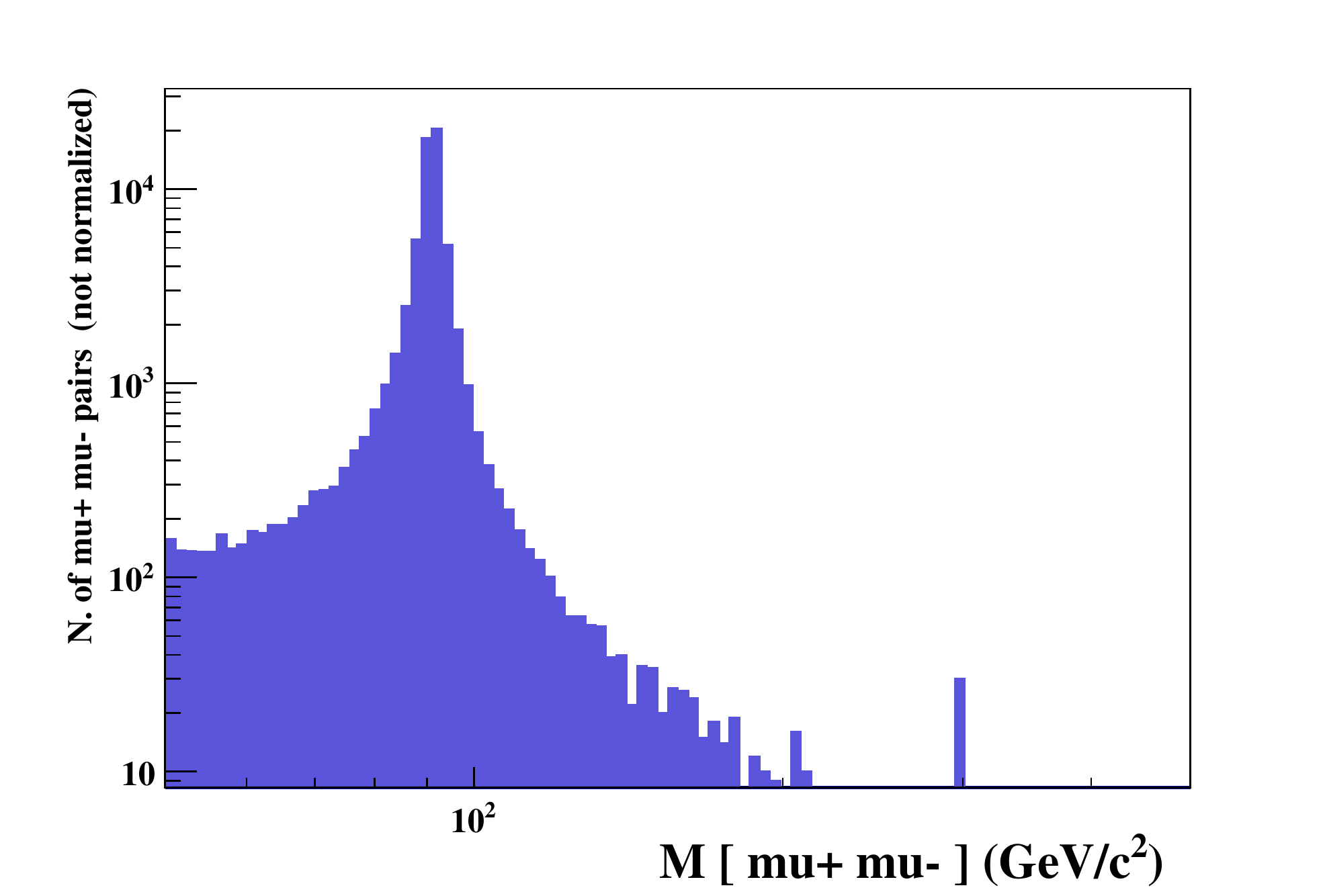}
\caption{Histogram generated by {\tt
    MadAnalysis}. \label{fig:histogram}}
\end{figure}

\vsp
\begin{lstlisting}[style=terminal]
ma5 > open DarkBSPlot/PDF
\end{lstlisting}
\vsp

In the last page of this pdf file (see Fig.~\ref{fig:histogram}) we
can see the histogram generated by {\tt MadAnalysis}. There are two
visible peaks, at $m_{\mu \mu} = 91$ GeV and $m_{\mu \mu} = 300$
GeV. We have \emph{discovered} two particles: the $Z$ and $Z^\prime$
bosons.

\subsection{Summary of the lecture}
\label{subsec:lecture4-summary}

We conclude the course here. In the last lecture we have reviewed our
previous lectures by applying what we have learned to a new model with
a few additional complications. After implementing the model in
\SARAH, we have used \SPheno to obtain numerical results for a
specific benchmark point, \MO to compute the relic density and \MG to
run a simple but interesting collider study.


\newpage

%% file: summary.tex
\section{Summary}
\label{sec:summary}

In this course we have focused on three computer tools, nowadays
widely used in particle physics:

\begin{itemize}

\item {\SARAH} (and \SPheno) : analytical and numerical exploration of a new physics model
\item {\MO} : dark matter studies
\item {\MG} (and {\tt Pythia}, {\tt PGS} and {\tt MadAnalysis}) : collider simulations

\end{itemize}

For each of these tools, we have learned the basics and trained with
some practical applications. For this purpose we have used the
scotogenic model as workbench: we computed mass matrices and vertices,
obtained numerical results for the particle spectrum, calculated
flavor observables, decay rates and the dark matter relic density, and
simulated $\eta_R \, \eta^+$ production at the LHC.

In the final lecture we reviewed the whole process with a slightly
more complicated model with an additional $U(1)_X$ gauge symmetry. We
paid attention to the particularities of the model and how they must
be implemented in \SARAH and used the resulting output to run the
model in \MO and \MG. This way we also computed the dark matter relic
density and performed a simple but illustrative LHC simulation.

Before concluding, let us repeat two important messages: (i) using
these computers tools is not so hard, and (ii) do not trust blindly in
a computer code. I hope I convinced you about the first point, whereas
for the second you will get convinced as soon as you find a weird
result caused by a bug in a code.

Finally, let me emphasize once more that the computer tools described
in this course offer many additional possibilities worth exploring. I
could only cover the most basic features of \SARAH, \MO and \MG, but
there are many more. Some are straightforward and some require to get
into technical details. Although getting started is easy, only with
frequent practice one can really master all of these computer tools.

%% file: appendix.tex
\appendix

\section{Models already implemented in \SARAH}
\label{sec:appendix1}

The following models are included in the public version of \SARAH. 

\subsection{Supersymmetric Models}
\begin{itemize}
 \item Minimal supersymmetric standard model
    \begin{itemize} 
     \item With general flavor and CP structure ({\tt MSSM})
     \item Without flavor violation ({\tt MSSM/NoFV}, {\tt MSSM/NoFV2})
     \item With explicit CP violation in the Higgs sector ({\tt MSSM/CPV})
     \item In SCKM basis ({\tt MSSM/CKM})
     \item With non-holomorphic soft terms ({\tt NHSSM})
     \item With color sextets ({\tt MSSM6C})
     \item With charge and color breaking minima ({\tt CCB-MSSM/SfermionVEVs}, {\tt CCB-MSSM/StauVEVs}, \\ {\tt CCB-MSSM/StopVEVs}, {\tt CCB-MSSM/StauStopVEVs})
    \end{itemize}
   \item Singlet extensions
   \begin{itemize}
    \item Next-to-minimal supersymmetric standard model ({\tt NMSSM}, {\tt NMSSM/NoFV}, {\tt NMSSM/NoFV2}, {\tt NMSSM/CPV}, {\tt NMSSM/CKM})
    \item near-to-minimal supersymmetric standard model  ({\tt near-MSSM})
    \item General singlet extended, supersymmetric standard model ({\tt SMSSM})
    \item DiracNMSSM ({\tt DiracNMSSM})
  \end{itemize}
  \item Triplet extensions 
  \begin{itemize} 
    \item Triplet extended MSSM ({\tt TMSSM})
    \item Triplet extended NMSSM ({\tt TNMSSM})
  \end{itemize}
  \item MSSM with additional vector-like quark superfields ({\tt Vectorlike/SusyTop}, {\tt Vectorlike/SusyTopSpectator})
  \item NMSSM with additional vector-like superfields ({\tt NMSSM+VL/VLtop}, {\tt NMSSM+VL/5plets}, {\tt NMSSM+VL/10plets}, {\tt NMSSM+VL/5+10plets}, {\tt NMSSM+VL/5plets+RpV})
   \item Models with $R$-parity violation
  \begin{itemize}
    \item Bilinear RpV ({\tt MSSM-RpV/Bi}) 
    \item Lepton number violation ({\tt MSSM-RpV/LnV})
    \item Only trilinear lepton number violation ({\tt MSSM-RpV/TriLnV})
    \item Baryon number violation ({\tt MSSM-RpV/BnV})  
    \item $\mu\nu$SSM ({\tt munuSSM})
  \end{itemize}
   \item $U(1)$ extensions 
  \begin{itemize}
    \item $U(1)$-extended MSSM ({\tt UMSSM})
    \item Secluded MSSM ({\tt secluded-MSSM})
    \item Minimal $B-L$ model ({\tt B-L-SSM})
    \item Minimal singlet-extended $B-L$ model ({\tt N-B-L-SSM})
    \item $U(1)_L \times U(1)_R$ supersymmetric standard model ({\tt BxL-SSM})
    \item $U(1)^\prime$-extended MSSM with vector-like superfields ({\tt MSSM+U1prime-VL})
    \item $U(1)_X$ supersymmetric standard model ({\tt U1xMSSM}, {\tt U1xMSSM3G})
  \end{itemize}
   \item SUSY-scale seesaw extensions
    \begin{itemize}
      \item Inverse seesaw ({\tt inverse-Seesaw})
      \item Linear seesaw ({\tt LinSeesaw})
      \item Singlet extended inverse seesaw ({\tt inverse-Seesaw-NMSSM})
      \item Inverse seesaw with $B-L$ gauge group ({\tt B-L-SSM-IS})
      \item Minimal $U(1)_R \times U(1)_{B-L}$ model with inverse seesaw  ({\tt BLRinvSeesaw})
\end{itemize}
 \item Models with Dirac Gauginos
   \begin{itemize}
    \item MSSM/NMSSM with Dirac Gauginos ({\tt DiracGauginos})
    \item Minimal $R$-Symmetric SSM ({\tt MRSSM}), with explicit breaking of the $R$-symmetry ({\tt brokenMRSSM})
    \item Minimal Dirac Gaugino supersymmetric standard model ({\tt MDGSSM})
   \end{itemize}
 \item High-scale extensions
\begin{itemize}
 \item Seesaw 1 - 3 ($SU(5)$ version) ,
 ({\tt Seesaw1}, {\tt Seesaw2}, {\tt Seesaw3})
 \item Left/right model ($\Omega$LR) ({\tt Omega}, {\tt Omega\_Short})
 \item Quiver model ({\tt QEW12}, {\tt  QEWmld2L3})
\end{itemize}
\item $E_6$ inspired model with extra $U(1)$ ({\tt E6SSMalt3I})
\end{itemize}

\subsection{Non-Supersymmetric Models}
\begin{itemize}
\item Standard Model (SM) ({\tt SM}), Standard model in CKM basis ({\tt SM/CKM}) 
\item Two Higgs doublet model 
\begin{itemize}
 \item Type-I ({\tt THDM})
 \item Type-II ({\tt THDM-II})
 \item Type-III ({\tt THDM-III})
 \item Type-I with CP violation ({\tt THDM-CPV})
 \item Flipped ({\tt THDM-Flipped})
 \item Lepton specific ({\tt THDM-LS})
 \item Inert Higgs doublet model ({\tt Inert})
\end{itemize}
\item Two Higgs doublet models with additional fields 
\begin{itemize}
 \item With SM-like vector-like fermions: Type-I ({\tt THDM+VL/Type-I-SM-like-VL}),\\ Type-II ({\tt THDM+VL/Type-II-SM-like-VL})
 \item With exotic vector-like fermions: Type-I ({\tt THDM+VL/Type-I-VL}), Type-II ({\tt THDM+VL/Type-II-VL})
 \item With colored vector-like fermions: color triplet ({\tt THDM+VL/min-3}), color octet ({\tt THDM+VL/min-8})
 \item With a scalar $SU(2)$ septuplet ({\tt THDM/ScalarSeptuplet})
\end{itemize}
\item $U(1)$ extensions 
\begin{itemize}
 \item B-L extended SM ({\tt B-L-SM})
 \item B-L extended SM with inverse seesaw ({\tt B-L-SM-IS})
 \item Dark $U(1)^\prime$ ({\tt U1Extensions/darkU1})
 \item Hidden $U(1)$ ({\tt U1Extensions/hiddenU1})
 \item Simple $U(1)$ ({\tt U1Extensions/simpleU1})
 \item Scotogenic $U(1)$ ({\tt U1Extensions/scotoU1})
 \item Unconventional $U(1)_{B-L}$ ({\tt U1Extensions/BL-VL})
 \item Sample of $U(1)^\prime$ ({\tt U1Extensions/VLsample})
 \item With flavor non-universal charges ({\tt U1Extensions/nonUniversalU1})
 \item Leptophobic $U(1)$ ({\tt U1Extensions/U1Leptophobic})
 \item With a $Z^\prime$ mimicking a scalar resonance ({\tt U1Extensions/trickingLY})
\end{itemize}
\item Leptoquark models
\begin{itemize}
 \item Single scalar leptoquark models ({\tt Leptoquarks/ScalarLeptoquarks})
 \item Two scalar leptoquark models ({\tt Leptoquarks/TwoScalarLeptoquarks})
\end{itemize}
\item Singlet extensions with vector-like fermions
\begin{itemize}
 \item CP-even singlet ({\tt SM+VL/CPevenS}), CP-odd singlet ({\tt SM+VL/CPoddS}), complex singlet ({\tt SM+VL/complexS})
 \item Portal DM ({\tt SM+VL/PortalDM})
 \item $SU(2)$ triplet quark model ({\tt SM+VL/TripletQuarks})
\end{itemize}
\item Non-Abelian gauge-group extensions
\begin{itemize}
 \item Left-Right models: without bidoublets ({\tt LRmodels/LR-VL}), with $U(1)_L \times U(1)_R$ ({\tt LRmodels/LRLR}), with triplets ({\tt LRmodels/tripletLR}), Dark LR ({\tt darkLR})
 \item 331 models: without exotic charges ({\tt 331/v1}), with exotic charges ({\tt 331/v2})
 \item Gauged THDM ({\tt GTHDM})
\end{itemize}
\item SM extended with vector-like quarks ({\tt Vectorlike/TopR}, {\tt Vectorlike/TopX}, {\tt Vectorlike/BottomR},  \\ {\tt Vectorlike/BottomY}, {\tt Vectorlike/TopBottomL}) 
\item Triplet extensions ({\tt SM+Triplet/Real}, {\tt SM+Triplet/Complex})
\item Georgi-Machacek model ({\tt Georgi-Machacek})
\item Singlet extended SM ({\tt SSM})
\item SM extended by a scalar color octet ({\tt SM-8C})
\item SM extended by a scalar singlet and a scalar color octet ({\tt SM-S-Octet})
\item Singlet Scalar DM ({\tt SSDM})
\item Singlet-Doublet DM ({\tt SDDM})
\end{itemize}

Some of these models were implemented in \SARAH in order to address
the diphoton excess observed by ATLAS and CMS in 2015. For more
information see \cite{Staub:2016dxq}.

\section{The Standard Model}
\label{sec:appendix2}

In order to set the notation and conventions used throughout this
course, let us introduce the SM.

{
\renewcommand{\arraystretch}{1.6}
\begin{table}
\centering
\begin{tabular}{|c|c|c|} 
\hline \hline 
Field & Group & Coupling \\ 
 \hline 
$B$ & $U(1)_Y$ & $g_1$ \\ 
$W$ & $SU(2)_L$ & $g_2$\\ 
$g$ & $SU(3)_c$ & $g_3$\\ 
\hline \hline
\end{tabular} 
\caption{Gauge sector of the Standard Model.}
\label{tab:SM1}
\end{table}

\begin{table}
\centering
\begin{tabular}{|c|c|c|c|c|} 
\hline \hline 
Field & Spin  & Generations & \((U(1)_Y \times\, SU(2)_L \times\, SU(3)_c)\) \\ 
\hline 
\(H\) & \(0\)  & 1 & \((\frac{1}{2},{\bf 2},{\bf 1}) \) \\ 
\(q\) & \(\frac{1}{2}\)  & 3 & \((\frac{1}{6},{\bf 2},{\bf 3}) \) \\ 
\(\ell\) & \(\frac{1}{2}\)  & 3 & \((-\frac{1}{2},{\bf 2},{\bf 1}) \) \\ 
\(d\) & \(\frac{1}{2}\)  & 3 & \((\frac{1}{3},{\bf 1},{\bf \overline{3}}) \) \\ 
\(u\) & \(\frac{1}{2}\)  & 3 & \((-\frac{2}{3},{\bf 1},{\bf \overline{3}}) \) \\ 
\(e\) & \(\frac{1}{2}\)  & 3 & \((1,{\bf 1},{\bf 1}) \) \\ 
\hline \hline
\end{tabular} 
\caption{Matter content in the Standard Model.}
\label{tab:SM2}
\end{table}
}

The SM gauge symmetry is $SU(3)_c \times SU(2)_L \times U(1)_Y$ and
the gauge fields with the corresponding gauge couplings are given in
Table \ref{tab:SM1}. The matter content with the corresponding gauge
charges are shown in Table \ref{tab:SM2}. Note that our definition of
hypercharge is $Q = T_{3L} + Y$. The $SU(2)_L$ doublets can be
decomposed as
\begin{equation}
H = \left( \begin{array}{c}
H^+ \\
H^0 \end{array} \right) \quad , \quad q = \left( \begin{array}{c}
u_L \\
d_L \end{array} \right) \quad , \quad \ell = \left( \begin{array}{c}
\nu_L \\
e_L \end{array} \right) \, ,
\end{equation}
whereas the $SU(2)_L$ singlets can be identified as $d \equiv
d_R^\ast$, $u \equiv u_R^\ast$ and $e \equiv e_R^\ast$.

The Yukawa terms in the Lagrangian are
\begin{equation}
\mathcal{L}_Y=
Y_d H^\dagger \, \bar d \, q + Y_e H^\dagger \, \bar e \, \ell + Y_u H \, \bar u \, q +\hc \, ,
\end{equation}
where we omit flavor indices for the sake of clarity. $Y_{d,e,u}$ are
three $3 \times 3$ general complex matrices. The scalar potential of
the model takes the form
\begin{equation}
\mathcal{V} = -m_H^2 H^\dagger H + \frac{\lambda}{2}\left( H^\dag H \right)^2 \, ,
\end{equation}
where we have defined the $m_H^2$ with a negative sign for practical
reasons. We assume that this scalar potential is such that the neutral
component of the Higgs doublet takes a non-zero VEV. In this case one
can decompose $H^0$ as
\begin{equation}
H^0 = \frac{1}{\sqrt{2}} \left( v + h + i A \right) \, .
\end{equation}
Here $\langle H^0 \rangle = v/\sqrt{2} = 174$ GeV is the usual Higgs
VEV, $h$ is the CP-even state, the physical Higgs boson discovered at
the LHC, and $A$ is the CP-odd state that is absorbed by the $Z^0$ and
becomes its longitudinal component.

After electroweak symmetry breaking (EWSB), the remnant symmetry is
$SU(3)_c \times U(1)_Q$ and mixing among the gauge bosons is induced
\begin{eqnarray}
\left\{ B , W_3 \right\} \, &\to& \, \left\{ \gamma , Z \right\} \, , \\
\left\{ W_{1} , W_{2} \right\} \, &\to& \, \left\{ W^+ , W^- \right\} \, ,
\end{eqnarray}
where $W^\pm$, $\gamma$ and $Z$ are the mass eigenstates (with
$m_\gamma = 0$). The relation between the gauge and mass eigenstates
is given by the unitary transformations $Z^Z$ and $Z^W$, defined as
\begin{align} 
\left(\begin{array}{c} 
B\\ 
W_{3}\end{array} \right) 
 = & \,Z^{Z}
\left(\begin{array}{c} 
\gamma\\ 
Z\end{array} \right) \\ 
\left(\begin{array}{c} 
W_{{1}}\\ 
W_{{2}}\end{array} \right) 
 = & \,Z^{W}
\left(\begin{array}{c} 
W^+\\ 
\left(W^-\right)^\ast \end{array} \right) \, ,
\end{align} 
and the mixing matrices are given by
\begin{align} \label{eq:ZZ}
Z^{Z}= \, \left( 
\begin{array}{cc} 
\cos\Theta_W  & - \sin\Theta_W   \\ 
 \sin\Theta_W  & \cos\Theta_W \end{array} 
\right) \\ 
Z^{W}= \, \left( 
\begin{array}{cc} 
\frac{1}{\sqrt{2}} & \frac{1}{\sqrt{2}} \\ 
\frac{i}{\sqrt{2}}  & -\frac{i}{\sqrt{2}} \end{array} 
\right) \, . \label{eq:ZW}
\end{align} 
Regarding the fermions, they also get masses after EWSB. In the
basis \( \left(f_{L}\right), \left(f^*_{R}\right) \), with $f =
d,u,e$, the resulting Dirac mass term is given by
\begin{equation}
m_f = - \frac{1}{\sqrt{2}} \, v \, Y_f^T \, ,
\end{equation}
and the gauge and mass eigenstates are related by the unitary
transformations $U^f_{L,R}$,
\begin{align} 
U_{L} = V_u \, u_{L} \, ,\\ 
U_{R} = U_u \, u_{R} \, ,\\ 
D_{L} = V_d \, d_{L} \, ,\\ 
D_{R} = U_d \, d_{R} \, ,\\ 
E_{L} = V_e \, e_{L} \, ,\\ 
E_{R} = U_e \, e_{R} \, ,
\end{align} 
where the fields in capital letters are the mass eigenstates.

\section{The scotogenic model}
\label{sec:appendix3}

The \emph{scotogenic model}~\cite{Ma:2006km} is a popular extension of
the SM proposed by Ernest Ma in 2006 that includes non-zero neutrino
masses and dark matter. One of the main reasons for its popularity is
the simplicity of the model which, with just a few ingredients,
addresses these two major issues in particle physics and cosmology.

\tip{The other reason for its popularity is its fancy name. The Greek word \emph{skotos} (\textsigma \textkappa \textomikron \texttau \textomikron \textvarsigma) means \emph{darkness} and refers to the fact that neutrino masses are induced in this model thanks to the presence of the dark matter particles. So, remember, if you want your model to become popular, give it a good name!}

In the scotogenic model, the SM particle content is extended with
three singlet fermions, $N_i$ ($i=1$-$3$), and one $SU(2)_L$ doublet,
$\eta$. In addition, a $\mathbb{Z}_2$ parity is imposed, under which
the new particles are odd and the SM ones are even. This symmetry not
only prevents flavor changing neutral currents but it also renders
stable the lightest odd particle in the spectrum, which becomes a dark
matter candidate. In this model, two particles can play the role of
dark matter: the neutral scalar (an inert Higgs) or the lightest
singlet fermion.

{
\renewcommand{\arraystretch}{1.6}
\begin{table}
\centering
\begin{tabular}{|c|c|c|c|c|c|} 
\hline \hline 
Field & Spin  & Generations & \((U(1)_Y \times\, SU(2)_L \times\, SU(3)_c)\) & $\mathbb{Z}_2$ \\ 
\hline 
\(\eta\) & \(0\)  & 1 & \((\frac{1}{2},{\bf 2},{\bf 1}) \) & $-$ \\ 
\(N\) & \(\frac{1}{2}\)  & 3 & \((0,{\bf 1},{\bf 1}) \) & $-$ \\
\hline \hline
\end{tabular}
\caption{New particle content in the scotogenic model.}
\label{tab:scotogenic}
\end{table}
}

The new Lagrangian terms involving the right-handed neutrinos can be
written as
\begin{equation}
\mathcal{L}_N=
\frac{M_{N}}{2} \, \overline{N^c} \, N+
Y_N \, \eta \, \overline{N} \, \ell + \hc \, .
\end{equation}
The right-handed neutrino mass matrix $M_{N}$ can be taken to be
diagonal without loss of generality and we will do so in the following
discussion. We do not write the kinetic term for the right-handed
neutrinos since it takes the canonical form. The matrix of Yukawa
couplings, $Y_N$, is an arbitrary $3 \times 3$ complex matrix. We notice
that the usual neutrino Yukawa couplings with the SM Higgs doublet are
not allowed due to the $\mathbb{Z}_2$ symmetry. The scalar potential
of the model is given by
\begin{eqnarray}
\mathcal{V}\!\!\!&=&\:
-m_{H}^2 H^\dag H+m_\eta^2\eta^\dag\eta+
\frac{\lambda_1}{2}\left( H^\dag H\right)^2+
\frac{\lambda_2}{2}\left(\eta^\dag\eta\right)^2+
\lambda_3\left( H^\dag H\right)\left(\eta^\dag\eta\right)\nonumber\\
&&\!\!\!+
\lambda_4\left( H^\dag\eta\right)\left(\eta^\dag H\right)+
\frac{\lambda_5}{2}\left[\left( H^\dag\eta\right)^2+
\left(\eta^\dag H\right)^2\right] \, .
\end{eqnarray}
In the scotogenic model, the $\mathbb{Z}_2$ parity is assumed to be
preserved after electroweak symmetry breaking. This is guaranteed by
choosing a set of parameters that leads to a vacuum with
$\langle \eta \rangle = 0$. Therefore, the only non-zero VEV of the
model is the standard SM Higgs VEV,
\begin{equation}
\langle H^0 \rangle = \frac{v}{\sqrt{2}} \, .
\end{equation}
After electroweak symmetry breaking, the masses of the charged
component $\eta^+$ and neutral component
$\eta^0=(\eta_R+i\eta_I)/\sqrt{2}$ are split to
\begin{eqnarray}
m_{\eta^+}^2&=&m_\eta^2+\lambda_3\langle H^0\rangle^2 \, , \label{eq:mass1}\\
m_R^2&=&m_{\eta}^2+\left(\lambda_3+\lambda_4+\lambda_5\right)\langle H^0\rangle^2 \, ,\label{eq:mass2}\\
m_I^2&=&m_{\eta}^2+\left(\lambda_3+\lambda_4-\lambda_5\right)\langle H^0\rangle^2 \, . \label{eq:mass3}
\end{eqnarray}
The mass difference between $\eta_R$ and $\eta_I$ (the CP-even and
CP-odd components of $\eta^0$, respectively) is
$m_R^2-m_I^2=2\lambda_5\langle H^0\rangle^2$.

\begin{figure}[t]
\centering
\includegraphics[scale=0.5]{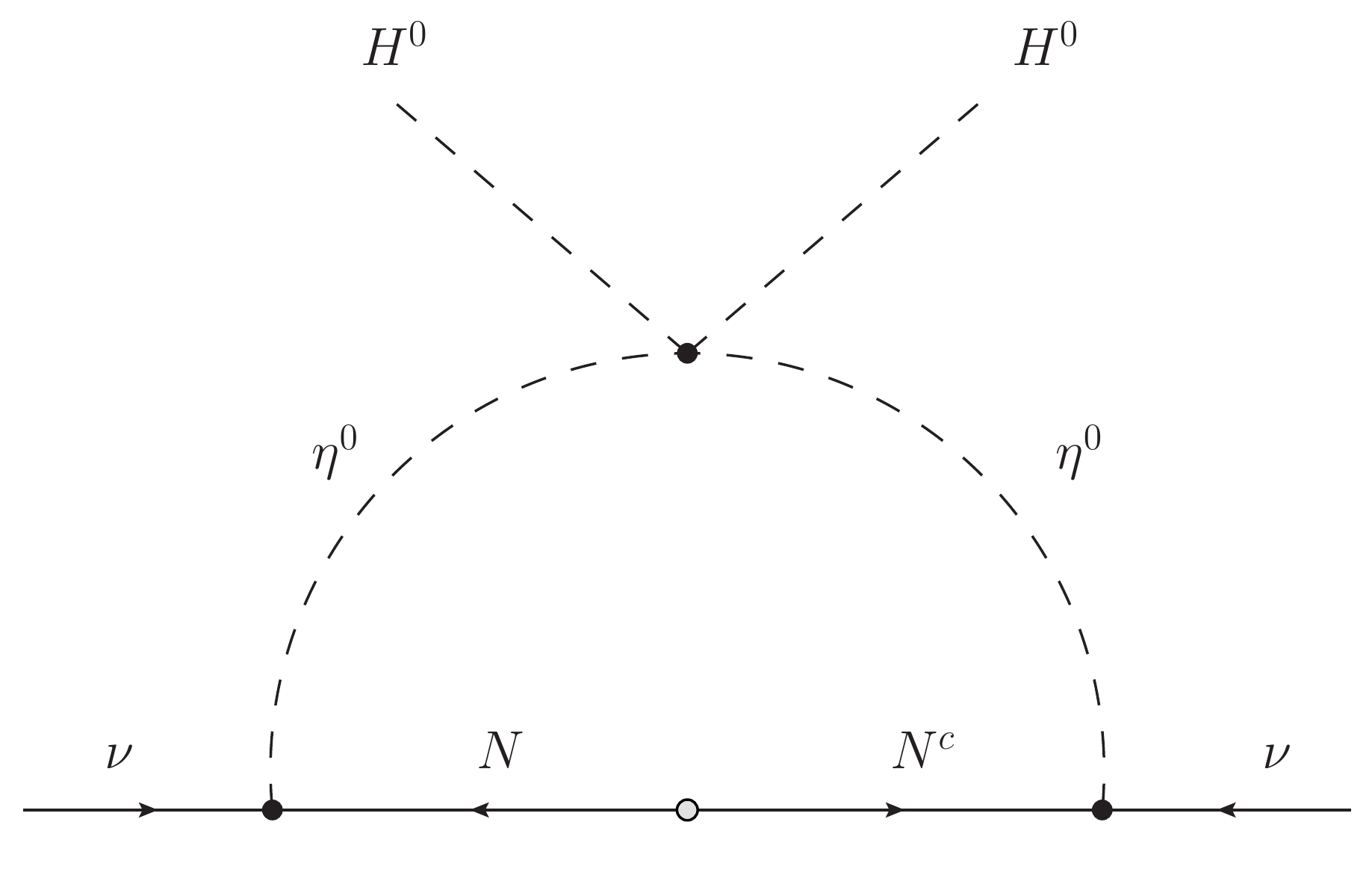}
\caption{1-loop neutrino masses in the scotogenic model. \label{fig:numass}}
\end{figure}
Inspecting the new terms in $\mathcal{L}_N$ and $\mathcal{V}$ one
finds that the presence of $\lambda_5 \neq 0$ breaks lepton number in
two units. Although the usual tree-level contribution to neutrino
masses is forbidden by the $\mathbb{Z}_2$ symmetry, these are induced
at the 1-loop level as shown in figure \ref{fig:numass}. This loop is
calculable and leads to the neutrino mass matrix~\footnote{We correct
this expression by including a factor of $1/2$ missing in all
references on the scotogenic model. I thank Takashi Toma for pointing
out this error in the literature. Notice also that the correct
expression was shown in version 1 of \cite{Merle:2015gea}.}
\begin{eqnarray}
\left(m_{\nu}\right)_{\alpha\beta}&=&
\sum_{i=1}^3\frac{\left(Y_N\right)_{i\alpha}\left(Y_N\right)_{i\beta}}{2(4\pi)^2} M_{N_i}
\left[\frac{m_R^2}{m_R^2-M_{N_i}^2}\log\left(\frac{m_R^2}{M_{N_i}^2}\right)
-\frac{m_I^2}{m_I^2-M_{N_i}^2}\log\left(\frac{m_I^2}{M_{N_i}^2}\right)\right]\nonumber\\
&\equiv&\left(Y_N^{T} \, \Lambda \, Y_N\right)_{\alpha\beta},
\label{eq:nu-mass}
\end{eqnarray}
where the $\Lambda$ matrix is defined as $\Lambda
= \text{diag}\left( \Lambda_1,\Lambda_2,\Lambda_3 \right)$, with
\begin{equation}
\Lambda_i=\frac{M_{N_i}}{2(4\pi)^2}
\left[\frac{m_R^2}{m_R^2-M_{N_i}^2}\log\left(\frac{m_R^2}{M_{N_i}^2}\right)
-\frac{m_I^2}{m_I^2-M_{N_i}^2}\log\left(\frac{m_I^2}{M_{N_i}^2}\right)\right] \, . 
\end{equation}
Simplified expressions can be obtained when $m_R^2 \approx
m_I^2 \equiv m_0^2$ ($\lambda_5\ll1$). In this case the mass matrix in
equation \eqref{eq:nu-mass} can be written as
\begin{equation} \label{eq:nu-mass2}
\left(m_\nu\right)_{\alpha\beta}\approx
\sum_{i=1}^3\frac{\lambda_5 \left(Y_N\right)_{i\alpha}\left(Y_N\right)_{i\beta}\langle H^0\rangle^2}
{(4\pi)^2 M_{N_i}}
\left[\frac{M_{N_i}^2}{m_{0}^2-M_{N_i}^2}
+\frac{M_{N_i}^4}{\left(m_{0}^2-M_{N_i}^2\right)^2}
\log\left(\frac{M_{N_i}^2}{m_0^2}\right)\right] \, .
\end{equation}
Compared to the standard seesaw formula, neutrino masses get an
additional suppression by roughly the factor $\sim \lambda_5 /
16 \pi^2$. Choosing $\lambda_5 \ll 1$, one can get the correct size
for neutrino masses, compatible with singlet fermions at the TeV scale
(or below) and sizable Yukawa couplings.

The conservation of $\mathbb{Z}_2$ leads to the existence of a stable
particle: the lightest particle charged under $\mathbb{Z}_2$. If
neutral, it will constitute a good dark matter candidate. There are,
therefore, two dark matter candidates in the scotogenic model: the
lightest singlet fermion $N_1$ and the lightest neutral $\eta$ scalar
($\eta_R$ or $\eta_I$).

\section{A model with a dark sector}
\label{sec:appendix4}

The model introduced in \cite{Sierra:2015fma} was motivated by some
anomalies in B meson decays recently found by the LHCb collaboration
(see \cite{LHCb:2015iha,Altmannshofer:2015sma} for some recent
references on the subject). These are not relevant for our course and
we will not discuss the details. However, the model will be used to go
a step beyond in complexity with respect to the scotogenic model
presented in Appendix \ref{sec:appendix3}.

{
\renewcommand{\arraystretch}{1.6}
\begin{table}
\centering
\begin{tabular}{|c|c|c|} 
\hline \hline 
Field & Group & Coupling \\ 
 \hline 
$B$ & $U(1)_Y$ & $g_1$ \\ 
$W$ & $SU(2)_L$ & $g_2$\\ 
$g$ & $SU(3)_c$ & $g_3$\\ 
$B_X$ & $U(1)_X$ & $g_X$ \\
\hline \hline
\end{tabular} 
\caption{Gauge sector of the model introduced in \cite{Sierra:2015fma}.}
\label{tab:DarkBS1}
\end{table}

\begin{table}
\centering
\begin{tabular}{|c|c|c|c|c|} 
\hline \hline 
Field & Spin  & Generations & \((U(1)_Y \times\, SU(2)_L \times\, SU(3)_c \times U(1)_X) \) \\ 
\hline 
\(\phi\) & \(0\)  & 1 & \((0,{\bf 1},{\bf 1},2) \) \\ 
\(\chi\) & \(0\)  & 1 & \((0,{\bf 1},{\bf 1},-1) \) \\ 
\(Q_{L,R}\) & \(\frac{1}{2}\)  & 3 & \((\frac{1}{6},{\bf 2},{\bf 3},2) \) \\ 
\(L_{L,R}\) & \(\frac{1}{2}\)  & 3 & \((-\frac{1}{2},{\bf 2},{\bf 1},2) \) \\ 
\hline \hline
\end{tabular} 
\caption{New particle content in the model of \cite{Sierra:2015fma}.}
\label{tab:DarkBS2}
\end{table}
}

We extend the SM gauge group with a new dark $U(1)_X$ factor (see
Table \ref{tab:DarkBS1} for details), under which all the SM particles
are assumed to be singlets. The only particles charged under the
$U(1)_X$ group are two pairs of vector-like fermions, $Q$ and $L$, as
well as the complex scalar fields, $\phi$ and $\chi$, as shown in
Table \ref{tab:DarkBS2}. $Q$ and $L$ are vector-like copies of the SM
doublets $q$ and $\ell$, and they can be decomposed as
\begin{equation}
Q_{L,R} = \left( \begin{array}{c}
U \\
D \end{array} \right)_{L,R} \quad , \quad L_{L,R} = \left( \begin{array}{c}
N \\
E \end{array} \right)_{L,R} \, .
\end{equation}

Besides canonical kinetic terms, the new vector-like fermions have
Dirac mass terms,
\begin{equation} \label{eq:VectorMass}
\mathcal L_m = m_Q \overline Q Q + m_L \overline L L \, ,
\end{equation}
as well as Yukawa couplings with the SM fermions
\begin{equation} \label{eq:VectorYukawa}
\mathcal L_Y = \lambda_Q \, \overline{Q_R} \, \phi \, q_L + \lambda_L \, \overline{L_R} \, \phi \, \ell_L + \hc \, ,
\end{equation}
where $\lambda_Q$ and $\lambda_L$ are $3$ component vectors. The
scalar potential takes the form
\begin{equation}
\mathcal V = \mathcal V_{\text{SM}} + \mathcal V\left( H , \phi , \chi \right) + \mathcal V\left( \phi , \chi \right) \, .
\end{equation}
Here $\mathcal V_{\text{SM}}$ is the SM scalar potential. The pieces
involving the $U(1)_X$ charged scalars are
\begin{equation}
\mathcal V\left( H , \phi , \chi \right) = \lambda_{H \phi} \, |H|^2 |\phi|^2 + \lambda_{H \chi} \, |H|^2 |\chi|^2
\end{equation}
and
\begin{eqnarray}
\mathcal V\left( \phi , \chi \right) &=& m_\phi^2 |\phi|^2 + m_\chi^2 |\chi|^2 + \frac{\lambda_\phi}{2} |\phi|^4 + \frac{\lambda_\chi}{2} |\chi|^4 \nonumber \\
&& + \lambda_{\phi \chi} \, |\phi|^2 |\chi|^2 + \left( \mu \, \phi \chi^2 + \text{h.c.} \right) \, .
\end{eqnarray}
We will assume that the scalar potential is such that only the
standard Higgs boson and the $\phi$ field acquire non-zero vacuum
expectation values,
\begin{equation}
\langle H^0 \rangle = \frac{v}{\sqrt{2}} \, , \qquad \langle \phi \rangle = \frac{v_\phi}{\sqrt{2}} \, .
\end{equation}
Therefore, the $\phi$ field will be responsible for the spontaneous
breaking of $U(1)_X$, which in turn results into a new massive gauge
boson, the $Z^\prime$ boson, with $m_{Z^\prime} = 2 g_X v_\phi$, a
mixed state of the neutral gauge bosons $B$, $W_3$ and
$B_X$. Moreover, the breaking of $U(1)_X$ also induces mixings between
the vector-like fermions and their SM counterparts thanks to the
Yukawa interactions in Eq. \eqref{eq:VectorYukawa}. And finally, after
spontaneous symmetry breaking, the resulting Lagrangian contains a
remnant $\mathbb{Z}_2$ symmetry, under which $\chi$ is odd and all the
other fields are even. Therefore, $\chi$ is a stable neutral scalar,
and thus a potentially valid DM candidate. It is worth noting that the
mechanism to stabilize the DM particle does not introduce additional
ad-hoc symmetries, but simply makes use of the original $U(1)_X$
symmetry of the model. This goal has been achieved by breaking the
continuous $U(1)_X$ symmetry to a remnant $\mathbb{Z}_2$, something
that can be easily accomplished with a proper choice of $U(1)_X$
charges.

Before concluding our review of the model we must comment on $U(1)$
mixing. It is well known that nothing prevents $U(1)$ factors from
mixing. In the model under consideration, this would be given by the
Lagrangian term
\begin{equation}
\mathcal L \supset \varepsilon \, F_{\mu \nu}^Y F^{\mu \nu}_X \, ,
\end{equation}
where $F_{\mu \nu}^{X,Y}$ are the usual field strength tensors for the
$U(1)_{X,Y}$ groups. In the presence of a non-zero $\varepsilon$
coupling, kinetic mixing between the $U(1)_X$ and $U(1)_Y$ gauge
bosons is induced.

\section{Installing {\tt ROOT}}
\label{sec:root}

In this Appendix we explain how to install {\tt ROOT}. The source code
can be downloaded from

\begin{center}
\url{https://root.cern.ch/downloading-root}
\end{center}

Once downloaded, we must untar the file and compile the code. For this purpose w will need to make use the {\tt CMake} interface. These are the steps:

\vsp
\begin{lstlisting}[style=terminal]
$ cp Download-Directory/root_vX.source.tar.gz $PATH/
$ cd $PATH
$ tar -xf root_vX.source.tar.gz
$ mkdir <builddir>
$ cd <builddir>
$ cmake ../root-X/
$ cmake --build
\end{lstlisting}
\vsp

Here {\tt X} is the {\tt ROOT} version we have downloaded and {\tt
<builddir>} is just the name, to be chosen by the user, for the
directory that we will use to build {\tt ROOT}. In case your computer
has more than one core you can replace the last command by

\vsp
\begin{lstlisting}[style=terminal]
$ cmake --build . -- -jn
\end{lstlisting}
\vsp

where {\tt n} is the number of cores in your computer. The compilation
of {\tt ROOT} can take a while, so we better find something to do
while waiting. Finally, when the compilation finished, we must tell
our system how to find {\tt ROOT} and its libraries. Depending on the
system, this must be done differently. For example, in systems with
Bash shell (such as most Linux distributions), this is done by adding
the lines

\vsp
\begin{lstlisting}[style=bashrc]
source $PATH/<builddir>/bin/thisroot.sh
\end{lstlisting}
\vsp

at the end of the {\tt \textasciitilde{}/.bashrc} file. Here {\tt
<builddir>} must be replaced by the name of the folder where {\tt
ROOT} was built. \\

For more information on how to install {\tt ROOT}:

\begin{center}
\url{https://root.cern.ch/building-root}
\end{center}

\section{\SARAH model files for the scotogenic model}
\label{sec:SARAH-scotogenic}

\begin{lstlisting}[style=scotogenic]
Off[General::spell]

Model`Name      = "Scotogenic";
Model`NameLaTeX = "Scotogenic Model";
Model`Authors   = "N. Rojas, A. Vicente";
Model`Date      = "2015-04-28";

(* "28-04-2015 (first implementation)" *)
(* "25-05-2015 (removed mixings in scalar sector)" *)
(* "10-06-2015 (fixed conventions)" *)

(*------------Particle Content---------------*)

(* Global Symmetries *)
Global[[1]] = {Z[2], Z2};

(*--------------Gauge Groups-----------------*)
Gauge[[1]]={B,   U[1], hypercharge, g1, False, 1};
Gauge[[2]]={WB, SU[2], left,        g2, True , 1};
Gauge[[3]]={G,  SU[3], color,       g3, False, 1};

(*--------------Matter Fields----------------*)
FermionFields[[1]] = {q , 3, {uL, dL},     1/6, 2,  3, 1};
FermionFields[[2]] = {l , 3, {vL, eL},    -1/2, 2,  1, 1};
FermionFields[[3]] = {d , 3, conj[dR],     1/3, 1, -3, 1};
FermionFields[[4]] = {u , 3, conj[uR],    -2/3, 1, -3, 1};
FermionFields[[5]] = {e , 3, conj[eR],       1, 1,  1, 1};
FermionFields[[6]] = {n , 3, conj[nR],       0, 1,  1,-1};

ScalarFields[[1]] =  {H,  1, {Hp, H0},     1/2, 2,  1,  1};
ScalarFields[[2]] =  {Et, 1, {etp,et0},    1/2, 2,  1, -1};

(*---------------DEFINITION------------------*)

NameOfStates={GaugeES, EWSB};

(* ----- Before EWSB ----- *)

DEFINITION[GaugeES][LagrangianInput]= 
{
  {LagFer   ,      {AddHC->True}},
  {LagNV    ,      {AddHC->True}},
  {LagH     ,      {AddHC->False}},
  {LagEt    ,      {AddHC->False}},
  {LagHEt   ,      {AddHC->False}},
  {LagHEtHC ,      {AddHC->True}}
};

LagFer   = Yd conj[H].d.q + Ye conj[H].e.l + Yu H.u.q + Yn Et.n.l;
LagNV    = Mn/2 n.n;
LagH     = -(- mH2 conj[H].H     + 1/2 lambda1 conj[H].H.conj[H].H );
LagEt    = -(+ mEt2 conj[Et].Et  + 1/2 lambda2 conj[Et].Et.conj[Et].Et );
LagHEt   = -(+ lambda3 conj[H].H.conj[Et].Et + lambda4 conj[H].Et.conj[Et].H );
LagHEtHC = -(+ 1/2 lambda5 conj[H].Et.conj[H].Et );

(* Gauge Sector *)

DEFINITION[EWSB][GaugeSector] =
{ 
  {{VB,VWB[3]},{VP,VZ},ZZ},
  {{VWB[1],VWB[2]},{VWp,conj[VWp]},ZW}
};

(* ----- VEVs ---- *)

DEFINITION[EWSB][VEVs]= 
{
  {H0,  {v, 1/Sqrt[2]}, {Ah, \[ImaginaryI]/Sqrt[2]}, {hh, 1/Sqrt[2]}},
  {et0, {0, 0}, {etI, \[ImaginaryI]/Sqrt[2]}, {etR, 1/Sqrt[2]}}
};

DEFINITION[EWSB][MatterSector]=
{
  {{conj[nR]},{X0, ZX}},
  {{vL}, {VL, Vv}},
  {{{dL}, {conj[dR]}}, {{DL,Vd}, {DR,Ud}}},
  {{{uL}, {conj[uR]}}, {{UL,Vu}, {UR,Uu}}},
  {{{eL}, {conj[eR]}}, {{EL,Ve}, {ER,Ue}}}
};

(*------------------------------------------------------*)
(* Dirac-Spinors *)
(*------------------------------------------------------*)

DEFINITION[EWSB][DiracSpinors]=
{
  Fd  -> {  DL, conj[DR]},
  Fe  -> {  EL, conj[ER]},
  Fu  -> {  UL, conj[UR]},
  Fv  -> {  VL, conj[VL]},
  Chi -> {  X0, conj[X0] }
};

DEFINITION[EWSB][GaugeES]=
{
  Fd1 ->{  FdL, 0},
  Fd2 ->{  0, FdR},
  Fu1 ->{  Fu1, 0},
  Fu2 ->{  0, Fu2},
  Fe1 ->{  Fe1, 0},
  Fe2 ->{  0, Fe2}
};
\end{lstlisting}

\begin{lstlisting}[style=parameters]
(* ::Package:: *)

ParameterDefinitions = { 

{g1,        { Description -> "Hypercharge-Coupling"}},
{g2,        { Description -> "Left-Coupling"}},
{g3,        { Description -> "Strong-Coupling"}},    

{AlphaS,    {Description -> "Alpha Strong"}},	
{e,         { Description -> "electric charge"}}, 
{Gf,        { Description -> "Fermi's constant"}},
{aEWinv,    { Description -> "inverse weak coupling constant at mZ"}},

{Yu,        { Description -> "Up-Yukawa-Coupling",
              DependenceNum ->  Sqrt[2]/v*{ {Mass[Fu,1],0,0},
					    {0,Mass[Fu,2],0},
					    {0,0,Mass[Fu,3]}}}}, 
{Yd,        { Description -> "Down-Yukawa-Coupling",
              DependenceNum ->  Sqrt[2]/v* {{Mass[Fd,1],0,0},
					    {0, Mass[Fd,2],0},
					    {0, 0, Mass[Fd,3]}}}},       									
{Ye,        { Description -> "Lepton-Yukawa-Coupling",
			  DependenceNum ->  Sqrt[2]/v* {{Mass[Fe,1],0,0},
             						{0, Mass[Fe,2],0},
             						{0, 0, Mass[Fe,3]}}}}, 
                                                                            
{ThetaW,    { Description -> "Weinberg-Angle",
              DependenceNum -> ArcSin[Sqrt[1 - Mass[VWp]^2/Mass[VZ]^2]]}},

{ZZ, {Description -> "Photon-Z Mixing Matrix"}},
{ZW, {Description -> "W Mixing Matrix", Dependence -> 1/Sqrt[2] {{1, 1},{I,-I}} }},
          
{Vu,        {Description ->"Left-Up-Mixing-Matrix"}},
{Vd,        {Description ->"Left-Down-Mixing-Matrix"}},
{Uu,        {Description ->"Right-Up-Mixing-Matrix"}},
{Ud,        {Description ->"Right-Down-Mixing-Matrix"}}, 
{Ve,        {Description ->"Left-Lepton-Mixing-Matrix"}},
{Ue,        {Description ->"Right-Lepton-Mixing-Matrix"}},

(* Scalar sector *)

{v,          { Description -> "EW-VEV",
               DependenceNum -> Sqrt[4*Mass[VWp]^2/(g2^2)],
               DependenceSPheno -> None  }},

{mH2,        { Description -> "SM Higgs Mass Parameter"}},

{mEt2, {LaTeX -> "m_\\eta^2",
	LesHouches -> {HDM,1},
	OutputName-> mEt2 }},

{lambda1,   {LaTeX -> "\\lambda_1",
	     LesHouches -> {HDM,2},
	     OutputName-> lam1 }},

{lambda2,   {LaTeX -> "\\lambda_2",
	     LesHouches -> {HDM,3},
	     OutputName-> lam2 }},

{lambda3,   {LaTeX -> "\\lambda_3",
	     LesHouches -> {HDM,4},
	     OutputName-> lam3 }},

{lambda4,   {LaTeX -> "\\lambda_4",
	     LesHouches -> {HDM,5},
	     OutputName-> lam4 }},

{lambda5,   {Real -> True,
	     LaTeX -> "\\lambda_5",
	     LesHouches -> {HDM,6},
	     OutputName-> lam5 }},

(* Fermion sector *)

{Yn,   {LaTeX -> "Y_N",
	LesHouches -> YN,
	OutputName->Yn }},

{Mn,   {LaTeX -> "M_N",
	LesHouches -> MN,
	OutputName->Mn }},

{ZX, {LaTeX -> "Z^{\\chi^0}",
      LesHouches -> ZXMIX,
      OutputName -> ZX }},

{Vv, {Description ->"Neutrino-Mixing-Matrix"}}

};
\end{lstlisting}

\begin{lstlisting}[style=particles]
(* ::Package:: *)
ParticleDefinitions[GaugeES] = {

      {H0,  {    PDG -> {0},
                 Width -> 0, 
                 Mass -> Automatic,
                 FeynArtsNr -> 1,
                 LaTeX -> "H^0",
                 OutputName -> "H0" }},

      {Hp,  {    PDG -> {0},
                 Width -> 0, 
                 Mass -> Automatic,
                 FeynArtsNr -> 2,
                 LaTeX -> "H^+",
                 OutputName -> "Hp" }}, 

      {et0, {    PDG -> {0},
                 Width -> 0, 
                 Mass -> Automatic,
                 LaTeX -> "\\eta^0",
                 OutputName -> "et0" }},

      {etp, {    PDG -> {0},
                 Width -> 0, 
                 Mass -> Automatic,
                 LaTeX -> "\\eta^+",
                 OutputName -> "etp" }}, 

      {VB,   { Description -> "B-Boson"}},
      {VG,   { Description -> "Gluon"}},
      {VWB,  { Description -> "W-Bosons"}},
      {gB,   { Description -> "B-Boson Ghost"}},
      {gG,   { Description -> "Gluon Ghost" }},
      {gWB,  { Description -> "W-Boson Ghost"}}

      };

  ParticleDefinitions[EWSB] = {

      {hh   ,  {  Description -> "Higgs",
                  PDG -> {25},
                  PDG.IX -> {101000001},
	          Mass -> Automatic }},                  
      {Ah   ,  {  Description -> "Pseudo-Scalar Higgs",
                  PDG -> {0},
                  PDG.IX ->{0},
                  Mass -> {0},
                  Width -> {0} }}, 
      {Hp,     {  Description -> "Charged Higgs", 
                  PDG -> {0},
                  PDG.IX ->{0},
                  Width -> {0}, 
                  Mass -> {0},
                  LaTeX -> {"H^+","H^-"},
                  OutputName -> {"Hp","Hm"} }},

      {etR,   {  Description -> "CP-even eta scalar", 
		 PDG -> {1001},
		 Mass -> LesHouches,
		 ElectricCharge -> 0,
		 LaTeX -> "\\eta_R",
		 OutputName -> "etR" }}, 
      {etI,   {  Description -> "CP-odd eta scalar", 
		 PDG -> {1002},
		 Mass -> LesHouches,
		 ElectricCharge -> 0,
		 LaTeX -> "\\eta_I",
		 OutputName -> "etI" }}, 
      {etp,   {  Description -> "Charged eta scalar", 
		 PDG -> {1003},
		 Mass -> LesHouches,
		 ElectricCharge -> 1,
		 LaTeX -> "\\eta^+",
                 OutputName -> "etp" }}, 

      {VP,   { Description -> "Photon"}}, 
      {VZ,   { Description -> "Z-Boson", Goldstone -> Ah }}, 
      {VWp,  { Description -> "W+ - Boson", Goldstone -> Hp}},
      {VG,   { Description -> "Gluon" }},

      {gP,   { Description -> "Photon Ghost"}},
      {gWp,  { Description -> "Positive W+ - Boson Ghost"}}, 
      {gWpC, { Description -> "Negative W+ - Boson Ghost" }}, 
      {gZ,   { Description -> "Z-Boson Ghost" }},
      {gG,   { Description -> "Gluon Ghost" }},
                 
      {Fd,   { Description -> "Down-Quarks"}},
      {Fu,   { Description -> "Up-Quarks"}},
      {Fe,   { Description -> "Leptons" }},
      {Fv,   { Description -> "Neutrinos" }},
      {Chi,  { Description -> "Singlet Fermions",
	       PDG -> {1012,1014,1016},
	       Mass -> LesHouches,
	       ElectricCharge -> 0,
	       LaTeX -> "N",
	       OutputName -> "N" }}

};

WeylFermionAndIndermediate =
{
   {H,      {LaTeX -> "H"}},
   {Et,     {LaTeX -> "\\eta"}},
   {dR,     {LaTeX -> "d_R" }},
   {eR,     {LaTeX -> "e_R" }},
   {lep,    {LaTeX -> "l" }},
   {uR,     {LaTeX -> "u_R" }},
   {q,      {LaTeX -> "q" }},
   {eL,     {LaTeX -> "e_L" }},
   {dL,     {LaTeX -> "d_L" }},
   {uL,     {LaTeX -> "u_L" }},
   {vL,     {LaTeX -> "\\nu_L" }},
   {DR,     {LaTeX -> "D_R" }},
   {ER,     {LaTeX -> "E_R" }},
   {UR,     {LaTeX -> "U_R" }},
   {EL,     {LaTeX -> "E_L" }},
   {DL,     {LaTeX -> "D_L" }},
   {UL,     {LaTeX -> "U_L" }},
   {X0,     {LaTeX -> "X^0"}},
   {VL,     {LaTeX -> "V_L" }},
   {n,      {LaTeX -> "N" }},
   {nR,     {LaTeX -> "\\nu_R" }}
};
\end{lstlisting}

\begin{lstlisting}[style=SPheno]
OnlyLowEnergySPheno = True;

MINPAR={
  {1,lambda1Input},
  {2,lambda2Input},
  {3,lambda3Input},
  {4,lambda4Input},
  {5,lambda5Input},
  {6,mEt2Input}
};

ParametersToSolveTadpoles = {mH2};

BoundaryLowScaleInput={
  {lambda1,lambda1Input},
  {lambda2,lambda2Input},
  {lambda3,lambda3Input},
  {lambda4,lambda4Input},
  {lambda5,lambda5Input},
  {mEt2,mEt2Input},
  {Yn, LHInput[Yn]},
  {Mn, LHInput[Mn]}
};

DEFINITION[MatchingConditions]= 
{{v, vSM}, 
 {Ye, YeSM},
 {Yd, YdSM},
 {Yu, YuSM},
 {g1, g1SM},
 {g2, g2SM},
 {g3, g3SM}};

ListDecayParticles = {Fu,Fe,Fd,Fv,VZ,VWp,hh,etR,etI,etp,Chi};
ListDecayParticles3B = {{Fu,"Fu.f90"},{Fe,"Fe.f90"},{Fd,"Fd.f90"}};
\end{lstlisting}

\section{\SARAH model files for the DarkBS model}
\label{sec:SARAH-darkbs}

\begin{lstlisting}[style=darkbs]
Off[General::spell]

Model`Name = "DarkBS";
Model`NameLaTeX ="DarkBS";
Model`Authors = "D. Sierra, F.Staub, A.Vicente";
Model`Date = "2015-03-11";

(*-------------------------------------------*)
(*   Particle Content*)
(*-------------------------------------------*)

(* Global Symmetries *)
Global[[1]] = {Z[2], Z2};

(* Gauge Groups *)

Gauge[[1]]={B,   U[1], hypercharge, g1,False,1};
Gauge[[2]]={WB, SU[2], left,        g2,True,1};
Gauge[[3]]={G,  SU[3], color,       g3,False,1};
Gauge[[4]]={Bp,  U[1], Uchi,        gX,False,1};

(* Matter Fields *)

FermionFields[[1]] = {q, 3, {uL, dL},     1/6, 2,  3, 0, 1};  
FermionFields[[2]] = {l, 3, {vL, eL},    -1/2, 2,  1, 0, 1};
FermionFields[[3]] = {d, 3, conj[dR],     1/3, 1, -3, 0, 1};
FermionFields[[4]] = {u, 3, conj[uR],    -2/3, 1, -3, 0, 1};
FermionFields[[5]] = {e, 3, conj[eR],       1, 1,  1, 0, 1};

FermionFields[[6]] = {lL, 1, {v4, e4},    -1/2, 2,  1,  2, 1};
FermionFields[[7]] = {lR, 1, {e5, v5},     1/2, 2,  1, -2, 1};
FermionFields[[8]] = {qL, 1, {u4, d4},     1/6, 2,  3,  2, 1};
FermionFields[[9]] = {qR, 1, {d5, u5},    -1/6, 2, -3, -2, 1};    

ScalarFields[[1]] =  {H, 1, {Hp, H0},   1/2, 2,  1,  0,  1};
ScalarFields[[2]] =  {Phi, 1, phi,        0, 1,  1,  2,  1};
ScalarFields[[3]] =  {Chi, 1, chi,        0, 1,  1, -1, -1};

(*----------------------------------------------*)
(*   DEFINITION                                 *)
(*----------------------------------------------*)

NameOfStates={GaugeES, EWSB};

(* ----- Before EWSB ----- *)

DEFINITION[GaugeES][LagrangianInput]= {
	{LagHC, {AddHC->True}},
	{LagNoHC,{AddHC->False}}
};

LagNoHC = -mH2 conj[H].H - mPhi2 Phi.conj[Phi] - mChi2 Chi.conj[Chi] - 1/2 \[Lambda] conj[H].H.conj[H].H \
-1/2 LamP Phi.conj[Phi].Phi.conj[Phi] -1/2 LamC Chi.conj[Chi].Chi.conj[Chi] \
- LamCP conj[Phi].Phi.conj[Chi].Chi - LamHP conj[H].H.conj[Phi].Phi - LamHC conj[H].H.conj[Chi].Chi;
LagHC =  -(Yd conj[H].d.q + Ye conj[H].e.l + Yu H.u.q + mQ qL.qR + mL lL.lR + lamQ Phi.qR.q + lamL Phi.lR.l );

(* Gauge Sector *)

DEFINITION[EWSB][GaugeSector] =
{ 
  {{VB,VWB[3],VBp},{VP,VZ,VZp},ZZ},
  {{VWB[1],VWB[2]},{VWp,conj[VWp]},ZW}
};

(* ----- VEVs ---- *)

DEFINITION[EWSB][VEVs]= 
{    {H0, {v, 1/Sqrt[2]}, {sigmaH, \[ImaginaryI]/Sqrt[2]},{phiH, 1/Sqrt[2]}},
     {phi, {vP, 1/Sqrt[2]}, {sigmaP, \[ImaginaryI]/Sqrt[2]},{phiP, 1/Sqrt[2]}}      };

DEFINITION[EWSB][MatterSector]=   {
    {{phiH,phiP},{hh,ZH}},
    {{sigmaH,sigmaP},{Ah,ZA}},
    {{{dL,d4}, {conj[dR],d5}}, {{DL,Vd}, {DR,Ud}}},
    {{{uL,u4}, {conj[uR],u5}}, {{UL,Vu}, {UR,Uu}}},
    {{{eL,e4}, {conj[eR],e5}}, {{EL,Ve}, {ER,Ue}}},
    {{vL,v4,v5},{VL,UV}}
};

(*------------------------------------------------------*)
(* Dirac-Spinors *)
(*------------------------------------------------------*)

DEFINITION[EWSB][DiracSpinors]={
 Fd ->{  DL, conj[DR]},
 Fe ->{  EL, conj[ER]},
 Fu ->{  UL, conj[UR]},
 Fv ->{  VL, conj[VL]}};

DEFINITION[EWSB][GaugeES]={
 Fd1 ->{  FdL, 0},
 Fd2 ->{  0, FdR},
 Fu1 ->{  Fu1, 0},
 Fu2 ->{  0, Fu2},
 Fe1 ->{  Fe1, 0},
 Fe2 ->{  0, Fe2}};
\end{lstlisting}

\begin{lstlisting}[style=parameters]
ParameterDefinitions = { 

{g1,        { Description -> "Hypercharge-Coupling"}},
{g2,        { Description -> "Left-Coupling"}},
{g3,        { Description -> "Strong-Coupling"}},

{gX,       {LaTeX -> "g_X",
             LesHouches -> {GAUGE,4},
             OutputName -> gX}},

{g1X,       {LaTeX -> "\\tilde{g}",
             LesHouches -> {GAUGE,10},
             OutputName -> g1X}},
{gX1,       {LaTeX -> "\\bar{g}",
             LesHouches -> {GAUGE,11},
             OutputName -> gX1}},

{AlphaS,    {Description -> "Alpha Strong"}},	
{e,         { Description -> "electric charge"}}, 

{Gf,        { Description -> "Fermi's constant"}},
{aEWinv,    { Description -> "inverse weak coupling constant at mZ"}},

{Yu,        { Description -> "Up-Yukawa-Coupling",
	      DependenceNum ->  Sqrt[2]/v* {{Mass[Fu,1],0,0},
             				    {0, Mass[Fu,2],0},
             				    {0, 0, Mass[Fu,3]}}}}, 
             									
{Yd,        { Description -> "Down-Yukawa-Coupling",
	      DependenceNum ->  Sqrt[2]/v* {{Mass[Fd,1],0,0},
					    {0, Mass[Fd,2],0},
					    {0, 0, Mass[Fd,3]}}}},
             									
{Ye,        { Description -> "Lepton-Yukawa-Coupling",
	      DependenceNum ->  Sqrt[2]/v* {{Mass[Fe,1],0,0},
					    {0, Mass[Fe,2],0},
					    {0, 0, Mass[Fe,3]}}}}, 
                                                                            
{\[Lambda],  { Description -> "SM Higgs Selfcouplings",
               DependenceNum -> Mass[hh]^2/(2 v^2)}},
{v,          { Description -> "EW-VEV",
               DependenceNum -> Sqrt[4*Mass[VWp]^2/(g2^2)],
               DependenceSPheno -> None  }},

{vP,   {LaTeX ->"v_\\phi",
        OutputName -> vP,
        LesHouches -> {DBS,20}}},

{mPhi2, {LaTeX -> "m_{\\phi}^2",
         OutputName->mPhi2,
         LesHouches ->  {DBS,1}}},

{mChi2, {LaTeX -> "m_{\\chi}^2",
         OutputName->mX2,
         LesHouches ->  {DBS,2}}},

{mQ, {LaTeX -> "m_Q",
      OutputName->mQ,
      LesHouches ->  {DBS,3}}},

{mL, {LaTeX -> "m_L",
      OutputName->mL,
      LesHouches ->  {DBS,4}}},

{LamP, {LaTeX -> "\\lambda_{\\phi}",
        OutputName->LamP,
        LesHouches ->  {DBS,10}}},

{LamC, {LaTeX -> "\\lambda_{\\chi}",
        OutputName->LamC,
        LesHouches ->  {DBS,11}}},

{LamCP, {LaTeX -> "\\lambda_{\\phi\\chi}",
         OutputName->LamCP,
         LesHouches ->  {DBS,12}}},

{LamHP, {LaTeX -> "\\lambda_{H\\phi}",
         OutputName->LamHP,
         LesHouches ->  {DBS,13}}},

{LamHC, {LaTeX -> "\\lambda_{H\\chi}",
         OutputName->LamHC,
         LesHouches ->  {DBS,14}}},

{lamQ, {LaTeX -> "\\lambda_Q",
        OutputName->lamQ,
        LesHouches ->  LAMQ}},

{lamL, {LaTeX -> "\\lambda_L",
        OutputName->lamL,
        LesHouches ->  LAML}},

{mH2,       { Description -> "SM Higgs Mass Parameter"}},

{ThetaW,    { Description -> "Weinberg-Angle",
              DependenceNum -> ArcSin[Sqrt[1 - Mass[VWp]^2/Mass[VZ]^2]]  }},

{ThetaWp,  { LaTeX -> "{\\Theta'}_W",
             Real ->True,
             DependenceSPheno -> ArcCos[Abs[ZZ[3,3]]],
             OutputName-> TWp,
             LesHouches -> {ANGLES,10}      }},

{ZH, { Description->"Scalar-Mixing-Matrix", 
       LaTeX -> "Z^H",
       Real -> True, 
       DependenceOptional ->   {{-Sin[\[Alpha]],Cos[\[Alpha]]},
                                {Cos[\[Alpha]],Sin[\[Alpha]]}}, 
       Value -> None, 
       LesHouches -> SCALARMIX,
       OutputName-> ZH     }},
             
{ZA, { Description->"Pseudo-Scalar-Mixing-Matrix", 
       LaTeX -> "Z^A",
       Real -> True,
       DependenceOptional -> {{-Cos[\[Beta]],Sin[\[Beta]]},
                              {Sin[\[Beta]],Cos[\[Beta]]}}, 
       Value -> None, 
       LesHouches -> PSEUDOSCALARMIX,
       OutputName-> ZA      }},

{\[Beta],   { Description -> "Pseudo Scalar mixing angle",
              DependenceSPheno -> ArcSin[Abs[ZH[1,2]]]  }},             

{\[Alpha],  { Description -> "Scalar mixing angle" }},  

{UV, { Description ->"Neutrino-Mixing-Matrix", 
       LaTeX -> "U^V",
       Dependence ->  None, 
       Value -> None, 
       LesHouches -> UVMIX,
       OutputName-> UV      }},

{ZZ, { Description -> "Photon-Z Mixing Matrix",
       Dependence->  {{Cos[ThetaW],-Sin[ThetaW] Cos[ThetaWp], Sin[ThetaW] Sin[ThetaWp]},                             
                      {Sin[ThetaW],Cos[ThetaW] Cos[ThetaWp],-Cos[ThetaW] Sin[ThetaWp]},
                      {0, Sin[ThetaWp], Cos[ThetaWp]}} }},

{ZW, { Description -> "W Mixing Matrix",
       Dependence ->   1/Sqrt[2] {{1, 1},
                                  {\[ImaginaryI],-\[ImaginaryI]}} }},

{Vu,        {Description ->"Left-Up-Mixing-Matrix"}},
{Vd,        {Description ->"Left-Down-Mixing-Matrix"}},
{Uu,        {Description ->"Right-Up-Mixing-Matrix"}},
{Ud,        {Description ->"Right-Down-Mixing-Matrix"}}, 
{Ve,        {Description ->"Left-Lepton-Mixing-Matrix"}},
{Ue,        {Description ->"Right-Lepton-Mixing-Matrix"}}

 }; 
\end{lstlisting}

\begin{lstlisting}[style=particles]
ParticleDefinitions[GaugeES] = {
      {H0,  {    PDG -> {0},
                 Width -> 0, 
                 Mass -> Automatic,
                 FeynArtsNr -> 1,
                 LaTeX -> "H^0",
                 OutputName -> "H0" }},

      {Hp,  {    PDG -> {0},
                 Width -> 0, 
                 Mass -> Automatic,
                 FeynArtsNr -> 2,
                 LaTeX -> "H^+",
                 OutputName -> "Hp" }},

      {VB,   { Description -> "B-Boson"}},                                                   
      {VG,   { Description -> "Gluon"}},          
      {VWB,  { Description -> "W-Bosons"}},          
      {gB,   { Description -> "B-Boson Ghost"}},                                                   
      {gG,   { Description -> "Gluon Ghost" }},          
      {gWB,  { Description -> "W-Boson Ghost"}}
      
      };
      
ParticleDefinitions[EWSB] = {

    {hh   ,  {  Description -> "Higgs",
                PDG -> {25,35},
                PDG.IX -> {101000001,101000002} }}, 
                 
    {Ah   ,  {  Description -> "Pseudo-Scalar Higgs",
                PDG -> {0,0},
                PDG.IX ->{0,0},
                Mass -> {0,0},
                Width -> {0,0} }},

     {Hp,     { Description -> "Charged Higgs", 
                PDG -> {0},
                PDG.IX ->{0},
                Width -> {0}, 
                Mass -> {0},
                LaTeX -> {"H^+","H^-"},
                OutputName -> {"Hp","Hm"} }},                                                   
      
      {VP,   { Description -> "Photon"}}, 
      {VZ,   { Description -> "Z-Boson",
      	       Goldstone -> Ah[{1}] }}, 
      {VZp,   { Description -> "Z'-Boson",
      	        Goldstone -> Ah[{2}] }}, 
      {VG,   { Description -> "Gluon" }},          
      {VWp,  { Description -> "W+ - Boson",
      	       Goldstone -> Hp }},         
      {gP,   { Description -> "Photon Ghost"}},                                                   
      {gWp,  { Description -> "Positive W+ - Boson Ghost"}}, 
      {gWpC, { Description -> "Negative W+ - Boson Ghost" }}, 
      {gZ,   { Description -> "Z-Boson Ghost" }},
      {gZp,  { Description -> "Z'-Ghost" }},
      {gG,   { Description -> "Gluon Ghost" }},      

      {chi, { LaTeX -> "\\chi",
              PDG -> {50},
              OutputName -> "chi",
              ElectricCharge->0}}, 
                 
      {Fd,   { Description -> "Down-Quarks",
               PDG -> {1,3,5,7},
               PDG.IX->{-110890201,-110890202,-110890203,-110890204} }},   
      {Fu,   { Description -> "Up-Quarks",
               PDG -> {2,4,6,8},
               PDG.IX->{110100401,110100402,110100403,110100404} }},   
      {Fe,   { Description -> "Leptons",
               PDG -> {11,13,15,17},
               PDG.IX -> {-110000601,-110000602,-110000603,-110000604} }},
      {Fv,   { Description -> "Neutrinos",
               PDG -> {12,14,16,18,20},
               PDG.IX ->{-110000001,-110000002,-110000003,-110000004,-110000005} }}                                                              
     
};

WeylFermionAndIndermediate = {
     
    {H,      {   PDG -> {0},
                 Width -> 0, 
                 Mass -> Automatic,
                 LaTeX -> "H",
                 OutputName -> "" }},

   {dR,     {LaTeX -> "d_R" }},
   {eR,     {LaTeX -> "e_R" }},
   {lep,    {LaTeX -> "l" }},
   {uR,     {LaTeX -> "u_R" }},
   {q,      {LaTeX -> "q" }},
   {eL,     {LaTeX -> "e_L" }},
   {dL,     {LaTeX -> "d_L" }},
   {uL,     {LaTeX -> "u_L" }},
   {vL,     {LaTeX -> "\\nu_L" }},

   {DR,     {LaTeX -> "D_R" }},
   {ER,     {LaTeX -> "E_R" }},
   {UR,     {LaTeX -> "U_R" }},
   {EL,     {LaTeX -> "E_L" }},
   {DL,     {LaTeX -> "D_L" }},
   {UL,     {LaTeX -> "U_L" }}
};
\end{lstlisting}

\begin{lstlisting}[style=SPheno]
OnlyLowEnergySPheno = True;

MINPAR={
  {1,LambdaInput},
  {2,LPInput},
  {3,LCInput},
  {4,LCPInput},
  {5,LHPInput},
  {6,LHCInput},
  {10, mChi2Input},
  {11, mQInput},
  {12, mLInput},
  {20, gXInput},
  {21, MZpMass}
};

ParametersToSolveTadpoles = {mH2, mPhi2};

BoundaryLowScaleInput={
  {lamQ,      LHInput[lamQ]},
  {lamL,      LHInput[lamL]},
  {\[Lambda], LambdaInput},
  {LamP,   LPInput},
  {LamC,   LCInput},
  {LamCP,  LCPInput},
  {LamHP,  LHPInput},
  {LamHC,  LHCInput},
  {mChi2,  mChi2Input},
  {mQ,     mQInput},
  {mL,     mLInput},
  {gX,     gXInput},
  {g1X, 0},
  {gX1, 0},
  {vP, MZpMass/(2*gX)}
};

DEFINITION[MatchingConditions]= 
{{v, vSM}, 
 {Ye, YeSM},
 {Yd, YdSM},
 {Yu, YuSM},
 {g1, g1SM},
 {g2, g2SM},
 {g3, g3SM}};

ListDecayParticles = {Fu, Fe, Fd, hh, VZp};
ListDecayParticles3B = {{Fu,"Fu.f90"},{Fe,"Fe.f90"},{Fd,"Fd.f90"}};
\end{lstlisting}